\documentclass[onecolumn,notitlepage,pra,showpacs,superscriptaddress,amssymb,amsmath,amsmath]{revtex4-1}

\usepackage{graphicx}
\usepackage{epstopdf}
\usepackage{bm}
\usepackage{hyperref}
\usepackage{comment}
\usepackage{braket}
\usepackage{xfrac}

\usepackage[normalem]{ulem}

\newcommand*{\beq}{\begin{equation}}
\newcommand*{\eeq}{\end{equation}}
\newcommand*{\lp}{\left(}
\newcommand*{\rp}{\right)}

\newcommand*{\half}{\frac{1}{2}}
\newcommand*{\benu}{\begin{enumerate}}
\newcommand*{\eenu}{\end{enumerate}}

\newcommand{\A}{\mathcal{A}}

\hypersetup{%
   pdfpagemode=None, 
   pdfstartpage=1,
   pdfmenubar=true,
   pdftoolbar=true,
   colorlinks = true,
   linkcolor=blue,
   citecolor=blue,
   urlcolor=blue,
   bookmarksopen=false
 }

\begin{document}
\title{Many interacting fermions in a one-dimensional harmonic trap:\\ a quantum-chemical treatment}

\author{Tomasz Grining}
\address{Faculty of Chemistry, University of Warsaw, Pasteura 1, 02-093 Warsaw, Poland}
\author{Micha\l~Tomza}
\affiliation{ICFO-Institut de Ci\`encies Fot\`oniques, The Barcelona Institute of Science and Technology, Av.~Carl Friedrich Gauss 3, 08860 Castelldefels (Barcelona), Spain}
\author{Micha\l~Lesiuk}
\address{Faculty of Chemistry, University of Warsaw, Pasteura 1, 02-093 Warsaw, Poland}   
\author{Micha\l~Przybytek}
\address{Faculty of Chemistry, University of Warsaw, Pasteura 1, 02-093 Warsaw, Poland}
\author{Monika~Musia{\l}}
\address{Institute of Chemistry, University of Silesia, Szkolna 9, 40-006 Katowice, Poland}
\author{Pietro Massignan}
\affiliation{ICFO-Institut de Ciencies Fotoniques, Barcelona Institute of Science and Technology, Av.~Carl Friedrich Gauss 3, 08860 Castelldefels (Barcelona), Spain}
\author{Maciej Lewenstein}
\affiliation{ICFO-Institut de Ciencies Fotoniques, Barcelona Institute of Science and Technology, Av.~Carl Friedrich Gauss 3, 08860 Castelldefels (Barcelona), Spain}
\address{ICREA-Instituci\'o Catalana de Recerca i Estudis Avan\c{c}ats, 08010 Barcelona, Spain}
\author{Robert Moszynski}
\address{Faculty of Chemistry, University of Warsaw, Pasteura 1, 02-093 Warsaw, Poland}

\date{\today}

\begin{abstract}

We employ \textit{ab initio} methods of quantum chemistry to investigate spin-1/2 fermions interacting via a two-body contact potential in a one-dimensional harmonic trap. The convergence of the total energy with the size of the one-particle basis set is analytically investigated for the two-body problem and the same form of the convergence formula is numerically confirmed to be valid for the many-body case. Benchmark calculations for two to six fermions with the full configuration interaction method equivalent to the exact diagonalization approach, and the coupled cluster method including single, double, triple, and quadruple excitations are presented. The convergence of the correlation energy with the level of excitations included in the coupled cluster model is analyzed. The range of the interaction strength for which single-reference coupled cluster methods work is examined. Next, the coupled cluster method restricted to single, 
double, and noniterative triple excitations, CCSD(T), is employed to study a two-component Fermi gas composed of 6 to 80 atoms in a one-dimensional harmonic trap. The density profiles of  trapped atomic clouds are also reported. Finally, a comparison with experimental results for few-fermion systems is presented. Upcoming possible applications and extensions of the presented approach are discussed. 

\end{abstract}

\pacs{}

\maketitle

\section{Introduction}

Ultracold gases are highly controllable systems ideal for investigating different phenomena of quantum many-body physics~\cite{Lewenstein2012,Bloch2008,HouchesVol94,Giorgini2008,Guan2013,Blume2012}. They can be prepared in a well-defined quantum state, carefully manipulated, and accurately measured, and thus can serve as a perfect tool for quantum simulations~\cite{Lewenstein2007}. On one hand, they can be used to realize highly non-trivial states of matter. On the other hand,  the problems of condensed-matter or other areas of physics can be mapped on and solved by such quantum simulators.
Reducing the dimensionality of the trapped gases to one dimension brings a plethora of new interesting possibilities~\cite{Giamarchi}. Experimental studies of the Tonks-Girardeau gas~\cite{Tilman1D,Weiss1D,Belen} and of the super Tonks-Girardeau gas~\cite{Haller2009} are the first and most eminent examples.

Levi Tonks (1897-1971) was the first to consider in 1936 \cite{Tonks1936} the problem of equation of state simultaneously for one, two, and three-dimensional gases of hard elastic spheres -- this has led him to the concept of a (classical) gas of impenetrable particles in 1D. Marvin Girardeau (1930-2015) was a real pioneer of the studies of the quantum impenetrable gas of bosons, known since then as Tonks-Girardeau gas. In particular, Girardeau investigated in 1960 the relation between   systems of impenetrable bosons and fermions in one dimension \cite{Marvin1960}. This seminal work attained an extreme importance in the cold atom era, and has led to numerous nearly equally seminal generalizations and developments: studies of  1D Coulomb gas \cite{Marvincoulomb}, studies of trapped bosonic gases in 1D \cite{Marvin2}, studies of quantum dynamics \index{Marvin3,Marvin3a} and quantum solitons \cite{Marvin4}, general theory of Fermi-Bose  \cite{Marvin5,Marvin6} and anyon-fermion mapping \cite{Marvinanyon}, investigations of  super-Tonks-Girardeau gas \cite{Marvinsuper,Marvinsuper1}, of 1D dipolar gases \cite{Marvindipol}, and much more. From the point of view of the present work the most important were more recent generalizations and applications of Girardeau's approach to soluble models of strongly interacting 1D ultracold gas mixtures \cite{Marvin7,Marvin7a}, and especially to spinor Fermi gases \cite{Marvin8,Marvin8a,Marvin8b,Marvin9}. Here we investigate precisely  the problem of 1D fermionic gas of atoms of spin 1/2, trapped in a harmonic potential and in the presence of strong interactions.    

Recently, experiments achieving a deterministic preparation of tunable several-fermion systems in a one-dimensional trap became possible~\cite{Serwane2011}. 
This opened the new area of ultracold research on systems with a complete control over all degrees of freedom: the particle number, the internal and motional states of the particles, and the strength of the interparticle interactions.
 The fermionization of two distinguishable fermions~\cite{Zurn2012}, the formation of a Fermi sea~\cite{Wenz2013}, pairing in few-fermion systems~\cite{Zurn2013}, two fermions in a double well potential~\cite{Murmann2015}, and antiferromagnetic Heisenberg spin chain of few cold atoms~\cite{Murmann2015b} have been investigated experimentally and are a promise of upcoming new and equally fascinating research. 

The aforementioned experiments allowed to observe for the first time the transition between the few-fermion limit and the many-fermion limit of trapped atoms at ultralow temperatures.
The emergence of the many-body properties of the physical systems is crucial across all areas of research. 
The Fermi gases in one dimension in the many-body regime have been studied intensively over the years~\cite{Recati2003,Juillet2004,Astrakharchik2004,Tokatly2004,Fuchs2004,Carusotto2004,
Astrakharchik2005,Orso2007,Hu2007,ColomeTatche2008}.  Recently, several papers used analytic methods and exact diagonalization to investigate in detail the few-body regime~\cite{Peotta2012,Gharashi2013,Sowinski2013,Astrakharchik2013,Deuretzbacher2014,Volosniev2014,
Lindgren2014,Levinsen2014,DAmico2015,Berger2015,Sowinski2015,Massignan2015}.
Nevertheless, new numerical approaches providing additional insight on the experimental findings and having the predictive power of proposing new experiments are always welcome. Especially methods that can cover both regimes of few and many fermions are of great interest.

The goal of the present work is to provide new numerical tools to investigate the Fermi systems of few to many cold atoms. More specifically, we consider spin-1/2 fermions confined in a one-dimensional harmonic trap and  interacting via a two-body contact potential. For this aim, we employ a quantum chemistry approach -- the coupled
cluster (CC) method~\cite{Coester1958,Cizek1966,Cizek1969,Cizek1971,PaldusPRA72,Bartlett1981,
Bartlett1989,Bishop1991,Paldus1999,Musial2007,Lyakh2012}. This method has successfully been applied to study various properties of atoms, molecules, and condensed phases -- see, for instance, Refs. \cite{McGuyerPRL13,McGuyerNaturePhys14} for applications to high-precision spectroscopy of ultracold molecules, Ref. \cite{BukowskiScience07} for simulations of  liquid water properties, or Ref. \cite{PodeszwaJCP14} to determine the structure and characteristics of molecular crystals.

In the condensed-matter physics, the coupled cluster method has up to now been successfully applied to ultracold gases of bosonic atoms in traps~\cite{Cederbaum2006,Alon2006} and allowed to describe correlations beyond the mean-field regime in a Bose-Einstein condensate of thousands of atoms. However, the ground state of bosonic systems and bosonic many-body wave functions have much simpler form than the fermionic counterparts, and therefore the advantage of the CC method was not fully pronounced in that case.
The CC approach has also found numerous applications in studies of spin-1/2 lattice models,  both in  one-dimensional chains and in two-dimensional square lattices (see, e.g., Refs.~\cite{BishopPRL94,BishopPRA11}).
In the following, we will show that the CC method proves to be ideally suited to study the problem of many fermionic atoms in one-dimensional traps. 


The plan of our paper is as follows. Section~\ref{sec:theory} describes the theoretical framework, including 
the definition of the many-body Hamiltonian in subsection~\ref{sec:ham}, and the discussion of the  exact solution for the two-body case, and convergence of the energy with the size of basis set in subsection~\ref{sec:exact}. This is followed by the summary of  employed many-body approaches in subsection~\ref{sec:many-body}. 
Section~\ref{sec:results} presents the results on the convergence with the size of the one-particle
basis set in subsection~\ref{sec:conv_basis}, the convergence with the excitation level included in the coupled cluster Ansatz for the wave function in subsection~\ref{sec:conv_ex}, the density profiles in subsection~\ref{sec:density}, 
and the comparison with experiments in subsection~\ref{sec:exp}. 
Section~\ref{sec:summary} summarizes our paper and discusses possible future applications of the developed approach and further developments. Details of the analytic solutions of the two-body case of subsection~\ref{sec:exact}, and the derivations of the corresponding convergence laws  with size of the basis set are presented in appendices~\ref{ap:two-body}--\ref{ap:conv_var}. 
\section{Theoretical framework}
\label{sec:theory}

\subsection{The many-body Hamiltonian}
\label{sec:ham}

The Hamiltonian describing a system of $N$ spin-1/2 fermions (atoms) in a one-dimensional harmonic trap reads
\beq
\label{hhhh}
\hat H = -\frac{\hbar^2}{2m} \sum_{i=1}^{N} \partial_{x_i}^2 + \frac 12 m \omega^2 \sum_{i=1}^{N}  x_i^2+ g \sum_{i<j}
\delta(x_i-x_j),
\eeq
where $x_i$ represents the coordinate of the $i$-th atom, $m$ is the atom mass, $\omega$ is the frequency of the
trap, and $g$ is the strength of the two-body contact interaction.
Throughout the paper we use units of energy and  interaction strength that correspond to $\omega=m=\hbar=1$. 
This amounts to measuring energies $E$ in units of $\hbar\omega$, lengths in units of the harmonic oscillator characteristic length $a_{ho}=\sqrt{\hbar/(m\omega)}$, and the interaction strength $g$ in units of $\hbar\omega a_{\rm ho}$.
Obviously, the Hamiltonian is symmetric
with respect to the transposition of two arbitrary space coordinates, implying that the eigenfunctions must transform according
to an  irreducible representation of the permutation group $S_N$. In this paper we limit ourselves to the spin-1/2 atoms, so that the wave function will be fully antisymmetric with respect to the simultaneous transposition of any space and spin coordinates
of two atoms. From now on we assume that $N=N_\uparrow+N_\downarrow$ describes the total number of atoms ($N_\uparrow$ atoms with the spin projection 1/2, and $N_\downarrow$ atom with the spin projection $-1/2$).
We will consider either systems with $N_\uparrow=N_\downarrow$, that is, in the simplest spin singlet $S=0$ state, or ones with $N_\uparrow \neq N_\downarrow$  such that the total spin $S = |N_\uparrow-N_\downarrow|$, that is, in a spin-stretched (high-spin) configuration.

In almost all many-body theories it is customary to use the so-called algebraic approximation \cite{Wilson1976}, i.e.
to use a finite set of one-particle functions. These one-particle functions are usually expanded in terms of some
known basis functions. In our case the most natural choice of the basis functions is obviously given by the
eigenfunctions of the one-dimensional harmonic oscillator:
\beq
\label{basis}
\varphi_n(x) = \frac{1}{\sqrt{2^n n!}} \frac{1}{\pi^{1/4}} e^{-\frac{x^2}{2}}
H_n\lp x\rp,
\eeq
where $H_n(x)$ are the Hermite polynomials. 
The integrals of the one-particle part of the Hamiltonian are diagonal, and given by the eigenenergies of the harmonic oscillator, while the two-particle integrals may be calculated numerically using the exact Gauss-Hermite quadratures.

Obviously, eigenfunctions of the harmonic oscillator provide the basis set that is the simplest for numerical applications. There are many other choices, that require more numerical effort, but should assure better numerical precision and convergence: an obvious example in the case of non-harmonic potential is to use the corresponding orbitals that are exact eigenfunctions of the one-particle Hamiltonian.  Explorations of these choices go, however,  beyond the scope of this paper.

\subsection{The two-body case and the convergence with the size of the one-particle basis set}
\label{sec:exact}

To investigate in detail the properties of the considered system it is useful to solve  analytically the two-body case first.
The exact solution of the two-body problem in a one-dimensional harmonic trap was found by Busch et al.~\cite{Busch1998} and by Franke-Arnold et al.~\cite{FrankeArnold2003}.
For convenience and to introduce the notation we decided to include a detailed description of the solution of the aforementioned Hamiltonian in appendix~\ref{ap:two-body}.
The ground state wave function is 
\begin{align}
\label{exact}
\Psi(X,x)= C_n C_\epsilon \,e^{-X^2/2-x^2/2} H_n(X)\, U(-\epsilon/2,1/2,x^2),
\end{align}
where $X=(x_1+x_2)/\sqrt{2}$ and $x=(x_1-x_2)/\sqrt{2}$ are the center of mass and relative coordinates, and $C_n$ and $C_\epsilon$ are the normalization constants for the $X$- and $x$-dependent portions. The corresponding total energy of the system (including explicitly the zero-point energies of the two particles) is then $E=\epsilon+n+1$, while $U(a,b,x)$ stands for the Tricomi function \cite{AbramowitzStegun}. The energy of the relative motion $\epsilon$ is determined from
\begin{align}
\label{eq:1/g}
\frac{1}{g}=\frac{\Gamma\left(\frac{2-\epsilon}{2}\right)}{\sqrt{2}\epsilon\Gamma\left(\frac{1-\epsilon}{2}\right)},
\end{align}
where $\Gamma(.)$ is the Gamma function. See appendix \ref{ap:two-body} for more details.

To address the problem of the energy convergence 
 with increasing number $n_b$ of one-particle functions in the basis set, we have pursued two complementary strategies.
 The first follows the one developed by Hill \cite{Hill1985} for the case of the helium atom, where the exact wavefunction, expanded in terms of an infinite sum over basis functions, 
 is truncated at a fixed number $n_b$ of basis set functions. This leads to an approximated energy $\epsilon_{n_b}$ converging to the exact value $\epsilon$ as (see appendix~\ref{ap:conv_anal} for details)  
\begin{align}
\label{econv3-intext}
 \epsilon_{n_b}-\epsilon = \frac{2^{5/2}C_\epsilon^2}{\Gamma(-\half \epsilon)^2}
 \left( \frac{1}{\sqrt{n_b}}+\frac{g}{\pi n_b}\right. 
\left. + \frac{15+8\epsilon}{48 \,n_b^{3/2}} \right) 
+ \mathcal{O}\left(n_b^{-2}\right).
\end{align}
Note that the formula found by Hill, although derived explicitly for the two-electron case, has been successfully applied across a wide range of atomic and molecular calculations (see e.g.~Ref.~\cite{HelgakerJCP97,HalkierCPL98}). Our aim here is similar, in that the above functional form of the two-body extrapolation formula will prove crucial for obtaining accurate results for systems with $N>2$, as discussed below. 

Since the approach outlined above does not allow for the optimization of the coefficients in a smaller Hilbert space, we also followed a second strategy, based on an actual variational minimization of the energy in the space spanned by the states contained in the basis set. 
The result reads (see appendix~\ref{ap:conv_var} for details)
\begin{equation}
\tilde \epsilon_{n_b}-\epsilon= \A
\left\{\frac{1}{\sqrt{n_b}}-\frac{\A x''}{2x'}\frac{1}{n_b}+\right.
\left.\left[\frac{5+4\epsilon}{24}+\frac{\A^2}{2}\left(\frac{x''}{x'}\right)^2-\frac{x'''}{3x'}\right]\frac{1}{n_b^{3/2}}\right\} +\mathcal{O}(n_b^{-2}),
\label{convDeltaE-intext}
\end{equation}
where  $x=g^{-1}(\epsilon)$, as given by Eq.~\eqref{eq:1/g}, $x'\equiv\partial_\epsilon [g^{-1}(\epsilon)]$, and $\mathcal{A}=-1/(\sqrt{2}\pi x')$.

Note that both approaches yield the same functional form of the convergence formula, differing in coefficients that are specific to the two-body case, and leading terms in Eq.~\eqref{econv3-intext} and Eq.~\eqref{convDeltaE-intext} are numerically confirmed to be the same.

\subsection{Summary of the many-body approaches}
\label{sec:many-body}
 
As already stated above, almost all many-body theories are based on the algebraic approximation~\cite{Wilson1976},
i.e., on the parametrization of the wave function by expansion in a finite set of basis functions. Since we deal with 
states with total spin $S=0$ or high-spin (spin stretched) states with the total spin $S=|N_\uparrow-N_\downarrow|$, the first approximation to the exact wave function will be the Slater determinant $\Phi$ built of one-particle functions obtained by
solving the Hartree-Fock equations of the mean-field theory (cf.~\cite{Szabo1996}).
 We will denote the $N$ functions occupied in the
reference Slater determinant by $\{\phi_\alpha\}_{\alpha=1}^{N}$, while the remaining one-particle functions
that are not occupied in the reference determinant will be denoted by $\{\phi_\rho\}_{\rho=N+1}^{2n_b}$, where
$n_b$ is the number of the harmonic oscillator functions used to expand the one-particle solutions of the Hartree-Fock
equations. Note that the total number of one-particle functions is $2n_b$ instead of $n_b$ since $\phi_\alpha$ and $\phi_\rho$ include the dependence on the spin coordinate through the spin functions. Obviously, the relation $n_b \gg N$ must hold. There is a theorem stating that if a set of one-particle
functions spans the one-particle  Hilbert space $L^2({\mathbb R}^3)$, then the set of all $N$-particle determinantal wave functions 
constructed from this one-particle set will span the antisymmetric part of the $N$ particle Hilbert space~\cite{Reed}. In other
words, the $N$ particle wave function can rigorously be expanded in terms of the Slater determinants built from the
complete set of one-particle functions. The same theorem holds in the algebraic approximation, and it is the basis of exact diagonalization, otherwise referred to as the full configuration interaction (FCI) method.  The difference between the mean-field and FCI energies is called the correlation energy. 

In the full configuration interaction method
the wave function of the many-fermion system is represented as
\beq
\label{fci1}
\Psi=(1+\hat{C})\Phi,
\eeq
where the CI operator $\hat{C}$ may be written as a linear combination of $l$-particle excitation operators,
\beq
\label{fci2}
\hat{C} = \sum_{l=1}^{N} \hat{C_l}, \; \; \; \; \;
\hat{C_l}= \frac{1}{(l!)^2}C_{\rho_1...\rho_l}^{\alpha_1...\alpha_l}e^{\rho_1...\rho_l}_{\alpha_1...\alpha_l},
\eeq
with coefficients $C_{\rho_1...\rho_l}^{\alpha_1...\alpha_l}$.
Summation over the repeated lower and upper indices is assumed. Note that the indices $\alpha$ and
$\rho$ refer to the one-particle functions that are occupied or empty in the reference Slater determinant $\Phi$, so Eq.~(\ref{fci1}) can be viewed as an operator representation of the expansion of the exact wave function in terms of the all possible $N$-particle determinantal wave functions constructed from the $\{\phi_\alpha\}_{\alpha=1}^{N} \oplus\{\phi_\rho\}_{\rho=N+1}^{2n_b}$ set of the one-particle functions.
The $l$-particle excitation operator $e^{\rho_1...\rho_l}_{\alpha_1...\alpha_l}$ is given by the product
of one-particle excitation operators $e^{\rho_1}_{\alpha_1}\cdots e^{\rho_l}_{\alpha_l}$. The one-particle excitation
operators are in turn defined in terms of the conventional fermionic creation and annihilation operators $e^\rho_\alpha
=a^\dagger_\rho a_\alpha$, where the creation and annihilation operators are defined with the respect to
the physical vacuum $|0\rangle$, but applied to  the Fermi
vacuum $\Phi$~\cite{PaldusAQC75}.
Note that the one-particle functions $\phi_\alpha$ and $\phi_\rho$ are always orthogonal,
so that the excitation operators commute. Note also that our approach is limited to the $N$ particle Hilbert
space (layer of the Fock space with the fixed number of spin-1/2 fermions).  This means that the algebra of the
excitation operators is not only commutative, but also nilpotent. The latter property easily follows from the fact
that we can replace at most $N$ one-particle functions $\phi_\alpha$ in the reference Slater determinant, and any
further action will give zero as the result. An important consequence of this property is that any transcendental function of the excitation operators that has a well defined Taylor
expansion will reduce to a finite polynomial.
The CI coefficients $C_{\rho_1...\rho_l}^{\alpha_1...\alpha_l}$ and the energy are obtained by diagonalizing the
Hamiltonian matrix constructed from the matrix elements ${\mathbb H}_{lm}=\langle e^{\rho_1...\rho_l}_{\alpha_1...\alpha_l}\Phi|
\hat{H}|e^{\rho_1...\rho_m}_{\alpha_1...\alpha_m}\Phi\rangle$. This explains why the FCI method is also referred to as the
exact diagonalization method.

The computational cost of the FCI method is prohibitive, as it scales as $N^2 (2n_b)^{N+2}$ for $N \ll n_b$ \cite{Olsen1988}, which limits this method to small
systems (small $N$) or small number of basis functions $n_b$. One possible way to cure this problem is to limit the
number of excitations included in Eq.~(\ref{fci2}), e.g., to single and double excitations. Such a truncation may lead to considerable savings of computer time, but unfortunately has a serious drawback. Any truncated CI method 
is no longer size-extensive, which means the energy of two non-interacting systems is not the sum of the energies of these two systems.

To overcome the size-extensivity problem of the limited CI expansions, the coupled cluster (CC) method was introduced,
first in the nuclear physics~\cite{Coester1958} and slightly later in quantum
chemistry~\cite{Cizek1966,Cizek1969} and in the electron gas theory \cite{Freeman1978}. In this method the wave function is given by the following exponential Ansatz
\beq
\label{cc1}
\Psi = e^{\hat{T}}\Phi,
\eeq
where the cluster operator $\hat{T}$ is given by
\beq
\label{cc2}
\hat{T} = \sum_{l=1}^{N} \hat{T_l}, \; \; \; \; \;
\hat{T_l} = \frac{1}{(l!)^2}t_{\rho_1...\rho_l}^{\alpha_1...\alpha_l}e^{\rho_1...\rho_l}_{\alpha_1...\alpha_l}.
\eeq
Note that by comparison with Eq.~(\ref{fci1}) the following relation between the FCI and cluster operator must
hold, $\hat{T} = \ln(1+\hat{C})$, but since the algebra of the excitation operators is nilpotent, the
logarithm, as well as the exponential function in Eq.~(\ref{cc1}), reduce to finite polynomials. The CC method is
non-variational, in the sense that it  does not involve any optimization over a set of variational parameters. The energy is given by the expression
\beq
\label{cc3}
E=\braket{\Phi | e^{-\hat{T}}\hat{H}e^{\hat{T}} \Phi},
\eeq
while the cluster amplitudes $t_{\rho_1...\rho_l}^{\alpha_1...\alpha_l}$ are obtained by solving the following system 
of nonlinear algebraic equations
\beq
\label{cc4}
0=\braket{e^{\rho_1...\rho_l}_{\alpha_1...\alpha_l}\Phi|e^{-\hat{T}}\hat{H}e^{\hat{T}} \Phi}, \; \; \; \; \; l=1,...,N.
\eeq
Note that by virtue of the Baker-Campbell-Hausdorff formula the exponential factor
$e^{-\hat{T}}\hat{H}e^{\hat{T}}$ reduces to multiple commutators of $\hat{H}$ and $\hat{T}$, and one can prove that this
commutator expansion is finite and contains at most four-fold commutators if only two-particle
interactions are present in the Hamiltonian~\cite{Coester1958}.

Obviously, the coupled cluster method including all excitations is fully equivalent to the FCI method, and
its computational cost is as high. However, the truncated CC methods are much less time consuming, and
unlike the truncated CI methods, remain size-extensive. Due to the exponential form of the Ansatz, Eq.~\eqref{cc1}, the CC method truncated to single and double excitations effectively includes triply, quadruply, and
higher excited determinants through the products, e.g. $\hat{T}_1\hat{T}_2$ or $\hat{T}_2^2$.
In the present work we will use the series of approximations with the cluster operator limited to single
and double (CCSD) \cite{Purvis1982}, single, double, and triple (CCSDT) \cite{Noga1987}, and  single, double, 
triple, and quadruple excitations (CCSDTQ) \cite{Kucharski1991,Oliphant1991,Kucharski1992}. The computational cost of these methods scales
as $n_b^6$, $n_b^8$, and $n_b^{10}$, respectively. 
For a system consisting of a dozen of atoms methods
including triple excitations cannot reasonably be used, while the experience gained for many-electron systems suggests that
the triple excitations have an important contribution to the correlation energy. To reduce the computational
cost of the CC method including triple excitations, the CCSD(T) method with the $n_b^7$ scaling was introduced \cite{Raghavachari1989}.
In this method the CCSD equations are solved iteratively, and the contribution of the triple excitations to the
correlation energy is evaluated from the expression based on the many-body perturbation theory. 
It turned out that
the CCSD(T) method is very accurate for many properties of atoms and molecules and, as for now, it is considered
as the golden standard of quantum chemistry. As we show in the next section, this method can also be
applied with success to the system of ultracold atoms in a one-dimensional harmonic trap interacting via a
short-range contact type potential.

In order to introduce the perturbation expansion of the correlation energy, often
referred to as many-body perturbation theory (MBPT)~\cite{Bartlett1981}, or M{\o}ller-Plesset
perturbation theory (MPPT)~\cite{Møller1934,Krishnan1980,Kucharski1989},
it is useful to rewrite the equations for the cluster operator $\hat{T}$ in the
following operator form~\cite{Jeziorski1981,Jeziorski1993}
\beq
\label{cc5}
\hat{T} = \hat{\mathcal{R}}\lp e^{-\hat{T}} \hat{W} e^{\hat{T}}\rp = \hat{\mathcal{R}}\Big( \hat{W} + [\hat{W},\hat{T}]+ \frac12 [\hat{W},\hat{T}]_2 + \frac16 [\hat{W},\hat{T}]_3 + \frac1{24} [\hat{W},\hat{T}]_4\Big)\, ,
\eeq
where $\hat{W}$ is the correlation (fluctuation) potential in the M{\o}ller-Plesset partitioning of
the Hamiltonian 
\beq
\label{MP}
\hat{H}=\hat{F}+\hat{W},
\eeq
$\hat{F}$ is the Fock operator, and $[\hat{W}, \hat{T}]_n$ denotes the $n$-fold nested commutator: $[\hat{W},\hat{T}]_0 = \hat{W}$,
$[\hat{W},\hat{\hat{T}}]_{n+1} = [[\hat{W},\hat{T}]_n,\hat{T}]$. The nested commutator expansion in Eq. (\ref{cc5}) terminates
after the quadruple commutators since in our case the operator $W$ contains
only two-particle interactions. The resolvent superoperator $\hat{\cal R}$ is defined for arbitrary
operator $\hat{X}$ as \cite{Jeziorski1981,Jeziorski1993}
\beq
\label{res}
\hat{\mathcal{R}} = \sum_{n=1}^{N} \hat{\mathcal{R}}_n\, , \; \; \; \; \;
\hat{\mathcal{R}}_n(\hat{X}) = \lp\frac1{n!}\rp^2 \braket{e^{\alpha_1\dots\alpha_n}_{\rho_1\dots\rho_n}\Phi |\hat{X}\Phi} \frac{e^{\rho_1\dots\rho_n}_{\alpha_1\dots\alpha_n}}{\epsilon^{\alpha_1\dots\alpha_n}_{\rho_1\dots\rho_n}}\, ,
\eeq
where
\beq
\epsilon^{\alpha_1\dots\alpha_n}_{\rho_1\dots\rho_n} = \epsilon_{\alpha_1} + \cdots + \epsilon_{\alpha_n} - \epsilon_{\rho_1} - \cdots - \epsilon_{\rho_n}\, ,
\eeq
and $\epsilon_\kappa$ denotes the one-particle energy associated with the one-particle function labeled by the index
$\kappa$. We assume that the energy of the highest occupied one-particle function is smaller than the energy
of the lowest unoccupied in the reference determinant $\Phi$, so the superoperators $\hat{\mathcal{R}}_n$ are always well defined. Note
that for a given $\hat{X}$ the operator $\hat{Y} = \hat{\mathcal{R}}_n(\hat{X})$ can be viewed as a formal solution of
the equation
\beq
\label{equres}
\braket{e^{\rho_1\dots\rho_n}_{\alpha_1\dots\alpha_n} \Phi | ([\hat{F},\hat{Y}] + \hat{X}) \Phi} = 0\, .
\eeq
Obviously, this solution is unique if we assume that $\hat{Y}$
belongs to the linear span of the excitation operators $e^{\alpha_1\dots\alpha_n}_{\rho_1\dots\rho_n}$.

The many-body perturbation expansion of the energy is obtained by introducing the following parametrization
of the correlation operator $\hat{W}$,
\beq
\label{param}
\hat{W} \rightarrow \lambda \hat{W},
\eeq
where the complex parameter $\lambda$ is introduced to derive the perturbation expansions of Eqs.~(\ref{cc3})
and (\ref{cc5}). Obviously, the physical value of $\lambda$ is equal to one. By substituting the parametrized
Eq.~(\ref{param}) into Eq.~(\ref{MP}) the cluster operator became dependent on $\lambda$, and can be expanded
as a power series in $\lambda$
\beq
\label{expanT}
\hat{T}=\sum_{k=1}^\infty \lambda^k \hat{T}^{(k)}.
\eeq
Substituting the above expansion in the expression for the energy (\ref{cc3}) and collecting all terms 
at $\lambda^n$ gives the expression for the $n$th-order correction to the energy in the many-body
perturbation theory. The cluster operators necessary to evaluate this expression are obtained by
substituting Eq.~(\ref{expanT}) and again  collecting all terms at $\lambda^n$.

It follows immediately from this short sketch of the derivation that the MBPT and CC theories are strongly
connected. The analysis reported in Ref. \cite{Jeziorski1993} shows that the CCSD, CCSDT, and CCSDTQ are valid through the
third, fourth, and fifth order of the many-body perturbation theory. The golden standard of quantum
chemistry, the CCSD(T) method, is also valid through the fourth order of MBPT, although it is computationally
less demanding than the full CCSDT theory.

One of the observables that can be obtained with the presented method is the density profile of the trapped atoms. At a point $x_0$ it may be obtained as the expectation value of the operator $\sum_{i=1}^{N}\delta(x_i-x_0)$ within the FCI calculations, and from the Hellmann-Feynman theorem as the first derivative of the energy in the presence of the perturbation given by the same operator in the CC calculations. In practice, we compute this derivative by using the finite-field approach with a perturbation of the form $\phi_i(x_0)\phi_j(x_0)$ added to the $ij$-th element of the one-particle Hamiltonian matrix. 

The full configuration interaction and coupled cluster calculations were performed with the customized versions of the HECTOR~\cite{HECTOR} and ACESII codes~\cite{ACESII}, respectively. The CC and truncated CI calculations use the reference state $\Phi$ built of orbitals obtained with the restricted Hartree-Fock method for the $N_\uparrow=N_\downarrow$ case, whereas orbitals obtained with the unrestricted Hartree-Fock method are employed for the $N_\uparrow\ne N_\downarrow$ case.

\section{Numerical results and discussion}
\label{sec:results}

\begin{figure}[t!]
\includegraphics[width=0.5\columnwidth]{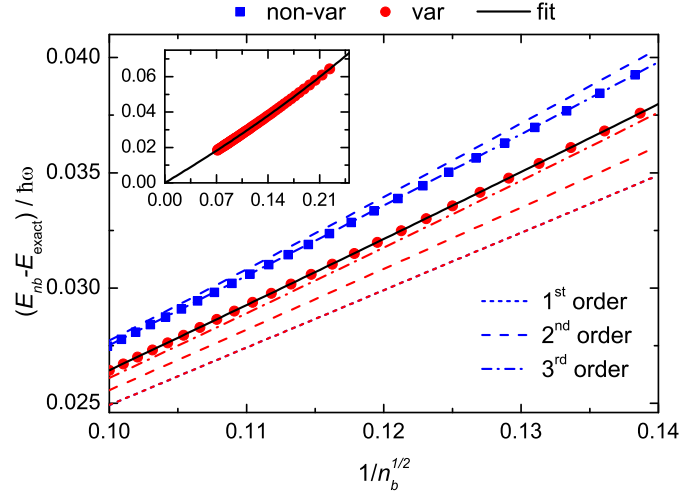}
\caption{Absolute error with respect to the exact result of the ground state energy of the 1+1 system for $g=5$.
Comparison of the non-variational energy (blue squares) obtained as the expectation value with the truncated exact wavefunction~[Eq.~\eqref{enb} in appendix~\ref{ap:conv_anal}] and variational energy (red dots) equivalent to exact diagonalization results with the asymptotic formulas of Eqs.~\eqref{econv3-intext} and~\eqref{convDeltaE-intext}.
Results with inclusion of the leading-order term only are given as the short-dashed lines, results with inclusion of the first two terms are given as the dashed lines, and with inclusion of all three terms are given as the dot-dashed lines. 
The solid black line is the fit of the variational results to the formula, $E_\infty+bn_b^{-1/2}+cn_b^{-1}+dn_b^{-3/2}$ and the inset shows the performance of this fit over full set of data.
}\label{fig:two-body}
\end{figure}

\subsection{Convergence with the size of the one-particle basis set}
\label{sec:conv_basis}

Examples of a slow convergence of the results with respect to the basis set
size are well-known in the literature. The conventional exact
diagonalization calculations for solving the electronic Schr{\"o}dinger equation converge
as $L^{-3}$, where $L$ is the highest angular momentum present in the
one-electron basis set \cite{Schwartz1962,Hill1985}. Even slower convergence ($L^{-1}$) was
observed for some relativistic corrections arising from the perturbative
approach based on the Breit-Pauli Hamiltonian \cite{Cencek2012}.

In order to check that the convergence formulas of Eq.~\eqref{econv3-intext} and Eq.~\eqref{convDeltaE-intext} for the two-body problem hold, we compare their predictions with the exact results in Fig.~\ref{fig:two-body}. Two sets of calculated energies relative to the exact value are presented as a function of the one-particle basis set size: non-variational energies obtained as the expectation value with the truncated exact wavefunction~[Eq.~\eqref{enb} in appendix~\ref{ap:conv_anal}] and variational energies equivalent to exact diagonalization results. The former one should be described by the asymptotic expansion of Eq.~\eqref{econv3-intext}, whereas the latter ones should coincide with the asymptotic expansion of Eq.~\eqref{convDeltaE-intext}. The asymptotic formulas with increasing number of terms are presented. The first terms of the asymptotic expansions give a reasonable estimate of the convergence rate and upon inclusion of the second and third terms the differences between the
analytical formulas and calculated values are greatly reduced. 
Therefore, the formulas~\eqref{econv3-intext} and~\eqref{convDeltaE-intext} are valid and can serve as a guide for further investigations. 

To evaluate the adequateness of the functional formula given by Eqs.~\eqref{econv3-intext} and~\eqref{convDeltaE-intext} for extrapolation of the variational energies to the complete basis set limit, Fig.~\ref{fig:two-body} presents also the fit of the variational results to the formula $E_\infty+A n_b^{-1/2}+B n_b^{-1}+C n_b^{-3/2}$.
This extrapolation formula behaves extremely well [see also the inset of Fig.~\ref{fig:two-body}].
For instance, the extrapolated energies in the complete basis set limit $E_\infty$ for $g=1$ and $g=5$ are $1.306744$ and $1.726780$, whilst the exact values equal to $1.306745$ and $1.726771$, respectively. Note that in the biggest basis set available, $n_b=200$, the calculated energies are $1.310545$ and $1.745325$, respectively, so that the accuracy gain of three order of magnitude is impressive and very important. 

Given the very slow convergence rate of the two-body energy with the number
of the basis functions $n_b$, it is very important to consider this convergence
in the many-body case. One can expect that the convergence pattern obtained for 
the two-body case is almost certainly valid for the many-body case. This fact
was observed in the electronic structure calculations on atoms and molecules,
and is related to the fact that the correlated pair functions of two electrons 
have the same behavior around the electrons coalescence points as in the 
discussed two-body example. Clearly, this analytic behavior determines the 
convergence rate of the results towards the exact value. A very similar 
situation is found for the partial wave expansion of the exact solution of 
the Schr\"odinger equation. The well-known $L^{-3}$ convergence pattern has 
rigorously been derived only for the helium atom \cite{Schwartz1962,Hill1985}, but was successfully
applied also for many-electron atoms/molecules \cite{Kutzelnigg1992,Schmidt1983} and is widely accepted to 
be universal (c.f. the discussion given by King \cite{King1996}).

\begin{figure}
\begin{center}
\includegraphics[width=0.8\columnwidth]{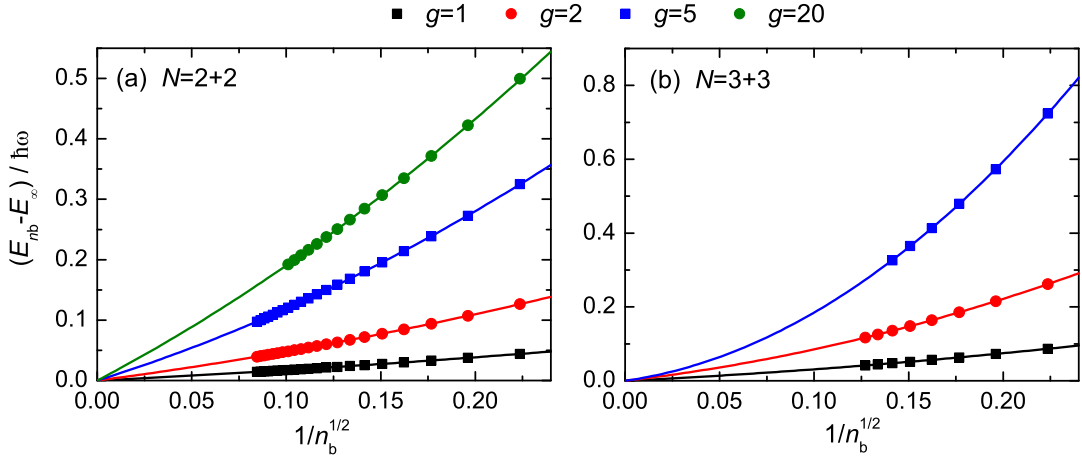}
\end{center}
\caption{Absolute error with respect to the complete basis set limit of the ground state energy of the 2+2~(a) and 3+3~(b) systems as a function of the one-particle basis set size for several interaction strengths obtained with the FCI method. Lines correspond to fits to the extrapolation function given by Eq.~\eqref{extrap}.}
\label{fig:conv_FCI}
\end{figure}

Following the discussion above and guided by the expressions (\ref{econv3-intext}) and (\ref{convDeltaE-intext}), we decided to adopt the following three-term 
extrapolation formula
\beq
\label{extrap}
E_{n_b}-E_\infty=\frac{A}{\sqrt{n_b}}+\frac{B}{n_b}
\eeq
where $E_\infty$ is the extrapolated energy to the infinite number of the
basis functions, $E_{n_b}$ is the energy computed with $n_b$ basis functions,
and $A$ and $B$ are the fit parameters.
To check the correctness of the above expression we performed FCI calculations
for the ground state of a balanced two-component Fermi gas. We have found that the
convergence pattern with the number of the basis functions exactly follows
our extrapolation formula.
The accuracy of our formula is illustrated in Fig.~\ref{fig:conv_FCI}.
There we show the absolute error with respect to the complete basis set limit in the 
ground state energy for the 2+2 and 3+3
systems ($N_\uparrow+N_\downarrow$ means $N_\uparrow$ atoms with the spin projection 1/2, and $N_\downarrow$ atom with the spin projection $-1/2$) as a function of the number of one-particle basis functions for several interaction 
strengths obtained with the FCI method, as well as the error plotted as a function of $n_b$
according to Eq.~(\ref{extrap}). The agreement between the computed points and the
analytical fits to the extrapolation function is excellent, supporting the correctness
of our extrapolation scheme.

\begin{figure}
\begin{center}
\includegraphics[width=0.8\columnwidth]{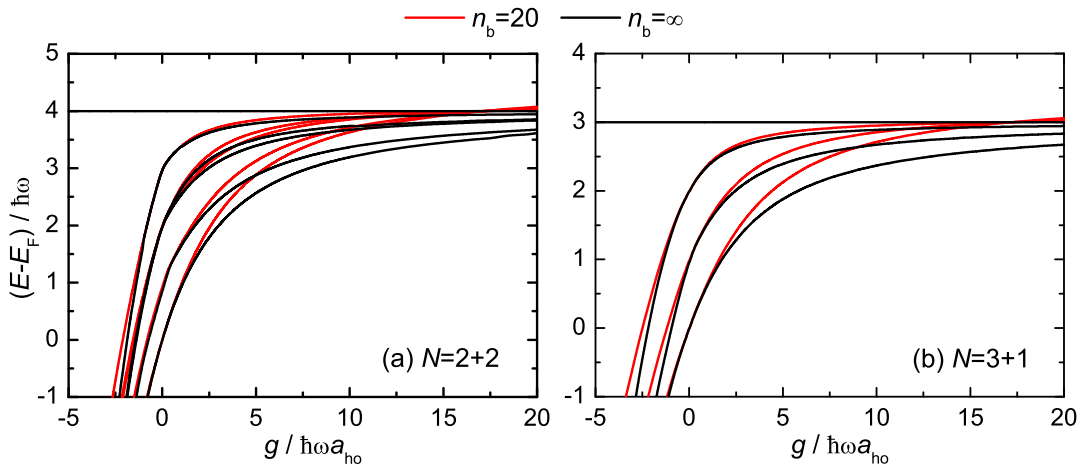}
\end{center}
\caption{Energy spectra (corrected by the ground-state energy of the noninteracting system $E_F$) for the 2+2~(a) and 3+1~(b) systems as a function of the interaction strength $g$ obtained with the FCI method in the basis of $n_b=20$ one-particle functions and in the complete basis set limit $n_b=\infty$.}
\label{fig:spectra}
\end{figure}

In order to get saturated results, careful studies of the convergence are necessary. As stated above,
the system of many identical fermions with spin-1/2 in a 1D trap interacting via the
contact potential has an even worse convergence than the one observed for the electronic Schr{\"o}dinger equation \cite{Schwartz1962,Hill1985,Cencek2012}, and apparent convergence may be observed.
This is illustrated in Fig.~\ref{fig:spectra} where we show the energy spectra 
for a few atoms in a trap obtained with a small number of the basis functions 
($n_b=20$ as in Ref.~\cite{Sowinski2013}) and in the complete basis set limit.
An inspection of this figure shows that the results obtained with 20 basis
functions are far from the complete basis set limit obtained from Eq.~(\ref{extrap}), especially in the limit of intermediate and large interaction strength.
The apparent convergence of the results observed by the Authors of Ref.~\cite{Sowinski2013}
is solely due to the pathological convergence pattern as $n_b^{-1/2}$. As will be shown
in the next section, a quantitative picture of the physics, and a quantitative
agreement with various experimental data, can only be obtained if the extrapolation
to the complete basis set limit is properly done.

As seen in Figs.~\ref{fig:conv_FCI} and \ref{fig:spectra}, both the absolute and relative errors due to the finite basis set size are increasing with the increasing interaction strength $g$. In the limit of very strong interactions, particularly approaching the unitary limit of Tonks-Girardeau (TG) gas (when the interaction strength is infinite, $g=\infty$), 
 this error can become  larger than the separation between the ground and excited states and will lead to artefacts in the description of physical phenomena both at zero and finite temperatures. 
The Authors of Ref.~\cite{Sowinski2013} investigated with the exact diagonalization method the few-fermion physics in the TG limit when the spectrum is quasi-degenerate, but the careful convergence analysis reveals that the results obtained with the basis set as small as $n_b=20$ give only a qualitative picture in this regime of interaction strength. 

\begin{figure}
\begin{center}
\includegraphics[width=0.5\columnwidth]{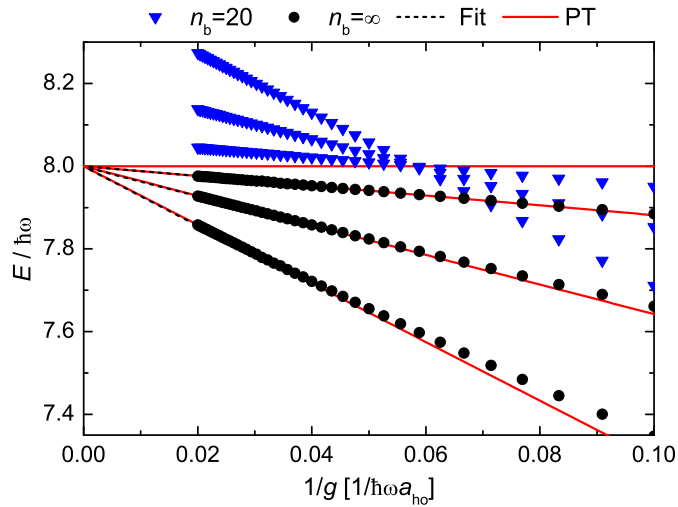}
\end{center}
\caption{Energy spectrum for the 3+1 system in the limit of the strong interaction as a function of $1/g$ obtained with the FCI method in the basis of $n_b=20$ one-particle functions and in the complete basis set limit $n_b=\infty$. The dashed black lines are the linear fits to the points in the complete basis set limit and the solid red lines are the analytilcal results obtained with the perturbation theory (PT) in Ref.~\cite{Levinsen2014}.}
\label{fig:1g}
\end{figure}

Figure~\ref{fig:1g} shows the energy spectrum for the example of the 3+1 system as a function of $1/g$ obtained with the FCI method in the limit of the strong interaction in the basis of $n_b=20$ one-particle functions and in the complete basis set limit $n_b=\infty$. The complete basis set limit results were obtained by extrapolation from the numerical results in basis of 30, 40, 50, and 60 functions.
In the TG limit the four states become degenerate. Unfortunately, any calculation with a finite number of one-particle basis set functions will always locate a crossing of these states at a finite $g$, potentially leading to incorrect conclusions. 
For $n_b=20$ the crossing appears at $g=17.1$ and corresponds to the large error of $0.4\,\omega$ in the ground-state energy. 
Therefore, the extrapolation to the complete basis set limit is not only needed to improve accuracy, but is necessary to have a quantitatively correct picture of the physics.
The numerical results in the complete basis set limit agree perfectly well with the analytical results obtained in the vicinity of the Tonks-Girardeau limit with the perturbation theory~\cite{Deuretzbacher2014,Levinsen2014}. The slopes of the curves fitted to the numerical data are $\{-7.14,-3.58,-1.18,0\}$, in a very good agreement with the exact contact coefficients $\{-7.08,-3.58,-1.18,0\}$. This confirms that the presented extrapolation scheme works also very well in the  regime of strong interactions allowing to approach the TG limit with the accurate finite basis set calculations. Interestingly, the contact coefficients corresponding to non-converged results, even in basis set as small as $n_b=20$, are surprisingly close to the correct values for the TG limit.

\begin{figure}[tb]
\begin{center}
\includegraphics[width=0.5\columnwidth]{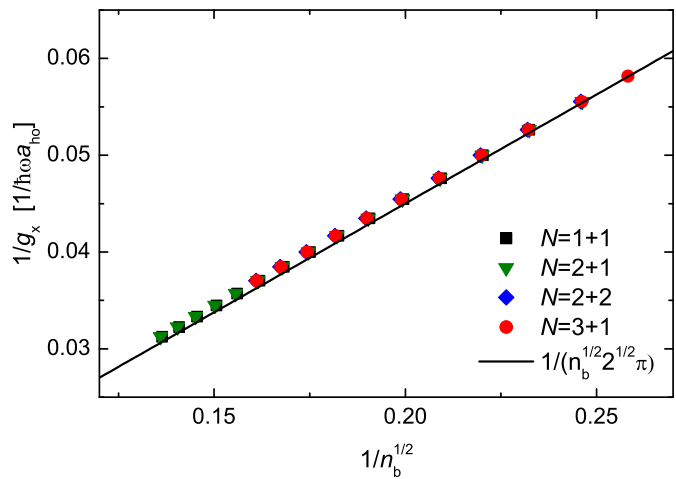}
\end{center}
\caption{Convergence of the critical coupling $1/g_x$ (at which states in the ground manifold become degenerate) as a function of the one-particle basis set size  obtained with the FCI method. The various symbols indicate systems with $N=N_\uparrow+N_\downarrow$ particles and the solid line is given by the first term of Eq.~\eqref{Deltax} in appendix~\ref{ap:conv_var}.}
\label{fig:1gx}
\end{figure}

As we have shown above, the energy spectrum in the TG limit obtained in the calculations with a single one-particle basis set has always an artificial crossing of the states at a finite value of the interaction strength $g$. In Fig.~\ref{fig:1gx} we plot the values of $1/g$ for which this crossing occurs as a function of the one-particle basis set size for a few systems. Interestingly, the interaction strength at which the states in the ground manifold become degenerate depends solely on the number of the basis functions and not on the number of atoms, neither on their state. This observation agrees with our prediction on the convergence of $1/g$ for a given energy (in this case the energy of the ferromagnetic state) with the size of the one-particle basis set to be independent in the leading order on both interaction strength and energy. The leading term of the convergence formula for the two-body problem, derived analytically in appendix~\ref{ap:conv_var}, is shown as a solid black line in Fig.~\ref{fig:1gx}. This observation suggests a second approach to get complete basis set limit results and to reach the TG limit with accurate finite basis set calculations, that is, instead of extrapolating the energy, one can extrapolate $1/g$ for fixed energies by using the universal convergence formula.
Exactly the same convergence coefficient valid for all presented few-body cases and observed universality suggest even a third approach to get accurately converged results with finite basis set calculations at a given interaction strength $g$. The idea is to use in the numerical calculations a renormalized coupling constant $g_{n_b}$ explicitly dependent on the high-energy cut-off set by $n_b$, which reproduces exact two-body result in a smaller Hilbert space spanned by $n_b$ basis functions. 
In this way, the inaccurate description of the short-distance two-body physics introduced by the high-energy cutoff may be cured order by order. A similar idea was discussed, e.g., in Refs.~\cite{Fischer2014,Rotureau2013}.

\subsection{Convergence with the excitation level included in the coupled cluster Ansatz for the wave function}
\label{sec:conv_ex}

\begin{figure}
\begin{center}
\includegraphics[width=0.8\columnwidth]{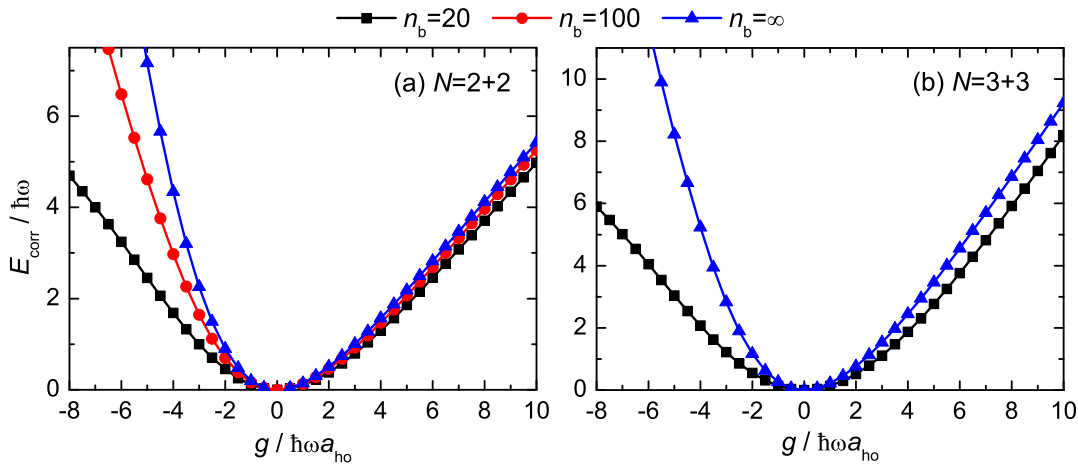}
\end{center}
\caption{Correlation energy for the 2+2~(a) and 3+3~(b) systems as a function of the interaction strength obtained with the FCI method for several one-particle basis set sizes and complete basis set limit.}
\label{fig:E_corr}
\end{figure}

Having solved the problem with the convergence of the results with the number of basis functions, we now
turn to the effect of the truncation of the excitation in the cluster operator, Eq.~\eqref{cc2}, on the correlation and total energies. We start the discussion of our results with Fig.~\ref{fig:E_corr} where 
we report the correlation energy for the 2+2 and 3+3 systems as a function of the 
interaction strength for two one-particle basis set sizes and the complete basis set limit.
An inspection of Fig.~\ref{fig:E_corr} shows that the basis set dependence of the energy is relatively
weak in the weakly correlated repulsive regime ($g>0$), and very pronounced in the strongly correlated attractive case ($g<0$).
These significantly different behaviors result from the fact that in the limit of the strong repulsion even distinguishable fermions tend to occupy different one-particle states and the total wave function approaches the structure of the ferromagnetic state (the Tonks-Girardeau gas limit). On the other hand, the fermions with the opposite spin projection tend to pair in the case of the attractive interaction and become tightly bound (hard-core) bosonic dimers in the limit of the strong attraction (the Lieb-Liniger gas limit). The description of these tightly bound pairs is challenging for calculations in the one-particle functions basis sets and explicitly correlated methods could potentially overcome the problem and significantly accelerate the convergence.
It is important to stress at this point that the mean-field energy converges very fast with the number
of one-particle functions, so the convergence problems are related to the basis saturation of the correlation energy.

The correlation energy presented in Fig.~\ref{fig:E_corr} diverges linearly with the interaction strength $g$
for both negative and positive values. One should note that in the weak interaction regime the distinction between the mean-field and correlation energies is reasonable. For intermediate and especially strong interaction regimes the mean-field description fails completely and diverging correlation energy compensates unphysically diverging mean-field energy, so no physical meaning should be attributed to this divergence. However, this behavior of the mean-field and correlation energies affects the performance of post-mean-field methods  in the strong interaction regime (including the standard coupled cluster method), which start from the mean-field solution and tend to recover correlation energy.

\begin{figure}
\begin{center}
\includegraphics[width=0.85\columnwidth]{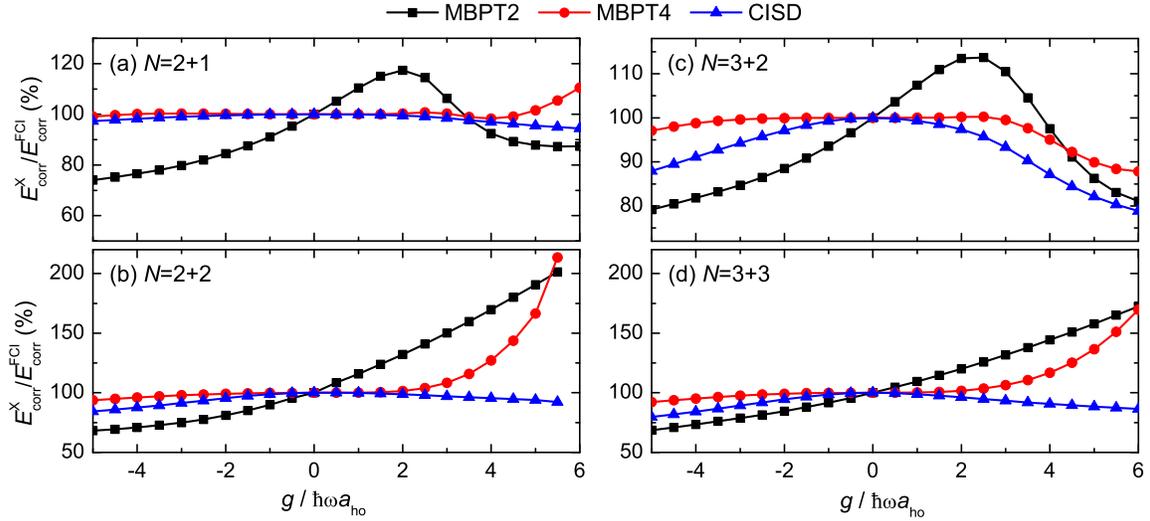}
\end{center}
\caption{The percentage of the ground state correlation energy reproduced with the many-body perturbation theory MBPT2, MBPT4, and the truncated configuration interaction method CISD in the 2+1~(a), 2+2~(b), 3+2~(c), and 3+3~(d) systems as a function of the interaction strength. Values are obtained for a selection of basis set sizes and extrapolated to the complete basis set limit.}
\label{fig:E_mp}
\end{figure}

\begin{figure}
\begin{center}
\includegraphics[width=0.85\columnwidth]{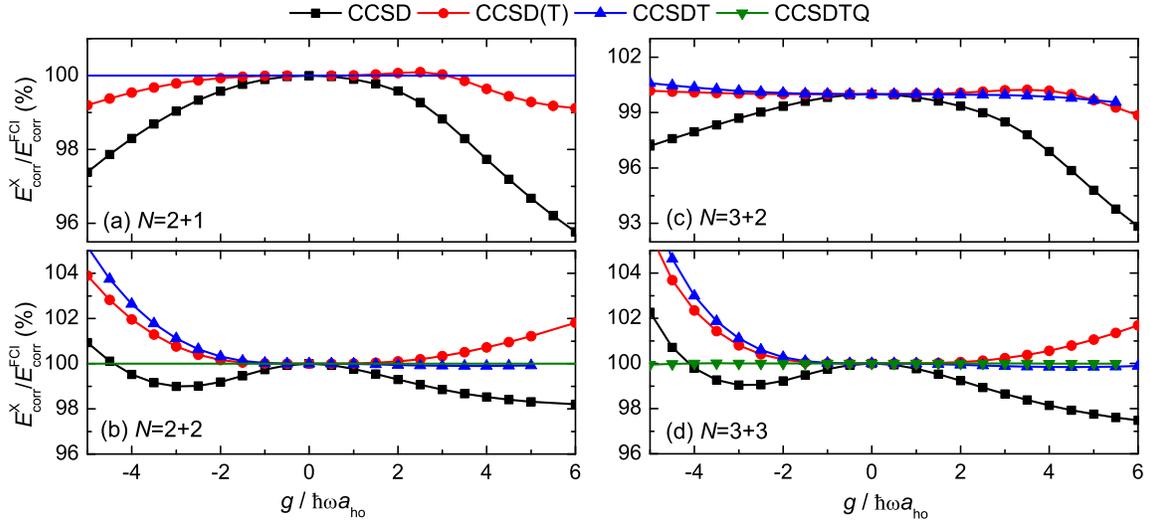}
\end{center}
\caption{The percentage of the ground state correlation energy reproduced at the CCSD, CCSD(T), CCSDT, and CCSDTQ levels of the coupled cluster theory in the 2+1~(a), 2+2~(b), 3+2~(c), and 3+3~(d) systems as a function of the interaction strength. Values are obtained for a selection of basis set sizes and extrapolated to the complete basis set limit.}
\label{fig:E_cc}
\end{figure}

We now turn to the applicability of the many-body perturbation theory and truncated configuration interaction
expansions for the energy calculations. In Fig.~\ref{fig:E_mp} we report the percentage of the ground state 
correlation energy reproduced with the second-order and fourth-order many-body perturbation theory, MBPT2 and MBPT4, 
and the configuration interaction method limited to single and double excitations, CISD, for the 2+1, 2+2, 3+2, and 3+3 
systems as functions of the interaction strength. As expected from the electronic structure calculations on atoms
and molecules the performance of the MBPT2 and MBPT4 methods is not very good, and quite erratic as a function of the
interaction strength. Given the fact that for atoms and molecules, the many-body perturbation theory is divergent \cite{Koch1996,Olsen1996,Leininger2000}
as perturbation theories applied in a different physical context \cite{Cwiok1992,Cwiok1992a}, it is not surprising that the
MBPT4 theory is not a big improvement over the MBPT2 approach, despite a much higher theoretical complexity and
computational time scaling with the size of the basis, $n_b^4$ for MBPT2 vs. $n_b^7$ for MBPT4.
Finally, we note that the variational CISD results do not offer a good advantage over the MBPT results.

Since the MBPT and limited CI theory perform badly, one may ask if a selective infinite-order summation of some MBPT diagrams
with different variants of the coupled cluster theory will improve the situation. This is indeed the case.
Fig.~\ref{fig:E_cc} shows the percentage of the ground state correlation energy reproduced with the CCSD, CCSD(T), CCSDT, 
and CCSDTQ methods for the 2+1, 2+2, 3+2, and 3+3 systems as functions of the interaction strength. First of all
note that for the 2+1 system CCSDT is equivalent to the FCI theory, and for the 2+2 system the same statement is valid
for CCSDTQ. This means that for these systems these particular methods will reproduce 100\% of the energy.
It may be surprising at first glance that some of the approximate variants of the coupled cluster theory
overestimate the total energy of the system. This is due to the fact that the truncated CC methods are not
variational, and one can obtain e.g. 101\% of the energy. An inspection of Fig.~\ref{fig:E_cc} shows that
all CC methods perform very well, although there is no clear convergence pattern with the excitation operators
included in the CC wave function. Indeed, the CCSD method tends to slightly underestimate the energy, while
methods including triple (and possibly quadrupole) excitations tend to slightly overestimate. In general,
the percentage errors are of the order of a few percent. Note also that the CCSDT method does not offer big advantage over the CCSD(T) approach, despite of much higher computational requirements. This is in line
with the results obtained for atoms and molecules, see for instance Refs.~\cite{Cencek2012,Lesiuk2015}.
Comparison of Figs.~\ref{fig:E_mp} and~\ref{fig:E_cc} shows that the cluster expansion of the wave function and the summation of the diagrams involving the single and double excitation in the MBPT theory to infinite order is crucial for an accurate description of many-atom systems in a one-dimensional harmonic trap. 

The discussion of the last two paragraphs is well summarized in Table~\ref{tab:numbers}, where we show the CC, MBPT, and CISD
results for selected values of the interaction strength. An analysis of the numerical results reported in this
table confirms a very good performance of various variants of the coupled cluster theory, as opposed to the erratic
behavior of the many-body perturbation theory, and the poor performance of the configuration interaction method limited
to single and double excitations.

\begin{table*}[t!]
\caption{The percentage of the ground state correlation energy reproduced at the CCSD, CCSD(T), CCSDT, and CCSDTQ levels of the coupled cluster theory and the many-body perturbation theory MBPT2, MBPT4, and the truncated configuration interaction methods CISD in the 2+1, 2+2,
3+2, and 3+3 systems for the different interaction strengths. N/D (no data) indicates that calculation was not feasible with used software.\label{tab:numbers}}
\begin{ruledtabular}
\begin{tabular}{lrrrrrrr}
system & CCSD & CCSD(T) & CCSDT & CCSDTQ & MBPT2 & MBPT4 & CISD \\
\hline
\multicolumn{8}{c}{$g=2$}\\
$N=2+1$ & 99.60 & 100.07 & 100 & - & 117.33 & 100.31 & 99.42 \\
$N=2+2$ & 99.30 & 100.10 & 99.95 & 100 & 115.26 & 101.49 & 98.70 \\
$N=3+2$ & 99.35 & 100.07 & 99.97 &  N/D  & 113.48 & 100.18 & 97.40 \\
$N=3+3$ & 99.23 &	100.06 & 99.94 & 99.999 & 120.07 & 101.65 & 96.17 \\
\hline
\multicolumn{8}{c}{$g=4$}\\
$N=2+1$ &  97.82 & 99.69 & 100 & - & 92.42 & 98.28 & 95.31 \\
$N=2+2$ &  98.53 & 100.72 & 99.90 & 100 &  153.09 & 127.21 & 95.39 \\
$N=3+2$ &  96.89 & 100.18 & 99.86 & N/D  & 97.54 & 95.05 & 87.18 \\
$N=3+3$ & 98.14 & 100.56 & 99.84 & 99.997
 & 144.33 & 116.83 & 90.64 \\
\hline
\multicolumn{8}{c}{$g=-4$}\\
$N=2+1$ &	98.33 & 99.46 & 100 & - & 76.51 & 100.08 & 98.24 \\
$N=2+2$ & 99.53 & 101.97 & 102.64 & 100 & 68.49 & 96.08 & 87.60 \\
$N=3+2$ & 97.96 & 100.10 & 100.35 & N/D & 97.54 &	95.05 & 87.18 \\
$N=3+3$ & 99.79 & 102.35 & 102.99 & 100.003 & 144.33 & 116.83 & 90.64 \\
\end{tabular}
\end{ruledtabular}
\end{table*}

\begin{figure}[tb]
\begin{center}
\includegraphics[width=0.85\columnwidth]{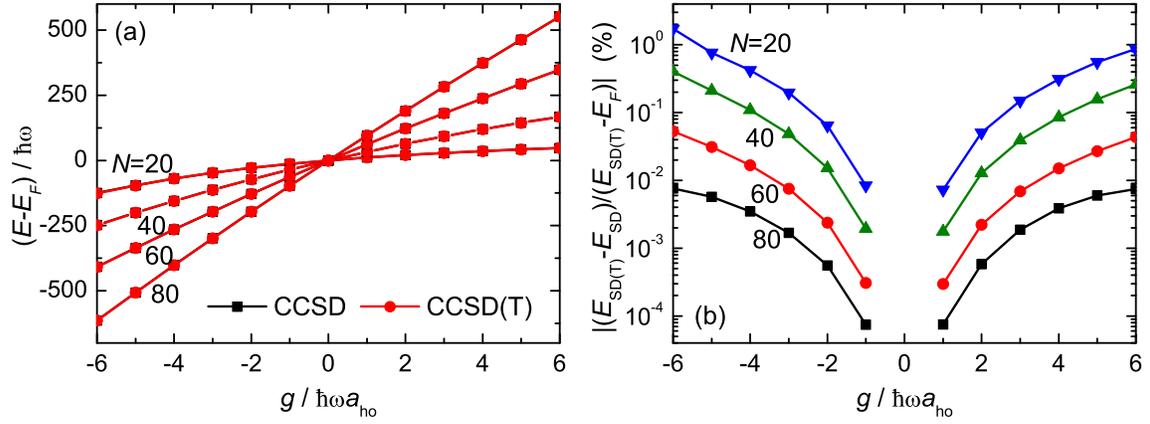}
\end{center}
\caption{(a): The ground state energy (relative to the ground-state energy of the noninteracting system $E_F$) of the 10+10, 20+20, 30+30, and 40+40 systems as a function of the interaction strength obtained at the CCSD and CCSD(T) levels of the coupled cluster theory extrapolated to the complete basis set limit. (b): The percentage of the ground state interaction energy 
accounted by the connected triples contribution calculated non-iteratively within many-body perturbation theory as a function of the interaction strength for the same systems.
}
\label{fig:CC_many}
\end{figure}

Finally, we note that the performance discussed above of the truncated CC methods shows a weak dependence on the
number of one-particle basis functions. Similarly, we do not observe any increase of the relative error with the number of atoms, although our comparison with the FCI results is restricted to the maximum of six atoms. For a larger number of atoms we can evaluate the contributions of the noniterative triple excitations in the CCSD(T) method. Figure~\ref{fig:CC_many}(a) shows the interaction energy in the ground state of the 10+10, 20+20, 30+30, and 40+40 systems obtained at the CCSD and CCSD(T) levels of the coupled cluster theory. Figure~\ref{fig:CC_many}(b) presents the percentage of the ground state interaction energy accounted by the difference between energy obtained with the CCSD(T) and CCSD methods, i.e.~by the connected triples contribution calculated noniteratively within the many-body perturbation theory. 
Interestingly, for a given interaction strength the contribution of the excitations higher than double to the ground-state interaction energy is becoming less important with increasing number of atoms in the system.
Based on this fact we predict that the performance of the CCSD(T) method for larger systems is as good as for the investigated small systems and the lack of higher excitations in the coupled cluster wave function does not introduce any significant errors.     

The energy spectrum of the investigated systems for strong repulsive interactions becomes quasi-degenerate (cf.~Figs.~\ref{fig:spectra} and \ref{fig:1g}) and the correlation energy diverges for both strongly attractive and repulsive interactions (cf.~Fig.~\ref{fig:E_corr}). 
These two effects lead to the convergence problem of the standard single-reference coupled cluster method starting from the antiferromagnetic reference state as used in the present study, restricting the interaction strength $g$ that can effectively be used in actual calculations to intermediate values between -6 and 6.
The possible solution to overcome this problem is the use of the multireference version of the coupled cluster theory or the ferromagnetic reference state for calculations in the limit of the strong repulsive interaction.

\subsection{Density profiles}
\label{sec:density}

The density profiles of the trapped clouds are important observables that can be measured experimentally and used to monitor the evolution of the interatomic interactions and resulting states of a many-body system.

In Ref.~\cite{Grining2015a} we have analyzed the density profiles for the 3+3 system with the FCI method and for the 15+15 system with the CC approach. We have shown that both methods perfectly reproduce the density profiles of the non-interacting gas, and that the FCI calculations allow to approach both the limit of strong attraction (the Lieb-Liniger gas) when the fermionic atoms of opposite spin projection pair into hard-core bosonic dimers, and the limit of strong repulsion (the Tonks-Girardeau gas) when even 
distinguishable fermions must occupy different one-particle levels. The CC method allows to describe the evolution of the  density profiles in the range of intermediate values of the interaction strength. The densities obtained with two methods agree with predictions of the local density approximation applied to the solution of the Gaudin-Yang integral equations describing a homogeneous gas~\cite{Astrakharchik2004} providing a further confirmation of the validity of the local density approximation for investigating a trapped 1D gas. Here, we present other examples for a few fermion systems obtained with the FCI method and for the many fermion systems calculated with the CC approach.

\begin{figure}
\begin{center}
\includegraphics[width=0.8\columnwidth]{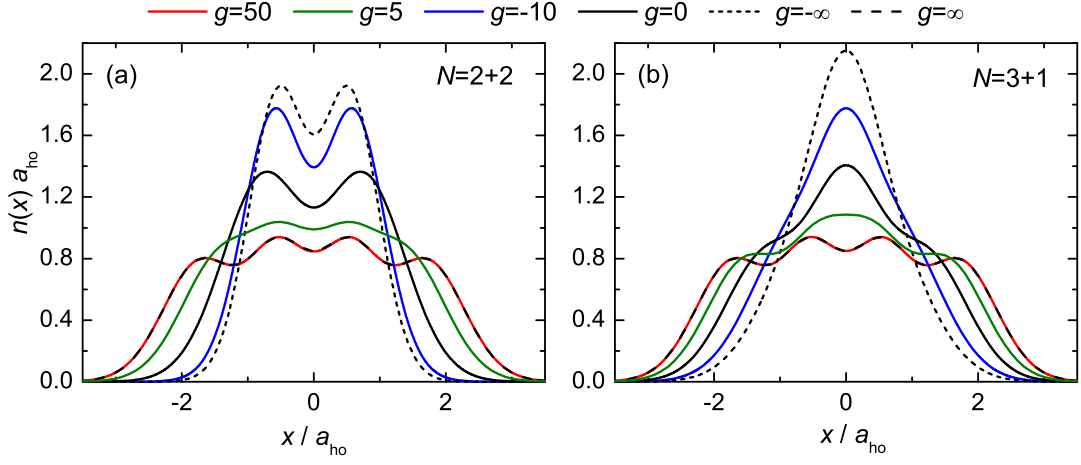}
\end{center}
\caption{Density profiles of a two-component Fermi gas obtained with the FCI method for the 2+2~(a) and 3+1~(b) systems for the three interaction strengths $g=50$, $g=5$ and $g=-10$. The analytic results for the limiting cases of strong attraction ($g=-\infty$), strong repulsion ($g=\infty$), and no interaction ($g=0$) are also presented.
}
\label{fig:densityFCI}
\end{figure}

\begin{figure}
\begin{center}
\includegraphics[width=0.8\columnwidth]{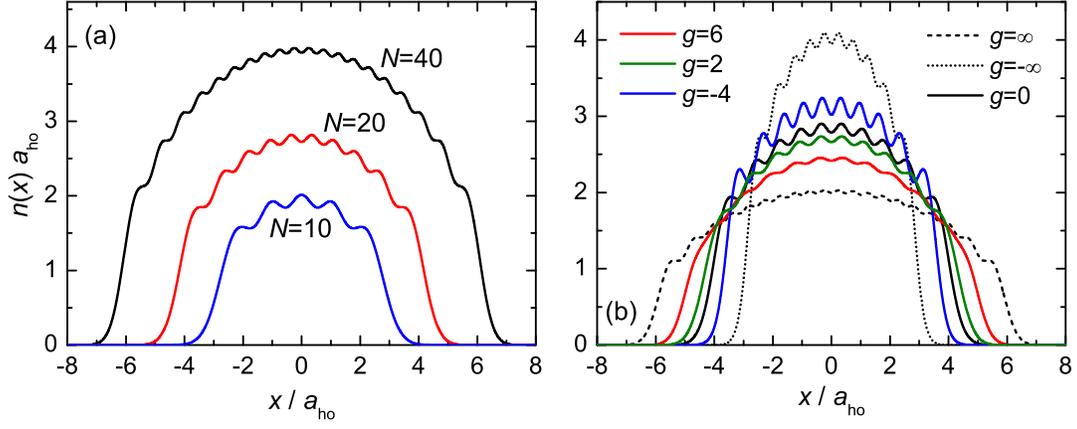}
\end{center}
\caption{Density profiles of a two-component Fermi gas obtained with the CCSD(T) method for the 5+5, 10+10, and 20+20 systems and the interaction strength $g=1$~(a), and for the 5+5 system with  different interaction strengths (b). The analytic results for the limiting cases of strong attraction ($g=-\infty$), strong repulsion ($g=\infty$), and no interaction ($g=0$) are also presented in panel (b).}
\label{fig:densityCC}
\end{figure}

Figure~\ref{fig:densityFCI} shows the density profiles for the 2+2 and 3+1 systems obtained with the FCI method for three interactions strengths: strongly repulsive ($g=50$), moderately repulsive ($g=5$) and strongly attractive ($g=-10$), together with the analytic results. In the balanced case, it is straightforward to find analytical expressions for the limiting cases of strong attraction ($n_{\rm LL}(x)=2\sum_{i=0}^{N/2-1}|\tilde\varphi_i(x)|^2$ for
the Lieb-Liniger gas of hard-core bosonic dimers at $g=-\infty$, with $\tilde\varphi_i(x)$ the $i$-th eigenfunction of the harmonic oscillator for a dimer of mass $2m$), strong repulsion ($n_{\rm TG}(x)=\sum_{i=0}^{N-1}|\varphi_i(x)|^2$ for the Tonks-Girardeau gas of "fermionized" fermions at $g=\infty$), and no interaction ($n_{0}(x)=2\sum_{i=0}^{N/2-1}|\varphi_i(x)|^2$ for $g=0$). In the case of strongly repulsive interaction, the FCI results are indistinguishable from the analytic result confirming perfect performance of the FCI method in this limit of the interaction strength. The regime of the strong attraction is much harder to describe due to the presence of strong two-body correlations when the hard-core bosonic dimers are formed. The presented results for $g=-10$ approach the analytic results but obtaining fully converged results for much more attractive interaction strengths, even with the extrapolation used within the paper, is very challenging. A possible solution to overcome the very slow convergence in the limit of strong attractive interaction can be to use the explicitly correlated method or to include the formation of the hard-core bosonic dimers directly in the structure of the wave function. 

Figure~\ref{fig:densityCC}(a) shows the density profiles for the 5+5, 10+10, and 20+20 systems obtained with the CCSD(T) method for the interaction strength $g=1$. With an increasing number of atoms the evolution of the size of the cloud can be observed.  
As we have shown in Ref.~\cite{Grining2015a}, the 
overall shape of the profiles can  
be described by the typical Thomas-Fermi profile of an inverted parabola. The relative size of the density oscillations are smaller and smaller with the increasing number of atoms, and the density profile approaches the exact shape given by the local density approximation in the thermodynamic limit.   
Figure~\ref{fig:densityCC}(b) shows the density profiles for the 5+5 system obtained with the CCSD(T) method for the three interactions strengths $g=-4$, $g=2$, and $g=6$. The present CC calculations are restricted to intermediate values of the interactions strength, which however extend far beyond the mean field regime \cite{Grining2015a}. Moreover, the accessible range is sufficient to observe significant modifications of the shape and the size of the cloud. The evolution of the cloud's shape towards the analytic results of the limiting cases of strong attraction ($g=-\infty$) and strong repulsion ($g=\infty$) is clearly visible in Fig.~\ref{fig:densityCC}(b). Finally, the CC method allows one to address much larger systems than the FCI approach.

\subsection{Comparison with the available experimental data}
\label{sec:exp}

\begin{figure}
\begin{center}
\includegraphics[width=0.5\columnwidth]{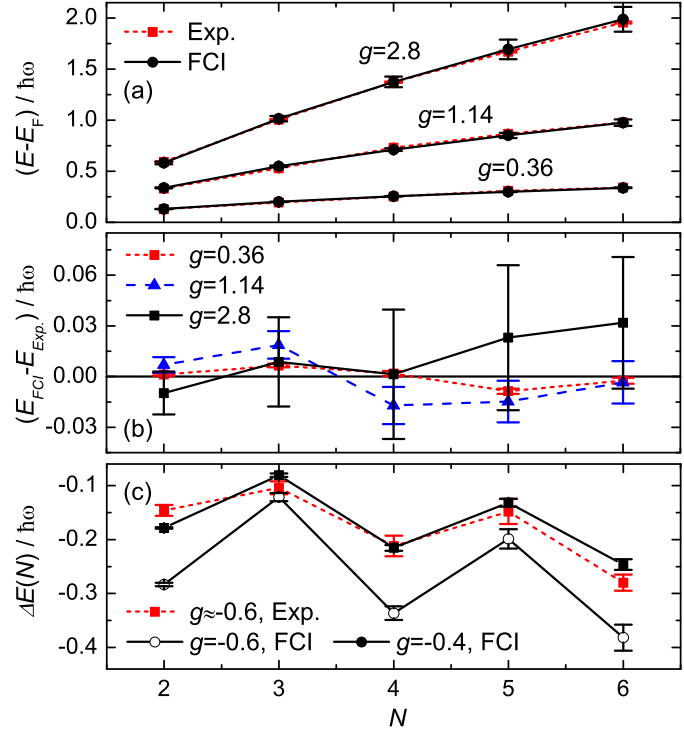}
\end{center}
\caption{(a): The ground state energy (relative to the ground-state energy of the noninteracting system $E_F$) of the $N$ atoms consisting of a single impurity with an opposite spin interacting with an increasing number of identical fermions for various interaction strengths $g$ obtained with the FCI method 
is compared to the measurements of Ref.~\cite{Wenz2013}. (b): The difference between theoretical and experimental values with conservatively estimated error bars of numerical results. (c): The separation energy obtained with the FCI method and measured in experiment~\cite{Zurn2013}.
}
\label{fig:exp}
\end{figure}

All the results reported thus far strongly suggest that the approximate variants of the coupled cluster method
combined with the efficient extrapolation schemes based on the rigorous analytical formulas derived for
the two-body case perform very well for a few-body and many-body systems of identical spin-1/2 atoms 
interacting via the short-range contact potential in a 1D harmonic trap. However, the most stringent 
test of the accuracy of any theoretical model is the comparison with precision experiments. 

We report such a comparison in Fig.~\ref{fig:exp}.
We decided to compare the best of our results, i.e.~the FCI
results, but any variant of the coupled cluster method that would include triple excitation would
lead to the same results within $1\,\%$, indistinguishable from the FCI values on the scale of the plot. Panel $(a)$ of this figure shows the comparison between the
measured~\cite{Wenz2013} and computed energies (with respect to the ground-state energy of the noninteracting system $E_F$) 
of a system of $N=N_\uparrow+1$ atoms consisting of a single impurity interacting with a Fermi sea of $N_\uparrow$ identical fermions
for various repulsive interactions.
An inspection of this figure shows that the agreement between theory and experiment is indeed excellent. 
Note parenthetically that
we include on this figure both the experimental error bars and the computational uncertainties $\Delta_E$ 
conservatively estimated from the extrapolated energies and the energies computed with the largest $n_b=200$ 
basis functions, $\Delta_E = E_{n_b=200}-E_\infty$. On the scale of this graph the experimental and theoretical
error bars are indistinguishable, so in the panel $(b)$ of this figure we report the differences between the
theoretical and experimental energies with the respective error bars. The agreement viewed in this way is
also very good, and the energy difference always lies within the combined theoretical and experimental
uncertainties. 

Less satisfactory is the agreement between theory and experiment for the separation energies
 $\Delta E(N)=\mu(N) - \mu^*(N)$, where $\mu(N)=E(N)-E(N-1)$ is the chemical potential, $E(N)$ is the extrapolated energy for the $N$-body system in the weakly attractive regime, and $\mu^*(N)$ is the chemical potential of the noniteracting system. This is illustrated in the panel~$(c)$ of Fig.~\ref{fig:exp}, where we
compare our theoretical results with the experimental data of Ref.~\cite{Zurn2013}. An inspection of this
figure shows that a relatively good agreement (but not within the experimental error bars) is observed
for odd values of $N$, and important disagreement by a factor of roughly 3/2 for even $N$. Note that
the experimental technique used in Ref.~\cite{Zurn2013} is different than in Ref.~\cite{Wenz2013}. 
The latter one is based on an accurate non-destructive measurement of the RF spectrum of the systems, whereas 
in the former one the system is probed by deforming the trapping potential and by observing the tunneling of particles out of the trap. We believe that the main source of disagreement in the second case comes from the perturbed character of the 1D harmonic shape of the trap during measurements and an approximate determination of the interaction strength in the experiment~\cite{Zurn2013}. Indeed, the results of the calculations with a smaller (in the absolute value) interaction strength $g$ are in a much better agreement for both
even and odd values of $N$, as it has also recently been shown by the authors of Ref.~\cite{DAmico2015}.

\section{Summary and conclusions}
\label{sec:summary}

In this paper we have reported the first application of the \textit{ab initio} methods of quantum chemistry to systems of many interacting fermions in a one-dimensional harmonic trap. Our results can be summarized as follows:
\begin{itemize}
\item The behavior of the two-body energy as a function of the number of single-particle functions $n_b$ included in the calculations for a fixed interaction strength has thoroughly been analyzed and an extrapolation formula has been derived. The convergence with $n_b$ is pathologically slow and is reported in the leading order as $n_b^{-1/2}$.
\item The convergence of the interaction strength as a function of the number of single-particle functions $n_b$ for calculations at fixed energy in the two-body case have also been analyzed and an extrapolation formula which does not depend on the energy and on the number of particles in the lowest order has been derived.
\item The convergence with the number of basis set functions $n_b$ in the many-body case has been analyzed and a suitable three-point extrapolation formula has been found.
\item The importance of the use of converged results to correctly describe the physics of the trapped Fermi systems has been 
pointed out. Shortcomings of the approaches lacking a proper analysis of the convergence issues have been shown.
\item Comparison of several quantum many-body methods has been reported. In particular, a careful consideration has been given to the level of the excitation present in the wavefunction in the coupled cluster method. An optimal method has been chosen and used throughout the rest of the study.
\item The limitations of the adopted coupled cluster implementation with the standard single-reference state have been discussed together with the analysis of the behavior of the decomposition of the total energy into the mean-field and correlation parts in the limits of strong repulsive and attractive interactions. 
\item The coupled cluster method restricted to single, double, and noniterative triple excitations CCSD(T) has been applied to describe systems of up to 80 fermionic atoms.
\item Density profiles have been computed and analyzed in the full spectrum of the interaction parameter and the number of atoms. The transition between the Lieb-Liniger gas of hard-core bosonic dimers and the Tonks-Girardeau gas has been observed with the FCI method.
\item Comparison with the available experimental data including estimates of the computational uncertainties has been provided. Very good agreement between the theory and experiment has been pointed out for the precision measurements based on the RF spectroscopy whereas 
the perturbed character of the 1D harmonic shape of the trap during measurements based on the observation of the tunneling of particles out of the trap
has been confirmed by confrontation with our accurate \textit{ab initio} calculations with estimated error bars.
\end{itemize}

The presented numerical approach allows us to get an important insight into the ongoing experiments on the trapped fermions. By extending the number of atoms from less than ten within the exact diagonalization approach (see, e.g., Ref.~\cite{Sowinski2015}) to many tens within the coupled cluster method, it is becoming possible to investigate the transition between few- and many-body systems. This is particularly important for a good understanding of the emergence of bulk matter properties, crucial across all areas of physics. 

Recently, we have applied the numerical approach developed here to investigate the properties of a two-component Fermi gas trapped in one dimension~\cite{Grining2015a}. We have addressed the question of how the observables, such as the energy, the chemical potential, the pairing gap, and the density profile, evolve as the number of particles increases from very few to many tens. We have found that the energy converges surprisingly rapidly to the many-body result for all interaction strengths between minus and plus infinity, covering the whole range from the molecular bosonic Tonks gas to the atomic (fermionic) one. On the other hand, we observed the emergence of a non-analytic behavior of the pairing gap only when a substantially larger number of particles is present in the trap.

We believe that the results presented here and in Ref.~\cite{Grining2015a} on several fermions in a harmonic trap will be followed by many applications of the proposed numerical approach to investigate interesting physics in different systems, geometries, and dimensions. We foresee that the coupled cluster method will describe atoms trapped in other potentials, such as double wells, arrays of microtraps, or even optical lattices equally well. The method should work also for more complex two-body interaction potentials, e.g., including van der Waals or long range 6dipole-dipole interactions. What is more, the method can be generalized to handle atoms trapped in two or three dimensions. The pathologically slow convergence with the number of single-particle basis functions included in the calculations can be overcome by using explicitly correlated basis functions in analogy with methods developed in quantum chemistry~\cite{Mitroy2013}. The use of the explicitly correlated methods can be crucial for an accurate description of systems in a higher number of dimensions. 
The range of interaction strengths which can be used in calculations with the single-reference coupled cluster method is restricted to intermediate values. This can be overcome by using a ferromagnetic (high-spin) reference state instead of the antiferromagnetic (low-spin) one which is typically used in the standard coupled cluster method. This would allow to describe many tens of interacting atoms in the strongly repulsive regime, and therefore would be ideal for describing atomic clouds in the Tonks-Girardeau limit. 
Finally, one can employ the coupled cluster method based on the high-spin reference state to describe many strongly-interacting bosonic atoms and investigate in detail the fermionization process in such a system.

\section{Acknowledgments}

We wish to thank Ignacio Cirac, Selim Jochim, Bogumi\l\ Jeziorski, Jesper Levinsen, and Meera Parish for fruitful discussions.
We acknowledge support from the EU grants ERC AdG OSYRIS, FP7 SIQS and EQuaM, FETPROACT QUIC, Spanish Ministry project FOQUS (FIS2013-46768-P) and the Generalitat de Catalunya project 2014 SGR 874, Fundaci\'o Cellex, the Ram\'on y Cajal programme, EU Marie Curie COFUND action (ICFOnest),  FNP START and MISTRZ programs, Polish National Science Centre (ST4/04929), and the PL-Grid Infrastructure.

\appendix

\section{The exact solution of the two-body case}
\label{ap:two-body}

Let us consider the Hamiltonian (\ref{hhhh}) with $N_\uparrow=N_\downarrow=1$. By introducing
new coordinates $x=\frac{1}{\sqrt{2}}(x_1-x_2)$ (relative motion) and $X=\frac{1}{\sqrt{2}}(x_1+x_2)$ (centre-of-mass
motion) one arrives at the formula
\begin{align}
\label{hrR}
\left[ -\half \partial_{x}^2-\half \partial_{X}^2 + \half x^2 + \half X^2 +
\bar{g}\delta(x)-E\right]\Psi(x,X)=0,
\end{align}
where $\bar{g}=g/\sqrt{2}$.
In these variables the exact wavefunction, $\Psi(x,X)$, separates into a product of functions $\phi$ and $\xi$
dependent only on $x$ and $X$, respectively. These functions obey the following differential equations
\begin{align}
\label{two}
\begin{split}
&-\half \phi''(x)+\half x^2 \phi(x)+\bar{g}\delta(x)\phi(x)=\lp\epsilon + \frac12\rp\phi(x),\\
&-\half \xi''(X)+\half X^2 \xi(X)=\lp E-\epsilon-\frac12\rp\xi(X),
\end{split}
\end{align}
where primes denote the usual differentiation over the corresponding variables. The equation for the
centre-of-mass motion can immediately be solved, as it coincides with the Schr\"{o}dinger equation for the quantum
harmonic oscillator. 

Further in the text we shall use the shorthand notation, $\hat{H}_x = -\half
\partial_{x}^2 + \half x^2+ \bar{g}\delta(x) - 1/2$, so that $\hat{H}_x\phi(x)=\epsilon \phi(x)$. The
Hamiltonian, $\hat{H}_x$, is invariant with respect to the spatial inversion, i.e., $x\rightarrow -x$. Therefore, its 
eigenfunctions possess a definite parity (even or odd). Eigenfunctions that are of odd parity must vanish at $x=0$.

A substitution $\phi(x)=e^{-x^2/2}f(x)$ in the first equation of (\ref{two}) gives the
corresponding differential equation for $f(x)$
\begin{align}
\label{inh}
f''(x)-2xf'(x)+2\epsilon f(x)=2\bar{g}\delta(x)f(x),
\end{align}
First, we solve the homogeneous differential
equation, i.e., without the Dirac delta term on the right-hand-side, which is well-known and
the general solution can be written as a linear combination of two functions
\begin{align}
C_U\, U(-\epsilon/2,1/2,x^2)+C_M\, M(-\epsilon/2,1/2,x^2),
\end{align}
where $C_U$ and $C_M$ are some constants necessary to satisfy the initial conditions, and $M$ and $U$ are the Kummer and
Tricomi
hypergeometric functions \cite{AbramowitzStegun}, respectively. Returning to the original (inhomogeneous) equation, we consider first the
odd eigenfunctions. As noted beforehand, they vanish at the origin ($x=0$) and are not affected by the presence of the
Dirac delta source. Therefore, for the odd states the solution of the quantum harmonic oscillator is simply obtained.

The problem of even states is more pronounced as the ground state is nodeless and even.
A detailed investigation of the differential equation (\ref{inh}) reveals that the even solutions are also written in
terms of $M$ and $U$, but the initial conditions need to be chosen properly to account for the presence of the Dirac
delta distribution. To find the required initial condition one integrates both sides of Eq. (\ref{inh}) on a small
interval around the origin, i.e., $(-\varepsilon,+\varepsilon)$. Subsequently, all terms that cannot contribute
in the small $\varepsilon$ limit are dropped. One can easily estimate which terms are significant by noting that both
solutions of the homogeneous equation are regular around $x=0$. Finally, the following result is obtained
\begin{align}
\label{init}
f'(+\varepsilon)-f'(-\varepsilon)=2\bar{g}f(0).
\end{align}
Let us recall that around $x=0$ the solutions of the homogeneous equation behave for $x>0$ like:
\begin{align}
&U(-\epsilon/2,1/2,x^2)=\frac{2^\epsilon \sqrt{\pi}}{\Gamma\left(\frac{1-\epsilon}{2}\right)}+\mathcal{O}(x),\\
&U'(-\epsilon/2,1/2,x^2)=\frac{\epsilon 2^\epsilon \sqrt{\pi}}{\Gamma\left(\frac{2-\epsilon}{2}\right)}+\mathcal{O}(x),\\
&M(-\epsilon/2,1/2,x^2)=1+\mathcal{O}(x^2),\\
&M'(-\epsilon/2,1/2,x^2)=-2\epsilon x+\mathcal{O}(x^2).
\end{align}
The functions $M$ are
unable to satisfy the above initial condition. Therefore, we must pick up the solution expressed through the Tricomi
functions $U$ and from Eq. (\ref{init}) we obtain
\begin{align}
\label{pen}
\frac{1}{\bar{g}}=\frac{\Gamma\left(\frac{2-\epsilon}{2}\right)}{\epsilon\Gamma\left(\frac{1-\epsilon}{2}\right)}.
\end{align}
The above expression cannot be solved with elementary methods. Nonetheless, for a given $\bar{g}$ the
solution is straightforward to obtain numerically. Finally, the exact wavefunction
of the Hamiltonian (\ref{hrR}) is given by
\begin{align}
\label{exact-app}
\Psi(x,X)=C_n C_\epsilon\,e^{-\half x^2-\half X^2} H_n(X)\, U(-\epsilon/2,1/2,x^2),
\end{align}
where $C_n$ and $C_\epsilon$ are the normalization constants for the $x$- and $X$-dependent portions, and the
corresponding total energy of the system is simply $E=\epsilon+n+1$. 

Having the exact wavefunction of the system, we can analyze the possible cusp-like conditions at the
particles coalescence points ($x=0$). The wavefunction around $x=0$ behaves as
\begin{align}
\Psi(x,X) \propto 1 - 2\frac{\Gamma(\half-\half \epsilon)}{\Gamma(-\half \epsilon)}\left|x\right| + \mathcal{O}(x^2),
\end{align}
which is a direct consequence of the properties of $U$. The above expression can considerably be simplified with the help
of Eq. (\ref{pen}) which leads to
\begin{align}
\label{cusp1}
\Psi(x,X) \propto 1 + \bar{g}\left|x\right| + \mathcal{O}(x^2).
\end{align}
Strikingly, the behavior of the exact wavefunction around $x=0$ depends only on the value of
$\bar{g}$. This is also the counterpart of the Kato's electron-electron cusp condition. In analogy, Eq. (\ref{cusp1})
can be written as
\begin{align}
\left. \frac{\partial \Psi(x,X)}{\partial x}\right|_{x=0^\pm} = \pm \bar{g}\, \Psi(0,X),
\end{align}
where the equality sign strictly holds for any $X$. Due to the requirement of even parity, the exact wavefunction possesses a derivative discontinuity at $x=0$ and a ``cusp'' at the origin. This discontinuity can be a considerable difficulty from the practical point of view. When the
wavefunction is modeled with smooth basis set functions, the cusp may be very difficult to reproduce.

\section{Energy convergence with the basis set size - truncated exact wave function}
\label{ap:conv_anal}

In the two-body case the expansion of the exact wavefunction in one-dimensional harmonic oscillator eigenfunctions can be written as
\begin{align}
\label{exp2}
\Psi(x,X) = \sum_{m=0}^\infty\sum_{n=0}^\infty c_{mn}\, \varphi_m(x)\, \varphi_n(X).
\end{align}
By recalling the form of the exact wave function, Eq. (\ref{exact}), one finds that the $X$-dependent part of the
wavefunction is automatically described exactly. Therefore, the actual task in the two-body calculations is to
approximate the $x$-dependent part of the wavefunction, $\phi(x)$, by the solutions of the quantum harmonic oscillator
eigenproblem, i.e.,
\begin{align}
\label{pw}
\phi(x) = C_\epsilon \,e^{-\half x^2} U(-\epsilon/2,1/2,x^2) = \sum_{n=0}^\infty c_n\,\varphi_n(x).
\end{align}
Due to the orthonormality of $\varphi_n(x)$ the coefficients $c_n$ obey the formal relationship
\begin{align}
\label{cn1}
c_n = C_\epsilon \int_{-\infty}^{+\infty} dx\, e^{-\half x^2} U(-\epsilon/2,1/2,x^2)\, \varphi_n(x).
\end{align}
Note that since the exact wavefunction is even, the coefficients vanish for odd $n$ by symmetry.

As the basis set, Eq.~\eqref{basis}, is complete in the second Sobolev space, we can construct systematic approximations
to the exact wavefunction by terminating the expansion (\ref{pw}) at some $n_b$. This gives a family of approximants
\begin{align}
\label{phinb}
\phi^{(n_b)}(x)= \sum_{n=0}^{n_b-1} c_n\,\varphi_n(x),
\end{align}
which are \emph{not} normalized to the unity. The energy connected with $\phi^{(n_b)}(x)$ for each $n_b$ is given by
\begin{align}
\label{enb}
\epsilon_{n_b}= \frac{\langle \phi^{(n_b)}|\hat{H}_x|\phi^{(n_b)}\rangle}{\langle \phi^{(n_b)}|\phi^{(n_b)} \rangle}.
\end{align}
Since the exact wavefunction, $\phi(x)$, is normalized, the exact energy of the relative motion is simply
$\epsilon=\langle \phi|\hat{H}_x|\phi
\rangle$. Let us also define the complementary function, 
$\phi^{(n_b)}_c(x)$, given for each $n_b$ by the expression
\begin{align}
\label{psic}
\phi^{(n_b)}_c(x)=\phi(x)-\phi^{(n_b)}(x)=\sum_{n=n_b}^\infty c_n\,\varphi_n(x).
\end{align}
We can now precisely state that we are interested in finding the asymptotic expansion of
$\epsilon_{n_b}-\epsilon$ as $n_b\rightarrow \infty$. At this point we must stress that the energy $\epsilon_{n_b}$ 
is not strictly equal to the energy calculated in the same basis set. In fact, in the actual finite basis set
calculations, the 
coefficients $c_n$ are determined variationally rather than by projection onto the exact wavefunction. Therefore, the 
variationally determined coefficients have an opportunity to relax and accommodate to the incompleteness of the basis
set.
However, for large $n_b$ this relaxation effect is small and virtually limited to the last coefficient, $c_{n_b-1}$, as shown by Carroll 
for the helium atom \cite{Carroll1979}. Therefore, the coefficients $c_n$ and the energies $\epsilon^{(n_b)}$ are expected to be very close
to the corresponding variational values.

Let us recall Eq. (\ref{cn1}) and insert the explicit form of $\varphi_n(x)$
\begin{align}
\label{cn2}
\begin{split}
c_n &= \frac{C_\epsilon \,\pi^{-1/4}}{\sqrt{2^n n!}}v_n,\\
v_n &= \int_{-\infty}^{+\infty} dx\, e^{-x^2} U(-\epsilon/2,1/2,x^2)\,H_n(x).
\end{split}
\end{align}
To evaluate the integrals $v_n$ we need to recall a particular integral representation of $U(a,b,z)$
\begin{align}
\label{uabzint}
U(a,b,z) = \frac{1}{\Gamma(a)}\int_0^\infty dt \,e^{-zt}\, t^{a-1} (1+t)^{b-a-1},
\end{align}
which is valid for $a>0$. Therefore, we confine ourselves here to the regime $\epsilon<0$ which roughly
corresponds to $\bar{g}<0$ (attractive potential). However, one can show that the final result derived here remains valid
also for $\epsilon>0$. By inserting the above expression into the definition of $v_n$ and exchanging the order of integrations, one finds
\begin{equation}
v_n = \frac{1}{\Gamma(-\half \epsilon)}\int_0^\infty dt\, t^{-\half \epsilon-1}(1+t)^{\half \epsilon-\half} \int_{-\infty}^{+\infty} dx\, e^{-x^2(1+t)} H_n(x).
\end{equation}
Let us consider only the inner integral for a moment, denoted by $I$. The change of variables
$u=x\sqrt{1+t}$ gives
\begin{align}
I=\frac{1}{\sqrt{1+t}}\int_{-\infty}^{+\infty} du\, e^{-u^2} H_n\left( \frac{u}{\sqrt{1+t}} \right).
\end{align}
To simplify the Hermite function under the integral sign one can make use of the following relation
\begin{align}
H_n(\gamma u) = \sum_{i=0}^{n/2}\gamma^{n-2i}(\gamma^2-1)^i { n \choose 2i } \frac{(2i)!}{i!} H_{n-2i}(x),
\end{align}
where $\gamma = (1+t)^{-\half}$. Upon inserting into the integral $I$ and making use of the orthogonality of the Hermite
polynomials one arrives at
\begin{align}
\begin{split}
I = \sqrt{\pi}\gamma (\gamma^2-1)^{n/2} \frac{n!}{(n/2)!}.
\end{split}
\end{align}
By returning to the definition of $v_n$ and slightly rearranging the following formula is obtained
\begin{align}
v_n = \frac{\sqrt{\pi}\,(-1)^{n/2}}{\Gamma(-\half \epsilon)}\frac{n!}{(n/2)!} \int_0^\infty dt\,\frac{t^{\half n-\half
\epsilon-1}}{(1+t)^{\half n-\half \epsilon+1}},
\end{align}
and the remaining integral is elementary, so that
\begin{align}
v_n = \frac{\sqrt{\pi}(-1)^{n/2}}{\Gamma(-\half \epsilon)}\frac{n!}{(n/2)!} \frac{2}{n-\epsilon}.
\end{align}
Therefore, the expression for $c_n$ reads
\begin{align}
\label{cn3}
c_n = C_\epsilon\frac{\pi^{1/4}(-1)^{n/2}}{\Gamma(-\half \epsilon)}\frac{\sqrt{n!}}{2^{n/2}(\half n)!} \frac{2}{n-\epsilon}.
\end{align}
and the coefficients vanish for odd $n$, i.e., $c_{2n+1}=0$.
The above expression is fairly difficult to handle because of the presence of the factorials. Fortunately, we
require only their asymptotic forms, i.e. approximately valid for large $n$. With the help of the Stirling formula,
$n!\rightarrow \sqrt{2\pi n}\, n^n e^{-n}$, one arrives at
\begin{align}
\label{cn4}
c_{2n} \rightarrow C_\epsilon \frac{2}{\Gamma(-\half \epsilon)} \frac{(-1)^{n}}{2n-\epsilon} \frac{1}{n^{1/4}},
\;\;\;\mbox{as}\;\;\; n\rightarrow \infty.
\end{align}

The first ingredient required for the presented derivation is the asymptotic formula for the square of the norm, $\langle \phi^{(n_b)}
| \phi^{(n_b)} \rangle$, for large value $n_b$. Taking advantage of the orthonormality of $\varphi_n$ and normalization of
the exact wavefunction one finds
\begin{align}
\label{shit}
\langle \phi^{(n_b)} | \phi^{(n_b)} \rangle = 1 - \sum_{n=n_b}^\infty c_n^2.
\end{align}
Since the second term of the above expression vanishes for large $n_b$, one can write that $\langle
\phi^{(n_b)} | \phi^{(n_b)} \rangle\rightarrow 1$ as $n_b\rightarrow \infty$ which is 
entirely sufficient for the present purposes.
One can also verify that a somewhat more accurate formula, accounting for the next term in the asymptotic expansion,
would include a term proportional to $1/n_b^{3/2}$. However, this term does not contribute to the final results and is 
omitted for brevity.

Because of the approximation derived for the denominator in Eq. (\ref{enb}) one can rewrite it for large $n_b$ as
\begin{align}
\epsilon_{n_b}-\epsilon \approx \langle \phi^{(n_b)}|\hat{H}_x|\phi^{(n_b)}\rangle - \langle \phi|\hat{H}_x|\phi \rangle \,.
\end{align}
By recalling the definition (\ref{psic}) and rearranging one gets
\begin{align}
\label{en1}
\epsilon_{n_b}-\epsilon \approx \langle \phi^{(n_b)}_c | \hat{H}_x | \phi^{(n_b)}_c \rangle - 2 \langle \phi | \hat{H}_x | \phi^{(n_b)}_c\rangle \,.
\end{align}
The first of the above matrix elements can be expanded to give
\begin{equation}
\label{syf2}
\langle \phi^{(n_b)}_c | \hat{H}_x | \phi^{(n_b)}_c \rangle =\sum_{n=n_b}^\infty c_n^2 \left(n+\half\right) + \bar{g} \left[ \sum_{n=n_b}^\infty c_n \varphi_n(0)  \right]^2\,.
\end{equation}
Similarly, for the second matrix element of Eq. (\ref{en1}) one obtains
\begin{equation}
\label{syf1}
\langle \phi | \hat{H}_x | \phi^{(n_b)}_c \rangle =
\sum_{n=n_b}^\infty c_n^2 \left(n+\half\right) + \bar{g}\, \phi(0) \sum_{n=n_b}^\infty c_n \varphi_n(0)\,,
\end{equation}
by noting that $\langle \phi | \varphi_n \rangle=c_n$ and $\phi(0)$ does not depend on $n_b$. 
Note that the following two infinite sums are necessary for the evaluation of Eqs. (\ref{syf1}) and (\ref{syf2})
\begin{align}
\label{t1}
&T_N^{(1)} = \sum_{n=n_b}^\infty c_n^2 \left(n+\half\right) = \sum_{n=n_b/2}^\infty c_{2n}^2\lp2n+\frac12\rp, \\
\label{t2}
&T_N^{(2)} = \sum_{n=n_b}^\infty c_n \varphi_n(0) = \sum_{n_b/2}^\infty c_{2n}\varphi_{2n}(0).
\end{align}
The simplest way to evaluate these sums for large $n_b$ is to exchange the summation indices to run only over even values
of $n$ and subsequently insert the asymptotic formula for $c_{2n}$, Eq. (\ref{cn4}). The resulting expressions are (showing only the first term of the Stirling series)
\begin{align}
\label{t1b}
&T_N^{(1)} \rightarrow \frac{2^{2}C_\epsilon^2}{\Gamma(-\half \epsilon)^2} \sum_{n=n_b/2}^\infty \frac{1}{(2n-\epsilon)^2}
\frac{2n+\half}{\sqrt{n}}, \\
\label{t2b}
&T_N^{(2)} \rightarrow \frac{2C_\epsilon}{\sqrt{\pi}\,\Gamma(-\half \epsilon)} \sum_{n=n_b/2}^\infty
\frac{1}{2n-\epsilon}
\frac{1}{\sqrt{n}},
\end{align}
where we have additionally used the relationship 
\begin{align}
 \varphi_{2n}(0) \rightarrow \frac{(-1)^{n}}{\sqrt{\pi}} n^{-1/4}\;\;\;\mbox{as}\;\;\; n\rightarrow \infty,
\end{align}
easily obtained from Eq. (\ref{basis}) by using the Stirling approximation. Finally, the above sums (taking into consideration all relevant terms in the Stirling series) can be evaluated
with the help of the Euler-MacLaurin resummation formula which leads to
\begin{align}
\label{t1c}
&T_N^{(1)} = \frac{2^{5/2}C_\epsilon^2}{\Gamma(-\half \epsilon)^2} \left(
\frac{1}{n_b^{1/2}}+\frac{8\epsilon+7}{48\,n_b^{3/2}} \right) + \mathcal{O}\left(n_b^{-\frac{5}{2}}\right),\\
&T_N^{(2)} = \frac{2^{3/2}C_\epsilon}{\sqrt{\pi}\,\Gamma(-\half
\epsilon)}\left(\frac{1}{n_b^{1/2}}+\frac{8\epsilon+11}{48\,n_b^{3/2}}\right) +
\mathcal{O}\left(n_b^{-\frac{5}{2}}\right).
\end{align}
Upon reinserting these expressions into Eqs.~(\ref{syf1}), (\ref{syf2}), and (\ref{en1}) one arrives at
\begin{align}
\label{econv}
 \epsilon_{n_b}-\epsilon = \frac{2^{5/2}C_\epsilon^2}{\Gamma(-\half \epsilon)^2}
 \left( \frac{1}{\sqrt{n_b}}+\frac{\bar{g}}{\pi} \frac{\sqrt{2}}{n_b}\right) + \mathcal{O}\left(n_b^{-\frac{3}{2}}\right).
\end{align}
Let us also mention that with the same methodology as presented above one can derive further terms in the asymptotic
expansion of $\epsilon_{n_b}-\epsilon$. For example, the expression including additionally terms proportional to ${n_b}^{-3/2}$ reads
\begin{equation}
\label{econv3}
 \epsilon_{n_b}-\epsilon = \frac{2^{5/2}C_\epsilon^2}{\Gamma(-\half \epsilon)^2}
 \left( \frac{1}{\sqrt{n_b}}+\frac{\bar{g}}{\pi} \frac{\sqrt{2}}{n_b}\right. 
\left. + \frac{15+8\epsilon}{48 \,n_b^{3/2}} \right) + \mathcal{O}\left(n_b^{-2}\right)\,.
\end{equation}
Note that the convergence proportional to $1/{\sqrt{n_b}}$ in the leading order is extremely slow. 
The latter convergence formula may be employed to describe numerical results in the weak and intermediate coupling regime. In the next appendix we will instead outline a different approach, which yields a convergence formula valid for all coupling strengths.

\section{Energy convergence with the basis set size - variational wave function}
\label{ap:conv_var}

In Appendix \ref{ap:conv_anal} we have considered the problem of the convergence of the calculations with the increasing basis set size
for the two-particle case by calculating the energy given by expectation value~\eqref{enb} with the truncated exact wave function~\eqref{phinb} for a finite interaction strength $g$. Now, we will present an alternative derivation, allowing for a variational relaxation of the coefficients $c_n$ in the truncated wave function~\eqref{phinb}. Obviously, since the functional that we minimize is quadratic, the resulting wave function  corresponds  to {\it  the exact ground state wave function of the truncated (projected) Hamiltonian}. 
In the following, we shall first consider the numerical convergence of the inverse coupling constant versus $n_b$, at fixed energy $\epsilon$, and finally derive the convergence of the energy at fixed coupling.

By minimizing the expectation value of the Hamiltonian $\hat{H}_x$ on the wavefunction~\eqref{phinb} with respect to its coefficients $c_n$, one finds
\begin{equation}
\label{minimize}
0=\frac{\partial}{\partial c_n^*}\bra{\varphi}\hat{H}_x-\epsilon\ket{\varphi}=(n-\epsilon_{n_b})c_n+g f_n\sum_{n'}c_{n'}f_{n'}\,,
\end{equation}
where $2^{1/4}f_n=\sum_k \bra{\varphi_n}k\rangle=\varphi_n(0)$ is the $n$-th harmonic oscillator wavefunction for the relative coordinate, evaluated at $x=0$.
We have $f_{2n+1}=0$, and
\beq
(f_{2n})^2=\frac{1}{\sqrt{2\pi}}\frac{(2n-1)!!}{2n!!}=\frac{1}{\sqrt{2\pi}}\frac{(2n)!}{2^{2n}(n!)^2}\,.
\eeq

By solving Eq.~\eqref{minimize} for $g$ in a procedure similar to that of Busch et al.~\cite{Busch1998}, one finds the result presented in Eq.~\eqref{eq:1/g} of the main text, 
\beq
g^{-1}(\epsilon)=\sum_{n=0}^\infty\frac{(f_{2n})^2}{\epsilon-2n}=\frac{\Gamma[(2-\epsilon)/2]}{\sqrt{2}\epsilon\Gamma[(1-\epsilon)/2]}\,.
\label{oneovergtwobody}
\eeq
Let us now define $x(\epsilon)\equiv g^{-1}(\epsilon)$.
In numerical calculations, one has to truncate the basis to a certain maximum number of states $n_b$, and therefore obtains only an approximate value,
\beq
x_{n_b}(\epsilon)=\frac{1}{g_{n_b}(\epsilon)}\equiv\sum_{n=0}^{n_b-1}\frac{(f_{2n})^2}{\epsilon-2n}\,.
\eeq
The convergence rate of this formula at a fixed energy $\epsilon$, as a function of the number of states included in the summation, $n_b$, is given by
\beq
\begin{split}
x_{n_b}(\epsilon)-x(\epsilon)&=\sum_{n=n_b}^\infty\frac{(f_{2n})^2}{2n-\epsilon}
=\frac{1}{\sqrt{2}\pi}\left(\frac{1}{n_b^{1/2}}+\frac{5+4\epsilon}{24n_b^{3/2}}\right)
\\&+\mathcal{O}(n_b^{-2})\,,
\label{Deltax}
\end{split}
\eeq
as may be found using Stirling's formula, and then expanding the fraction inside Eq.\ \eqref{oneovergtwobody} for $2n\gg |\epsilon|$.

Let us now consider the numerical error $\Delta \epsilon_{n_b}=\tilde\epsilon_{n_b}-\epsilon$ obtained by computing the energy at a fixed $g$ with a basis set containing up to and including $n_b$ functions. The error is given by the solution of the implicit equation
\beq
x(\epsilon)=x_{n_b}(\tilde\epsilon_{n_b})\,,
\eeq
which can be rewritten as 
\begin{equation}
\Delta x_{n_b}(\epsilon+\Delta \epsilon_{n_b})=-[x(\epsilon+\Delta \epsilon_{n_b})-x(\epsilon)]\,.
\label{convergenceAtFixedInteraction}
\end{equation}
On the left-hand-side of the above equation we can now use Eq.~\eqref{Deltax}. 
 Assuming that $\Delta \epsilon_{n_b}$ can be expanded in powers of $1/\sqrt{n_b}$ and equating the coefficients on both sides, we obtain
\begin{equation}
\tilde\epsilon_{n_b}-\epsilon \approx \A
\left\{\frac{1}{n_b^{1/2}}-\frac{\A x''}{2x'}\frac{1}{n_b}+\right.
\left.\left[\frac{5+4\epsilon}{24}+\frac{\A^2}{2}\left(\frac{x''}{x'}\right)^2-\frac{x'''}{3x'}\right]\frac{1}{n_b^{3/2}} \right\}+\mathcal{O}(n_b^{-2})\,,
\label{convDeltaE}
\end{equation}
with $\mathcal{A}=-1/(\sqrt{2}\pi x')$, and $x'\equiv\partial_\epsilon [g^{-1}(\epsilon)]$.
In the vicinity of the Tonks-Girardeau point where $\epsilon\approx1$ we have $\epsilon=1-g^{-1}\sqrt{8/\pi}+\ldots$, so that $\mathcal{A}=2/\pi^{3/2}$. In the weak coupling limit $\epsilon\approx 0$ instead, we have $\epsilon=g/\sqrt{2\pi}+\ldots$ so that $\mathcal{A}=g^2/(2\pi^{3/2})$. Both these limits coincide with the analytical results of Ref.~\cite{Levinsen2014}.

\bibliography{fermions_CC}

\begin{thebibliography}{117}%
\makeatletter
\providecommand \@ifxundefined [1]{%
 \@ifx{#1\undefined}
}%
\providecommand \@ifnum [1]{%
 \ifnum #1\expandafter \@firstoftwo
 \else \expandafter \@secondoftwo
 \fi
}%
\providecommand \@ifx [1]{%
 \ifx #1\expandafter \@firstoftwo
 \else \expandafter \@secondoftwo
 \fi
}%
\providecommand \natexlab [1]{#1}%
\providecommand \enquote  [1]{``#1''}%
\providecommand \bibnamefont  [1]{#1}%
\providecommand \bibfnamefont [1]{#1}%
\providecommand \citenamefont [1]{#1}%
\providecommand \href@noop [0]{\@secondoftwo}%
\providecommand \href [0]{\begingroup \@sanitize@url \@href}%
\providecommand \@href[1]{\@@startlink{#1}\@@href}%
\providecommand \@@href[1]{\endgroup#1\@@endlink}%
\providecommand \@sanitize@url [0]{\catcode `\\12\catcode `\$12\catcode
  `\&12\catcode `\#12\catcode `\^12\catcode `\_12\catcode `\%12\relax}%
\providecommand \@@startlink[1]{}%
\providecommand \@@endlink[0]{}%
\providecommand \url  [0]{\begingroup\@sanitize@url \@url }%
\providecommand \@url [1]{\endgroup\@href {#1}{\urlprefix }}%
\providecommand \urlprefix  [0]{URL }%
\providecommand \Eprint [0]{\href }%
\providecommand \doibase [0]{http://dx.doi.org/}%
\providecommand \selectlanguage [0]{\@gobble}%
\providecommand \bibinfo  [0]{\@secondoftwo}%
\providecommand \bibfield  [0]{\@secondoftwo}%
\providecommand \translation [1]{[#1]}%
\providecommand \BibitemOpen [0]{}%
\providecommand \bibitemStop [0]{}%
\providecommand \bibitemNoStop [0]{.\EOS\space}%
\providecommand \EOS [0]{\spacefactor3000\relax}%
\providecommand \BibitemShut  [1]{\csname bibitem#1\endcsname}%
\let\auto@bib@innerbib\@empty
\bibitem [{\citenamefont {Lewenstein}\ \emph {et~al.}(2012)\citenamefont
  {Lewenstein}, \citenamefont {Sanpera},\ and\ \citenamefont
  {Ahufinger}}]{Lewenstein2012}%
  \BibitemOpen
  \bibfield  {author} {\bibinfo {author} {\bibfnamefont {M.}~\bibnamefont
  {Lewenstein}}, \bibinfo {author} {\bibfnamefont {A.}~\bibnamefont {Sanpera}},
  \ and\ \bibinfo {author} {\bibfnamefont {V.}~\bibnamefont {Ahufinger}},\
  }\href@noop {} {\emph {\bibinfo {title} {Ultracold atoms in Optical Lattices:
  simulating quantum many body physics}}}\ (\bibinfo  {publisher} {Oxford
  University Press},\ \bibinfo {address} {Oxford},\ \bibinfo {year}
  {2012})\BibitemShut {NoStop}%
\bibitem [{\citenamefont {Bloch}\ \emph {et~al.}(2008)\citenamefont {Bloch},
  \citenamefont {Dalibard},\ and\ \citenamefont {Zwerger}}]{Bloch2008}%
  \BibitemOpen
  \bibfield  {author} {\bibinfo {author} {\bibfnamefont {I.}~\bibnamefont
  {Bloch}}, \bibinfo {author} {\bibfnamefont {J.}~\bibnamefont {Dalibard}}, \
  and\ \bibinfo {author} {\bibfnamefont {W.}~\bibnamefont {Zwerger}},\ }\href
  {\doibase 10.1103/RevModPhys.80.885} {\bibfield  {journal} {\bibinfo
  {journal} {Rev. Mod. Phys.}\ }\textbf {\bibinfo {volume} {80}},\ \bibinfo
  {pages} {885} (\bibinfo {year} {2008})}\BibitemShut {NoStop}%
\bibitem [{\citenamefont {Salomon}\ \emph {et~al.}(2012)\citenamefont
  {Salomon}, \citenamefont {Shlyapnikov},\ and\ \citenamefont
  {Cugliandolo}}]{HouchesVol94}%
  \BibitemOpen
  \bibinfo {editor} {\bibfnamefont {C.}~\bibnamefont {Salomon}}, \bibinfo
  {editor} {\bibfnamefont {G.~V.}\ \bibnamefont {Shlyapnikov}}, \ and\ \bibinfo
  {editor} {\bibfnamefont {L.~F.}\ \bibnamefont {Cugliandolo}},\ eds.,\
  \href@noop {} {\emph {\bibinfo {title} {Many-Body Physics with Ultracold
  Gases}}},\ \bibinfo {series} {Lecture Notes of Les Houches Summer School},
  Vol.~\bibinfo {volume} {94}\ (\bibinfo  {publisher} {Oxford University
  Press},\ \bibinfo {year} {2012})\BibitemShut {NoStop}%
\bibitem [{\citenamefont {Giorgini}\ \emph {et~al.}(2008)\citenamefont
  {Giorgini}, \citenamefont {Pitaevskii},\ and\ \citenamefont
  {Stringari}}]{Giorgini2008}%
  \BibitemOpen
  \bibfield  {author} {\bibinfo {author} {\bibfnamefont {S.}~\bibnamefont
  {Giorgini}}, \bibinfo {author} {\bibfnamefont {L.~P.}\ \bibnamefont
  {Pitaevskii}}, \ and\ \bibinfo {author} {\bibfnamefont {S.}~\bibnamefont
  {Stringari}},\ }\href {\doibase 10.1103/RevModPhys.80.1215} {\bibfield
  {journal} {\bibinfo  {journal} {Rev. Mod. Phys.}\ }\textbf {\bibinfo {volume}
  {80}},\ \bibinfo {pages} {1215} (\bibinfo {year} {2008})}\BibitemShut
  {NoStop}%
\bibitem [{\citenamefont {Guan}\ \emph {et~al.}(2013)\citenamefont {Guan},
  \citenamefont {Batchelor},\ and\ \citenamefont {Lee}}]{Guan2013}%
  \BibitemOpen
  \bibfield  {author} {\bibinfo {author} {\bibfnamefont {X.-W.}\ \bibnamefont
  {Guan}}, \bibinfo {author} {\bibfnamefont {M.~T.}\ \bibnamefont {Batchelor}},
  \ and\ \bibinfo {author} {\bibfnamefont {C.}~\bibnamefont {Lee}},\ }\href
  {\doibase 10.1103/RevModPhys.85.1633} {\bibfield  {journal} {\bibinfo
  {journal} {Rev. Mod. Phys.}\ }\textbf {\bibinfo {volume} {85}},\ \bibinfo
  {pages} {1633} (\bibinfo {year} {2013})}\BibitemShut {NoStop}%
\bibitem [{\citenamefont {Blume}(2012)}]{Blume2012}%
  \BibitemOpen
  \bibfield  {author} {\bibinfo {author} {\bibfnamefont {D.}~\bibnamefont
  {Blume}},\ }\href {\doibase 10.1088/0034-4885/75/4/046401} {\bibfield
  {journal} {\bibinfo  {journal} {Rep. Prog. Phys.}\ }\textbf {\bibinfo
  {volume} {75}},\ \bibinfo {pages} {046401} (\bibinfo {year}
  {2012})}\BibitemShut {NoStop}%
\bibitem [{\citenamefont {Lewenstein}\ \emph {et~al.}(2007)\citenamefont
  {Lewenstein}, \citenamefont {Sanpera}, \citenamefont {Ahufinger},
  \citenamefont {Damski}, \citenamefont {Sen(De)},\ and\ \citenamefont
  {Sen}}]{Lewenstein2007}%
  \BibitemOpen
  \bibfield  {author} {\bibinfo {author} {\bibfnamefont {M.}~\bibnamefont
  {Lewenstein}}, \bibinfo {author} {\bibfnamefont {A.}~\bibnamefont {Sanpera}},
  \bibinfo {author} {\bibfnamefont {V.}~\bibnamefont {Ahufinger}}, \bibinfo
  {author} {\bibfnamefont {B.}~\bibnamefont {Damski}}, \bibinfo {author}
  {\bibfnamefont {A.}~\bibnamefont {Sen(De)}}, \ and\ \bibinfo {author}
  {\bibfnamefont {U.}~\bibnamefont {Sen}},\ }\href {\doibase
  10.1080/00018730701223200} {\bibfield  {journal} {\bibinfo  {journal} {Adv.
  Phys.}\ }\textbf {\bibinfo {volume} {56}},\ \bibinfo {pages} {243} (\bibinfo
  {year} {2007})}\BibitemShut {NoStop}%
\bibitem [{\citenamefont {Giamarchi}(2004)}]{Giamarchi}%
  \BibitemOpen
  \bibfield  {author} {\bibinfo {author} {\bibfnamefont {T.}~\bibnamefont
  {Giamarchi}},\ }\href@noop {} {\emph {\bibinfo {title} {Quantum Physics in
  One Dimension}}}\ (\bibinfo  {publisher} {Clarendon Press},\ \bibinfo
  {address} {Oxford},\ \bibinfo {year} {2004})\BibitemShut {NoStop}%
\bibitem [{\citenamefont {Moritz}\ \emph {et~al.}(2003)\citenamefont {Moritz},
  \citenamefont {St\"oferle}, \citenamefont {K\"ohl},\ and\ \citenamefont
  {Esslinger}}]{Tilman1D}%
  \BibitemOpen
  \bibfield  {author} {\bibinfo {author} {\bibfnamefont {H.}~\bibnamefont
  {Moritz}}, \bibinfo {author} {\bibfnamefont {T.}~\bibnamefont {St\"oferle}},
  \bibinfo {author} {\bibfnamefont {M.}~\bibnamefont {K\"ohl}}, \ and\ \bibinfo
  {author} {\bibfnamefont {T.}~\bibnamefont {Esslinger}},\ }\href {\doibase
  10.1103/PhysRevLett.91.250402} {\bibfield  {journal} {\bibinfo  {journal}
  {Phys. Rev. Lett.}\ }\textbf {\bibinfo {volume} {91}},\ \bibinfo {pages}
  {250402} (\bibinfo {year} {2003})}\BibitemShut {NoStop}%
\bibitem [{\citenamefont {Kinoshita}\ \emph {et~al.}(2004)\citenamefont
  {Kinoshita}, \citenamefont {Wenger},\ and\ \citenamefont {Weiss}}]{Weiss1D}%
  \BibitemOpen
  \bibfield  {author} {\bibinfo {author} {\bibfnamefont {T.}~\bibnamefont
  {Kinoshita}}, \bibinfo {author} {\bibfnamefont {T.}~\bibnamefont {Wenger}}, \
  and\ \bibinfo {author} {\bibfnamefont {D.~S.}\ \bibnamefont {Weiss}},\ }\href
  {\doibase 10.1126/science.1100700} {\bibfield  {journal} {\bibinfo  {journal}
  {Science}\ }\textbf {\bibinfo {volume} {305}},\ \bibinfo {pages} {1125}
  (\bibinfo {year} {2004})}\BibitemShut {NoStop}%
\bibitem [{\citenamefont {Paredes}\ \emph {et~al.}(2004)\citenamefont
  {Paredes}, \citenamefont {Widera}, \citenamefont {Murg}, \citenamefont
  {Mandel}, \citenamefont {Foelling}, \citenamefont {Cirac}, \citenamefont
  {Shlyapnikov}, \citenamefont {Hansch},\ and\ \citenamefont {Bloch}}]{Belen}%
  \BibitemOpen
  \bibfield  {author} {\bibinfo {author} {\bibfnamefont {B.}~\bibnamefont
  {Paredes}}, \bibinfo {author} {\bibfnamefont {A.}~\bibnamefont {Widera}},
  \bibinfo {author} {\bibfnamefont {V.}~\bibnamefont {Murg}}, \bibinfo {author}
  {\bibfnamefont {O.}~\bibnamefont {Mandel}}, \bibinfo {author} {\bibfnamefont
  {S.}~\bibnamefont {Foelling}}, \bibinfo {author} {\bibfnamefont
  {I.}~\bibnamefont {Cirac}}, \bibinfo {author} {\bibfnamefont {G.~V.}\
  \bibnamefont {Shlyapnikov}}, \bibinfo {author} {\bibfnamefont {T.~W.}\
  \bibnamefont {Hansch}}, \ and\ \bibinfo {author} {\bibfnamefont
  {I.}~\bibnamefont {Bloch}},\ }\href {\doibase 10.1038/nature02530} {\bibfield
   {journal} {\bibinfo  {journal} {Nature}\ }\textbf {\bibinfo {volume}
  {429}},\ \bibinfo {pages} {277} (\bibinfo {year} {2004})}\BibitemShut
  {NoStop}%
\bibitem [{\citenamefont {Haller}\ \emph {et~al.}(2009)\citenamefont {Haller},
  \citenamefont {Gustavsson}, \citenamefont {Mark}, \citenamefont {Danzl},
  \citenamefont {Hart}, \citenamefont {Pupillo},\ and\ \citenamefont
  {N\"agerl}}]{Haller2009}%
  \BibitemOpen
  \bibfield  {author} {\bibinfo {author} {\bibfnamefont {E.}~\bibnamefont
  {Haller}}, \bibinfo {author} {\bibfnamefont {M.}~\bibnamefont {Gustavsson}},
  \bibinfo {author} {\bibfnamefont {M.~J.}\ \bibnamefont {Mark}}, \bibinfo
  {author} {\bibfnamefont {J.~G.}\ \bibnamefont {Danzl}}, \bibinfo {author}
  {\bibfnamefont {R.}~\bibnamefont {Hart}}, \bibinfo {author} {\bibfnamefont
  {G.}~\bibnamefont {Pupillo}}, \ and\ \bibinfo {author} {\bibfnamefont
  {H.-C.}\ \bibnamefont {N\"agerl}},\ }\href {\doibase 10.1126/science.1175850}
  {\bibfield  {journal} {\bibinfo  {journal} {Science}\ }\textbf {\bibinfo
  {volume} {325}},\ \bibinfo {pages} {1224} (\bibinfo {year}
  {2009})}\BibitemShut {NoStop}%
\bibitem [{\citenamefont {Tonks}(1936)}]{Tonks1936}%
  \BibitemOpen
  \bibfield  {author} {\bibinfo {author} {\bibfnamefont {L.}~\bibnamefont
  {Tonks}},\ }\href {\doibase 10.1103/PhysRev.50.955} {\bibfield  {journal}
  {\bibinfo  {journal} {Phys. Rev.}\ }\textbf {\bibinfo {volume} {50}},\
  \bibinfo {pages} {955} (\bibinfo {year} {1936})}\BibitemShut {NoStop}%
\bibitem [{\citenamefont {Girardeau}(1960)}]{Marvin1960}%
  \BibitemOpen
  \bibfield  {author} {\bibinfo {author} {\bibfnamefont {M.}~\bibnamefont
  {Girardeau}},\ }\href {\doibase 10.1063/1.1703687} {\bibfield  {journal}
  {\bibinfo  {journal} {J. Math. Phys.}\ }\textbf {\bibinfo {volume} {1}},\
  \bibinfo {pages} {516} (\bibinfo {year} {1960})}\BibitemShut {NoStop}%
\bibitem [{\citenamefont {Astrakharchik}\ and\ \citenamefont
  {Girardeau}(2011)}]{Marvincoulomb}%
  \BibitemOpen
  \bibfield  {author} {\bibinfo {author} {\bibfnamefont {G.~E.}\ \bibnamefont
  {Astrakharchik}}\ and\ \bibinfo {author} {\bibfnamefont {M.~D.}\ \bibnamefont
  {Girardeau}},\ }\href {\doibase 10.1103/PhysRevB.83.153303} {\bibfield
  {journal} {\bibinfo  {journal} {Phys. Rev. B}\ }\textbf {\bibinfo {volume}
  {83}},\ \bibinfo {pages} {153303} (\bibinfo {year} {2011})}\BibitemShut
  {NoStop}%
\bibitem [{\citenamefont {Girardeau}\ \emph {et~al.}(2001)\citenamefont
  {Girardeau}, \citenamefont {Wright},\ and\ \citenamefont
  {Triscari}}]{Marvin2}%
  \BibitemOpen
  \bibfield  {author} {\bibinfo {author} {\bibfnamefont {M.~D.}\ \bibnamefont
  {Girardeau}}, \bibinfo {author} {\bibfnamefont {E.~M.}\ \bibnamefont
  {Wright}}, \ and\ \bibinfo {author} {\bibfnamefont {J.~M.}\ \bibnamefont
  {Triscari}},\ }\href {\doibase 10.1103/PhysRevA.63.033601} {\bibfield
  {journal} {\bibinfo  {journal} {Phys. Rev. A}\ }\textbf {\bibinfo {volume}
  {63}},\ \bibinfo {pages} {033601} (\bibinfo {year} {2001})}\BibitemShut
  {NoStop}%
\bibitem [{\citenamefont {Girardeau}\ and\ \citenamefont
  {Wright}(2000)}]{Marvin4}%
  \BibitemOpen
  \bibfield  {author} {\bibinfo {author} {\bibfnamefont {M.~D.}\ \bibnamefont
  {Girardeau}}\ and\ \bibinfo {author} {\bibfnamefont {E.~M.}\ \bibnamefont
  {Wright}},\ }\href {\doibase 10.1103/PhysRevLett.84.5691} {\bibfield
  {journal} {\bibinfo  {journal} {Phys. Rev. Lett.}\ }\textbf {\bibinfo
  {volume} {84}},\ \bibinfo {pages} {5691} (\bibinfo {year}
  {2000})}\BibitemShut {NoStop}%
\bibitem [{\citenamefont {Yukalov}\ and\ \citenamefont
  {Girardeau}(2005)}]{Marvin5}%
  \BibitemOpen
  \bibfield  {author} {\bibinfo {author} {\bibfnamefont {V.}~\bibnamefont
  {Yukalov}}\ and\ \bibinfo {author} {\bibfnamefont {M.~D.}\ \bibnamefont
  {Girardeau}},\ }\href {\doibase 10.1002/lapl.200510011} {\bibfield  {journal}
  {\bibinfo  {journal} {Laser Phys. Lett.}\ }\textbf {\bibinfo {volume} {2}},\
  \bibinfo {pages} {375} (\bibinfo {year} {2005})}\BibitemShut {NoStop}%
\bibitem [{\citenamefont {Girardeau}\ \emph {et~al.}(2004)\citenamefont
  {Girardeau}, \citenamefont {Nguyen},\ and\ \citenamefont
  {Olshanii}}]{Marvin6}%
  \BibitemOpen
  \bibfield  {author} {\bibinfo {author} {\bibfnamefont {M.~D.}\ \bibnamefont
  {Girardeau}}, \bibinfo {author} {\bibfnamefont {H.}~\bibnamefont {Nguyen}}, \
  and\ \bibinfo {author} {\bibfnamefont {M.}~\bibnamefont {Olshanii}},\ }\href
  {\doibase 10.1016/j.optcom.2004.09.079} {\bibfield  {journal} {\bibinfo
  {journal} {Opt. Commun.}\ }\textbf {\bibinfo {volume} {243}},\ \bibinfo
  {pages} {3} (\bibinfo {year} {2004})}\BibitemShut {NoStop}%
\bibitem [{\citenamefont {Girardeau}(2006)}]{Marvinanyon}%
  \BibitemOpen
  \bibfield  {author} {\bibinfo {author} {\bibfnamefont {M.~D.}\ \bibnamefont
  {Girardeau}},\ }\href {\doibase 10.1103/PhysRevLett.97.100402} {\bibfield
  {journal} {\bibinfo  {journal} {Phys. Rev. Lett.}\ }\textbf {\bibinfo
  {volume} {97}},\ \bibinfo {pages} {100402} (\bibinfo {year}
  {2006})}\BibitemShut {NoStop}%
\bibitem [{\citenamefont {Girardeau}\ and\ \citenamefont
  {Astrakharchik}(2010)}]{Marvinsuper}%
  \BibitemOpen
  \bibfield  {author} {\bibinfo {author} {\bibfnamefont {M.~D.}\ \bibnamefont
  {Girardeau}}\ and\ \bibinfo {author} {\bibfnamefont {G.~E.}\ \bibnamefont
  {Astrakharchik}},\ }\href {\doibase 10.1103/PhysRevA.81.061601} {\bibfield
  {journal} {\bibinfo  {journal} {Phys. Rev. A}\ }\textbf {\bibinfo {volume}
  {81}},\ \bibinfo {pages} {061601} (\bibinfo {year} {2010})}\BibitemShut
  {NoStop}%
\bibitem [{\citenamefont {Girardeau}(2011)}]{Marvinsuper1}%
  \BibitemOpen
  \bibfield  {author} {\bibinfo {author} {\bibfnamefont {M.~D.}\ \bibnamefont
  {Girardeau}},\ }\href {\doibase 10.1103/PhysRevA.83.011601} {\bibfield
  {journal} {\bibinfo  {journal} {Phys. Rev. A}\ }\textbf {\bibinfo {volume}
  {83}},\ \bibinfo {pages} {011601} (\bibinfo {year} {2011})}\BibitemShut
  {NoStop}%
\bibitem [{\citenamefont {Girardeau}\ and\ \citenamefont
  {Astrakharchik}(2012)}]{Marvindipol}%
  \BibitemOpen
  \bibfield  {author} {\bibinfo {author} {\bibfnamefont {M.~D.}\ \bibnamefont
  {Girardeau}}\ and\ \bibinfo {author} {\bibfnamefont {G.~E.}\ \bibnamefont
  {Astrakharchik}},\ }\href {\doibase 10.1103/PhysRevLett.109.235305}
  {\bibfield  {journal} {\bibinfo  {journal} {Phys. Rev. Lett.}\ }\textbf
  {\bibinfo {volume} {109}},\ \bibinfo {pages} {235305} (\bibinfo {year}
  {2012})}\BibitemShut {NoStop}%
\bibitem [{\citenamefont {Girardeau}\ and\ \citenamefont
  {Olshanii}(2004)}]{Marvin7}%
  \BibitemOpen
  \bibfield  {author} {\bibinfo {author} {\bibfnamefont {M.~D.}\ \bibnamefont
  {Girardeau}}\ and\ \bibinfo {author} {\bibfnamefont {M.}~\bibnamefont
  {Olshanii}},\ }\href {\doibase 10.1103/PhysRevA.70.023608} {\bibfield
  {journal} {\bibinfo  {journal} {Phys. Rev. A}\ }\textbf {\bibinfo {volume}
  {70}},\ \bibinfo {pages} {023608} (\bibinfo {year} {2004})}\BibitemShut
  {NoStop}%
\bibitem [{\citenamefont {Girardeau}\ and\ \citenamefont
  {Minguzzi}(2007)}]{Marvin7a}%
  \BibitemOpen
  \bibfield  {author} {\bibinfo {author} {\bibfnamefont {M.~D.}\ \bibnamefont
  {Girardeau}}\ and\ \bibinfo {author} {\bibfnamefont {A.}~\bibnamefont
  {Minguzzi}},\ }\href {\doibase 10.1103/PhysRevLett.99.230402} {\bibfield
  {journal} {\bibinfo  {journal} {Phys. Rev. Lett.}\ }\textbf {\bibinfo
  {volume} {99}},\ \bibinfo {pages} {230402} (\bibinfo {year}
  {2007})}\BibitemShut {NoStop}%
\bibitem [{\citenamefont {Girardeau}\ and\ \citenamefont
  {Minguzzi}(2006)}]{Marvin8}%
  \BibitemOpen
  \bibfield  {author} {\bibinfo {author} {\bibfnamefont {M.~D.}\ \bibnamefont
  {Girardeau}}\ and\ \bibinfo {author} {\bibfnamefont {A.}~\bibnamefont
  {Minguzzi}},\ }\href {\doibase 10.1103/PhysRevLett.96.080404} {\bibfield
  {journal} {\bibinfo  {journal} {Phys. Rev. Lett.}\ }\textbf {\bibinfo
  {volume} {96}},\ \bibinfo {pages} {080404} (\bibinfo {year}
  {2006})}\BibitemShut {NoStop}%
\bibitem [{\citenamefont {Girardeau}(2010)}]{Marvin8a}%
  \BibitemOpen
  \bibfield  {author} {\bibinfo {author} {\bibfnamefont {M.~D.}\ \bibnamefont
  {Girardeau}},\ }\href {\doibase 10.1103/PhysRevA.82.011607} {\bibfield
  {journal} {\bibinfo  {journal} {Phys. Rev. A}\ }\textbf {\bibinfo {volume}
  {82}},\ \bibinfo {pages} {011607} (\bibinfo {year} {2010})}\BibitemShut
  {NoStop}%
\bibitem [{\citenamefont {Minguzzi}\ and\ \citenamefont
  {Girardeau}(2006)}]{Marvin8b}%
  \BibitemOpen
  \bibfield  {author} {\bibinfo {author} {\bibfnamefont {A.}~\bibnamefont
  {Minguzzi}}\ and\ \bibinfo {author} {\bibfnamefont {M.~D.}\ \bibnamefont
  {Girardeau}},\ }\href {\doibase 10.1103/PhysRevA.73.063614} {\bibfield
  {journal} {\bibinfo  {journal} {Phys. Rev. A}\ }\textbf {\bibinfo {volume}
  {73}},\ \bibinfo {pages} {063614} (\bibinfo {year} {2006})}\BibitemShut
  {NoStop}%
\bibitem [{\citenamefont {del Campo}\ \emph {et~al.}(2007)\citenamefont {del
  Campo}, \citenamefont {Muga},\ and\ \citenamefont {Girardeau}}]{Marvin9}%
  \BibitemOpen
  \bibfield  {author} {\bibinfo {author} {\bibfnamefont {A.}~\bibnamefont {del
  Campo}}, \bibinfo {author} {\bibfnamefont {J.~G.}\ \bibnamefont {Muga}}, \
  and\ \bibinfo {author} {\bibfnamefont {M.~D.}\ \bibnamefont {Girardeau}},\
  }\href {\doibase 10.1103/PhysRevA.76.013615} {\bibfield  {journal} {\bibinfo
  {journal} {Phys. Rev. A}\ }\textbf {\bibinfo {volume} {76}},\ \bibinfo
  {pages} {013615} (\bibinfo {year} {2007})}\BibitemShut {NoStop}%
\bibitem [{\citenamefont {Serwane}\ \emph {et~al.}(2011)\citenamefont
  {Serwane}, \citenamefont {Z\"urn}, \citenamefont {Lompe}, \citenamefont
  {Ottenstein}, \citenamefont {Wenz},\ and\ \citenamefont
  {Jochim}}]{Serwane2011}%
  \BibitemOpen
  \bibfield  {author} {\bibinfo {author} {\bibfnamefont {F.}~\bibnamefont
  {Serwane}}, \bibinfo {author} {\bibfnamefont {G.}~\bibnamefont {Z\"urn}},
  \bibinfo {author} {\bibfnamefont {T.}~\bibnamefont {Lompe}}, \bibinfo
  {author} {\bibfnamefont {T.~B.}\ \bibnamefont {Ottenstein}}, \bibinfo
  {author} {\bibfnamefont {A.~N.}\ \bibnamefont {Wenz}}, \ and\ \bibinfo
  {author} {\bibfnamefont {S.}~\bibnamefont {Jochim}},\ }\href {\doibase
  10.1126/science.1201351} {\bibfield  {journal} {\bibinfo  {journal}
  {Science}\ }\textbf {\bibinfo {volume} {332}},\ \bibinfo {pages} {336}
  (\bibinfo {year} {2011})}\BibitemShut {NoStop}%
\bibitem [{\citenamefont {Z\"urn}\ \emph {et~al.}(2012)\citenamefont {Z\"urn},
  \citenamefont {Serwane}, \citenamefont {Lompe}, \citenamefont {Wenz},
  \citenamefont {Ries}, \citenamefont {Bohn},\ and\ \citenamefont
  {Jochim}}]{Zurn2012}%
  \BibitemOpen
  \bibfield  {author} {\bibinfo {author} {\bibfnamefont {G.}~\bibnamefont
  {Z\"urn}}, \bibinfo {author} {\bibfnamefont {F.}~\bibnamefont {Serwane}},
  \bibinfo {author} {\bibfnamefont {T.}~\bibnamefont {Lompe}}, \bibinfo
  {author} {\bibfnamefont {A.~N.}\ \bibnamefont {Wenz}}, \bibinfo {author}
  {\bibfnamefont {M.~G.}\ \bibnamefont {Ries}}, \bibinfo {author}
  {\bibfnamefont {J.~E.}\ \bibnamefont {Bohn}}, \ and\ \bibinfo {author}
  {\bibfnamefont {S.}~\bibnamefont {Jochim}},\ }\href {\doibase
  10.1103/PhysRevLett.108.075303} {\bibfield  {journal} {\bibinfo  {journal}
  {Phys. Rev. Lett.}\ }\textbf {\bibinfo {volume} {108}},\ \bibinfo {pages}
  {075303} (\bibinfo {year} {2012})}\BibitemShut {NoStop}%
\bibitem [{\citenamefont {Wenz}\ \emph {et~al.}(2013)\citenamefont {Wenz},
  \citenamefont {Z{\"u}rn}, \citenamefont {Murmann}, \citenamefont {Brouzos},
  \citenamefont {Lompe},\ and\ \citenamefont {Jochim}}]{Wenz2013}%
  \BibitemOpen
  \bibfield  {author} {\bibinfo {author} {\bibfnamefont {A.}~\bibnamefont
  {Wenz}}, \bibinfo {author} {\bibfnamefont {G.}~\bibnamefont {Z{\"u}rn}},
  \bibinfo {author} {\bibfnamefont {S.}~\bibnamefont {Murmann}}, \bibinfo
  {author} {\bibfnamefont {I.}~\bibnamefont {Brouzos}}, \bibinfo {author}
  {\bibfnamefont {T.}~\bibnamefont {Lompe}}, \ and\ \bibinfo {author}
  {\bibfnamefont {S.}~\bibnamefont {Jochim}},\ }\href {\doibase
  10.1126/science.1240516} {\bibfield  {journal} {\bibinfo  {journal}
  {Science}\ }\textbf {\bibinfo {volume} {342}},\ \bibinfo {pages} {457}
  (\bibinfo {year} {2013})}\BibitemShut {NoStop}%
\bibitem [{\citenamefont {Z\"urn}\ \emph {et~al.}(2013)\citenamefont {Z\"urn},
  \citenamefont {Wenz}, \citenamefont {Murmann}, \citenamefont {Bergschneider},
  \citenamefont {Lompe},\ and\ \citenamefont {Jochim}}]{Zurn2013}%
  \BibitemOpen
  \bibfield  {author} {\bibinfo {author} {\bibfnamefont {G.}~\bibnamefont
  {Z\"urn}}, \bibinfo {author} {\bibfnamefont {A.~N.}\ \bibnamefont {Wenz}},
  \bibinfo {author} {\bibfnamefont {S.}~\bibnamefont {Murmann}}, \bibinfo
  {author} {\bibfnamefont {A.}~\bibnamefont {Bergschneider}}, \bibinfo {author}
  {\bibfnamefont {T.}~\bibnamefont {Lompe}}, \ and\ \bibinfo {author}
  {\bibfnamefont {S.}~\bibnamefont {Jochim}},\ }\href {\doibase
  10.1103/PhysRevLett.111.175302} {\bibfield  {journal} {\bibinfo  {journal}
  {Phys. Rev. Lett.}\ }\textbf {\bibinfo {volume} {111}},\ \bibinfo {pages}
  {175302} (\bibinfo {year} {2013})}\BibitemShut {NoStop}%
\bibitem [{\citenamefont {Murmann}\ \emph
  {et~al.}(2015{\natexlab{a}})\citenamefont {Murmann}, \citenamefont
  {Bergschneider}, \citenamefont {Klinkhamer}, \citenamefont {Z\"urn},
  \citenamefont {Lompe},\ and\ \citenamefont {Jochim}}]{Murmann2015}%
  \BibitemOpen
  \bibfield  {author} {\bibinfo {author} {\bibfnamefont {S.}~\bibnamefont
  {Murmann}}, \bibinfo {author} {\bibfnamefont {A.}~\bibnamefont
  {Bergschneider}}, \bibinfo {author} {\bibfnamefont {V.~M.}\ \bibnamefont
  {Klinkhamer}}, \bibinfo {author} {\bibfnamefont {G.}~\bibnamefont {Z\"urn}},
  \bibinfo {author} {\bibfnamefont {T.}~\bibnamefont {Lompe}}, \ and\ \bibinfo
  {author} {\bibfnamefont {S.}~\bibnamefont {Jochim}},\ }\href {\doibase
  10.1103/PhysRevLett.114.080402} {\bibfield  {journal} {\bibinfo  {journal}
  {Phys. Rev. Lett.}\ }\textbf {\bibinfo {volume} {114}},\ \bibinfo {pages}
  {080402} (\bibinfo {year} {2015}{\natexlab{a}})}\BibitemShut {NoStop}%
\bibitem [{\citenamefont {Murmann}\ \emph
  {et~al.}(2015{\natexlab{b}})\citenamefont {Murmann}, \citenamefont
  {Deuretzbacher}, \citenamefont {Z\"urn}, \citenamefont {Bjerlin},
  \citenamefont {Reimann}, \citenamefont {Santos}, \citenamefont {Lompe},\ and\
  \citenamefont {Jochim}}]{Murmann2015b}%
  \BibitemOpen
  \bibfield  {author} {\bibinfo {author} {\bibfnamefont {S.}~\bibnamefont
  {Murmann}}, \bibinfo {author} {\bibfnamefont {F.}~\bibnamefont
  {Deuretzbacher}}, \bibinfo {author} {\bibfnamefont {G.}~\bibnamefont
  {Z\"urn}}, \bibinfo {author} {\bibfnamefont {J.}~\bibnamefont {Bjerlin}},
  \bibinfo {author} {\bibfnamefont {S.~M.}\ \bibnamefont {Reimann}}, \bibinfo
  {author} {\bibfnamefont {L.}~\bibnamefont {Santos}}, \bibinfo {author}
  {\bibfnamefont {T.}~\bibnamefont {Lompe}}, \ and\ \bibinfo {author}
  {\bibfnamefont {S.}~\bibnamefont {Jochim}},\ }\href {\doibase
  10.1103/PhysRevLett.115.215301} {\bibfield  {journal} {\bibinfo  {journal}
  {Phys. Rev. Lett.}\ }\textbf {\bibinfo {volume} {115}},\ \bibinfo {pages}
  {215301} (\bibinfo {year} {2015}{\natexlab{b}})}\BibitemShut {NoStop}%
\bibitem [{\citenamefont {Recati}\ \emph {et~al.}(2003)\citenamefont {Recati},
  \citenamefont {Fedichev}, \citenamefont {Zwerger},\ and\ \citenamefont
  {Zoller}}]{Recati2003}%
  \BibitemOpen
  \bibfield  {author} {\bibinfo {author} {\bibfnamefont {A.}~\bibnamefont
  {Recati}}, \bibinfo {author} {\bibfnamefont {P.~O.}\ \bibnamefont
  {Fedichev}}, \bibinfo {author} {\bibfnamefont {W.}~\bibnamefont {Zwerger}}, \
  and\ \bibinfo {author} {\bibfnamefont {P.}~\bibnamefont {Zoller}},\ }\href
  {\doibase 10.1103/PhysRevLett.90.020401} {\bibfield  {journal} {\bibinfo
  {journal} {Phys. Rev. Lett.}\ }\textbf {\bibinfo {volume} {90}},\ \bibinfo
  {pages} {020401} (\bibinfo {year} {2003})}\BibitemShut {NoStop}%
\bibitem [{\citenamefont {Juillet}\ \emph {et~al.}(2004)\citenamefont
  {Juillet}, \citenamefont {Gulminelli},\ and\ \citenamefont
  {Chomaz}}]{Juillet2004}%
  \BibitemOpen
  \bibfield  {author} {\bibinfo {author} {\bibfnamefont {O.}~\bibnamefont
  {Juillet}}, \bibinfo {author} {\bibfnamefont {F.}~\bibnamefont {Gulminelli}},
  \ and\ \bibinfo {author} {\bibfnamefont {P.}~\bibnamefont {Chomaz}},\ }\href
  {\doibase 10.1103/PhysRevLett.92.160401} {\bibfield  {journal} {\bibinfo
  {journal} {Phys. Rev. Lett.}\ }\textbf {\bibinfo {volume} {92}},\ \bibinfo
  {pages} {160401} (\bibinfo {year} {2004})}\BibitemShut {NoStop}%
\bibitem [{\citenamefont {Astrakharchik}\ \emph {et~al.}(2004)\citenamefont
  {Astrakharchik}, \citenamefont {Blume}, \citenamefont {Giorgini},\ and\
  \citenamefont {Pitaevskii}}]{Astrakharchik2004}%
  \BibitemOpen
  \bibfield  {author} {\bibinfo {author} {\bibfnamefont {G.~E.}\ \bibnamefont
  {Astrakharchik}}, \bibinfo {author} {\bibfnamefont {D.}~\bibnamefont
  {Blume}}, \bibinfo {author} {\bibfnamefont {S.}~\bibnamefont {Giorgini}}, \
  and\ \bibinfo {author} {\bibfnamefont {L.~P.}\ \bibnamefont {Pitaevskii}},\
  }\href {\doibase 10.1103/PhysRevLett.93.050402} {\bibfield  {journal}
  {\bibinfo  {journal} {Phys. Rev. Lett.}\ }\textbf {\bibinfo {volume} {93}},\
  \bibinfo {pages} {050402} (\bibinfo {year} {2004})}\BibitemShut {NoStop}%
\bibitem [{\citenamefont {Tokatly}(2004)}]{Tokatly2004}%
  \BibitemOpen
  \bibfield  {author} {\bibinfo {author} {\bibfnamefont {I.~V.}\ \bibnamefont
  {Tokatly}},\ }\href {\doibase 10.1103/PhysRevLett.93.090405} {\bibfield
  {journal} {\bibinfo  {journal} {Phys. Rev. Lett.}\ }\textbf {\bibinfo
  {volume} {93}},\ \bibinfo {pages} {090405} (\bibinfo {year}
  {2004})}\BibitemShut {NoStop}%
\bibitem [{\citenamefont {Fuchs}\ \emph {et~al.}(2004)\citenamefont {Fuchs},
  \citenamefont {Recati},\ and\ \citenamefont {Zwerger}}]{Fuchs2004}%
  \BibitemOpen
  \bibfield  {author} {\bibinfo {author} {\bibfnamefont {J.~N.}\ \bibnamefont
  {Fuchs}}, \bibinfo {author} {\bibfnamefont {A.}~\bibnamefont {Recati}}, \
  and\ \bibinfo {author} {\bibfnamefont {W.}~\bibnamefont {Zwerger}},\ }\href
  {\doibase 10.1103/PhysRevLett.93.090408} {\bibfield  {journal} {\bibinfo
  {journal} {Phys. Rev. Lett.}\ }\textbf {\bibinfo {volume} {93}},\ \bibinfo
  {pages} {090408} (\bibinfo {year} {2004})}\BibitemShut {NoStop}%
\bibitem [{\citenamefont {Carusotto}\ and\ \citenamefont
  {Castin}(2004)}]{Carusotto2004}%
  \BibitemOpen
  \bibfield  {author} {\bibinfo {author} {\bibfnamefont {I.}~\bibnamefont
  {Carusotto}}\ and\ \bibinfo {author} {\bibfnamefont {Y.}~\bibnamefont
  {Castin}},\ }\href {\doibase 10.1016/j.optcom.2004.04.062} {\bibfield
  {journal} {\bibinfo  {journal} {Opt Commun}\ }\textbf {\bibinfo {volume}
  {243}},\ \bibinfo {pages} {81 } (\bibinfo {year} {2004})}\BibitemShut
  {NoStop}%
\bibitem [{\citenamefont {Astrakharchik}(2005)}]{Astrakharchik2005}%
  \BibitemOpen
  \bibfield  {author} {\bibinfo {author} {\bibfnamefont {G.~E.}\ \bibnamefont
  {Astrakharchik}},\ }\href {\doibase 10.1103/PhysRevA.72.063620} {\bibfield
  {journal} {\bibinfo  {journal} {Phys. Rev. A}\ }\textbf {\bibinfo {volume}
  {72}},\ \bibinfo {pages} {063620} (\bibinfo {year} {2005})}\BibitemShut
  {NoStop}%
\bibitem [{\citenamefont {Orso}(2007)}]{Orso2007}%
  \BibitemOpen
  \bibfield  {author} {\bibinfo {author} {\bibfnamefont {G.}~\bibnamefont
  {Orso}},\ }\href {\doibase 10.1103/PhysRevLett.98.070402} {\bibfield
  {journal} {\bibinfo  {journal} {Phys. Rev. Lett.}\ }\textbf {\bibinfo
  {volume} {98}},\ \bibinfo {pages} {070402} (\bibinfo {year}
  {2007})}\BibitemShut {NoStop}%
\bibitem [{\citenamefont {Hu}\ \emph {et~al.}(2007)\citenamefont {Hu},
  \citenamefont {Liu},\ and\ \citenamefont {Drummond}}]{Hu2007}%
  \BibitemOpen
  \bibfield  {author} {\bibinfo {author} {\bibfnamefont {H.}~\bibnamefont
  {Hu}}, \bibinfo {author} {\bibfnamefont {X.-J.}\ \bibnamefont {Liu}}, \ and\
  \bibinfo {author} {\bibfnamefont {P.~D.}\ \bibnamefont {Drummond}},\ }\href
  {\doibase 10.1103/PhysRevLett.98.070403} {\bibfield  {journal} {\bibinfo
  {journal} {Phys. Rev. Lett.}\ }\textbf {\bibinfo {volume} {98}},\ \bibinfo
  {pages} {070403} (\bibinfo {year} {2007})}\BibitemShut {NoStop}%
\bibitem [{\citenamefont {Colom\'e-Tatch\'e}(2008)}]{ColomeTatche2008}%
  \BibitemOpen
  \bibfield  {author} {\bibinfo {author} {\bibfnamefont {M.}~\bibnamefont
  {Colom\'e-Tatch\'e}},\ }\href {\doibase 10.1103/PhysRevA.78.033612}
  {\bibfield  {journal} {\bibinfo  {journal} {Phys. Rev. A}\ }\textbf {\bibinfo
  {volume} {78}},\ \bibinfo {pages} {033612} (\bibinfo {year}
  {2008})}\BibitemShut {NoStop}%
\bibitem [{\citenamefont {Peotta}\ \emph {et~al.}(2012)\citenamefont {Peotta},
  \citenamefont {Rossini}, \citenamefont {Silvi}, \citenamefont {Vignale},
  \citenamefont {Fazio},\ and\ \citenamefont {Polini}}]{Peotta2012}%
  \BibitemOpen
  \bibfield  {author} {\bibinfo {author} {\bibfnamefont {S.}~\bibnamefont
  {Peotta}}, \bibinfo {author} {\bibfnamefont {D.}~\bibnamefont {Rossini}},
  \bibinfo {author} {\bibfnamefont {P.}~\bibnamefont {Silvi}}, \bibinfo
  {author} {\bibfnamefont {G.}~\bibnamefont {Vignale}}, \bibinfo {author}
  {\bibfnamefont {R.}~\bibnamefont {Fazio}}, \ and\ \bibinfo {author}
  {\bibfnamefont {M.}~\bibnamefont {Polini}},\ }\href {\doibase
  10.1103/PhysRevLett.108.245302} {\bibfield  {journal} {\bibinfo  {journal}
  {Phys. Rev. Lett.}\ }\textbf {\bibinfo {volume} {108}},\ \bibinfo {pages}
  {245302} (\bibinfo {year} {2012})}\BibitemShut {NoStop}%
\bibitem [{\citenamefont {Gharashi}\ and\ \citenamefont
  {Blume}(2013)}]{Gharashi2013}%
  \BibitemOpen
  \bibfield  {author} {\bibinfo {author} {\bibfnamefont {S.~E.}\ \bibnamefont
  {Gharashi}}\ and\ \bibinfo {author} {\bibfnamefont {D.}~\bibnamefont
  {Blume}},\ }\href {\doibase 10.1103/PhysRevLett.111.045302} {\bibfield
  {journal} {\bibinfo  {journal} {Phys. Rev. Lett.}\ }\textbf {\bibinfo
  {volume} {111}},\ \bibinfo {pages} {045302} (\bibinfo {year}
  {2013})}\BibitemShut {NoStop}%
\bibitem [{\citenamefont {Sowi\'nski}\ \emph {et~al.}(2013)\citenamefont
  {Sowi\'nski}, \citenamefont {Grass}, \citenamefont {Dutta},\ and\
  \citenamefont {Lewenstein}}]{Sowinski2013}%
  \BibitemOpen
  \bibfield  {author} {\bibinfo {author} {\bibfnamefont {T.}~\bibnamefont
  {Sowi\'nski}}, \bibinfo {author} {\bibfnamefont {T.}~\bibnamefont {Grass}},
  \bibinfo {author} {\bibfnamefont {O.}~\bibnamefont {Dutta}}, \ and\ \bibinfo
  {author} {\bibfnamefont {M.}~\bibnamefont {Lewenstein}},\ }\href {\doibase
  10.1103/PhysRevA.88.033607} {\bibfield  {journal} {\bibinfo  {journal} {Phys.
  Rev. A}\ }\textbf {\bibinfo {volume} {88}},\ \bibinfo {pages} {033607}
  (\bibinfo {year} {2013})}\BibitemShut {NoStop}%
\bibitem [{\citenamefont {Astrakharchik}\ and\ \citenamefont
  {Brouzos}(2013)}]{Astrakharchik2013}%
  \BibitemOpen
  \bibfield  {author} {\bibinfo {author} {\bibfnamefont {G.~E.}\ \bibnamefont
  {Astrakharchik}}\ and\ \bibinfo {author} {\bibfnamefont {I.}~\bibnamefont
  {Brouzos}},\ }\href {\doibase 10.1103/PhysRevA.88.021602} {\bibfield
  {journal} {\bibinfo  {journal} {Phys. Rev. A}\ }\textbf {\bibinfo {volume}
  {88}},\ \bibinfo {pages} {021602} (\bibinfo {year} {2013})}\BibitemShut
  {NoStop}%
\bibitem [{\citenamefont {Deuretzbacher}\ \emph {et~al.}(2014)\citenamefont
  {Deuretzbacher}, \citenamefont {Becker}, \citenamefont {Bjerlin},
  \citenamefont {Reimann},\ and\ \citenamefont {Santos}}]{Deuretzbacher2014}%
  \BibitemOpen
  \bibfield  {author} {\bibinfo {author} {\bibfnamefont {F.}~\bibnamefont
  {Deuretzbacher}}, \bibinfo {author} {\bibfnamefont {D.}~\bibnamefont
  {Becker}}, \bibinfo {author} {\bibfnamefont {J.}~\bibnamefont {Bjerlin}},
  \bibinfo {author} {\bibfnamefont {S.~M.}\ \bibnamefont {Reimann}}, \ and\
  \bibinfo {author} {\bibfnamefont {L.}~\bibnamefont {Santos}},\ }\href
  {\doibase 10.1103/PhysRevA.90.013611} {\bibfield  {journal} {\bibinfo
  {journal} {Phys. Rev. A}\ }\textbf {\bibinfo {volume} {90}},\ \bibinfo
  {pages} {013611} (\bibinfo {year} {2014})}\BibitemShut {NoStop}%
\bibitem [{\citenamefont {{Volosniev}}\ \emph {et~al.}(2014)\citenamefont
  {{Volosniev}}, \citenamefont {{Fedorov}}, \citenamefont {{Jensen}},
  \citenamefont {{Valiente}},\ and\ \citenamefont {{Zinner}}}]{Volosniev2014}%
  \BibitemOpen
  \bibfield  {author} {\bibinfo {author} {\bibfnamefont {A.~G.}\ \bibnamefont
  {{Volosniev}}}, \bibinfo {author} {\bibfnamefont {D.~V.}\ \bibnamefont
  {{Fedorov}}}, \bibinfo {author} {\bibfnamefont {A.~S.}\ \bibnamefont
  {{Jensen}}}, \bibinfo {author} {\bibfnamefont {M.}~\bibnamefont
  {{Valiente}}}, \ and\ \bibinfo {author} {\bibfnamefont {N.~T.}\ \bibnamefont
  {{Zinner}}},\ }\href {\doibase 10.1038/ncomms6300} {\bibfield  {journal}
  {\bibinfo  {journal} {Nat.~Commun.}\ }\textbf {\bibinfo {volume} {5}},\
  \bibinfo {pages} {5300} (\bibinfo {year} {2014})}\BibitemShut {NoStop}%
\bibitem [{\citenamefont {Lindgren}\ \emph {et~al.}(2014)\citenamefont
  {Lindgren}, \citenamefont {Rotureau}, \citenamefont {Forss\'{e}n},
  \citenamefont {Volosniev},\ and\ \citenamefont {Zinner}}]{Lindgren2014}%
  \BibitemOpen
  \bibfield  {author} {\bibinfo {author} {\bibfnamefont {E.~J.}\ \bibnamefont
  {Lindgren}}, \bibinfo {author} {\bibfnamefont {J.}~\bibnamefont {Rotureau}},
  \bibinfo {author} {\bibfnamefont {C.}~\bibnamefont {Forss\'{e}n}}, \bibinfo
  {author} {\bibfnamefont {A.~G.}\ \bibnamefont {Volosniev}}, \ and\ \bibinfo
  {author} {\bibfnamefont {N.~T.}\ \bibnamefont {Zinner}},\ }\href {\doibase
  10.1088/1367-2630/16/6/063003} {\bibfield  {journal} {\bibinfo  {journal}
  {New J. Phys.}\ }\textbf {\bibinfo {volume} {16}},\ \bibinfo {pages} {063003}
  (\bibinfo {year} {2014})}\BibitemShut {NoStop}%
\bibitem [{\citenamefont {{Levinsen}}\ \emph {et~al.}(2015)\citenamefont
  {{Levinsen}}, \citenamefont {{Massignan}}, \citenamefont {{Bruun}},\ and\
  \citenamefont {{Parish}}}]{Levinsen2014}%
  \BibitemOpen
  \bibfield  {author} {\bibinfo {author} {\bibfnamefont {J.}~\bibnamefont
  {{Levinsen}}}, \bibinfo {author} {\bibfnamefont {P.}~\bibnamefont
  {{Massignan}}}, \bibinfo {author} {\bibfnamefont {G.~M.}\ \bibnamefont
  {{Bruun}}}, \ and\ \bibinfo {author} {\bibfnamefont {M.~M.}\ \bibnamefont
  {{Parish}}},\ }\href {\doibase 10.1126/sciadv.1500197} {\bibfield  {journal}
  {\bibinfo  {journal} {Sci. Adv.}\ }\textbf {\bibinfo {volume} {1}},\ \bibinfo
  {pages} {e1500197} (\bibinfo {year} {2015})}\BibitemShut {NoStop}%
\bibitem [{\citenamefont {D'Amico}\ and\ \citenamefont
  {Rontani}(2015)}]{DAmico2015}%
  \BibitemOpen
  \bibfield  {author} {\bibinfo {author} {\bibfnamefont {P.}~\bibnamefont
  {D'Amico}}\ and\ \bibinfo {author} {\bibfnamefont {M.}~\bibnamefont
  {Rontani}},\ }\href {\doibase 10.1103/PhysRevA.91.043610} {\bibfield
  {journal} {\bibinfo  {journal} {Phys. Rev. A}\ }\textbf {\bibinfo {volume}
  {91}},\ \bibinfo {pages} {043610} (\bibinfo {year} {2015})}\BibitemShut
  {NoStop}%
\bibitem [{\citenamefont {Berger}\ \emph {et~al.}(2015)\citenamefont {Berger},
  \citenamefont {Anderson},\ and\ \citenamefont {Drut}}]{Berger2015}%
  \BibitemOpen
  \bibfield  {author} {\bibinfo {author} {\bibfnamefont {C.~E.}\ \bibnamefont
  {Berger}}, \bibinfo {author} {\bibfnamefont {E.~R.}\ \bibnamefont
  {Anderson}}, \ and\ \bibinfo {author} {\bibfnamefont {J.~E.}\ \bibnamefont
  {Drut}},\ }\href {\doibase 10.1103/PhysRevA.91.053618} {\bibfield  {journal}
  {\bibinfo  {journal} {Phys. Rev. A}\ }\textbf {\bibinfo {volume} {91}},\
  \bibinfo {pages} {053618} (\bibinfo {year} {2015})}\BibitemShut {NoStop}%
\bibitem [{\citenamefont {Sowi\'nski}\ \emph {et~al.}(2015)\citenamefont
  {Sowi\'nski}, \citenamefont {Gajda},\ and\ \citenamefont
  {Rza{\.z}ewski}}]{Sowinski2015}%
  \BibitemOpen
  \bibfield  {author} {\bibinfo {author} {\bibfnamefont {T.}~\bibnamefont
  {Sowi\'nski}}, \bibinfo {author} {\bibfnamefont {M.}~\bibnamefont {Gajda}}, \
  and\ \bibinfo {author} {\bibfnamefont {K.}~\bibnamefont {Rza{\.z}ewski}},\
  }\href {http://stacks.iop.org/0295-5075/109/i=2/a=26005} {\bibfield
  {journal} {\bibinfo  {journal} {Europhys. Lett.}\ }\textbf {\bibinfo {volume}
  {109}},\ \bibinfo {pages} {26005} (\bibinfo {year} {2015})}\BibitemShut
  {NoStop}%
\bibitem [{\citenamefont {Massignan}\ \emph {et~al.}(2015)\citenamefont
  {Massignan}, \citenamefont {Levinsen},\ and\ \citenamefont
  {Parish}}]{Massignan2015}%
  \BibitemOpen
  \bibfield  {author} {\bibinfo {author} {\bibfnamefont {P.}~\bibnamefont
  {Massignan}}, \bibinfo {author} {\bibfnamefont {J.}~\bibnamefont {Levinsen}},
  \ and\ \bibinfo {author} {\bibfnamefont {M.~M.}\ \bibnamefont {Parish}},\
  }\href {\doibase 10.1103/PhysRevLett.115.247202} {\bibfield  {journal}
  {\bibinfo  {journal} {Phys. Rev. Lett.}\ }\textbf {\bibinfo {volume} {115}},\
  \bibinfo {pages} {247202} (\bibinfo {year} {2015})}\BibitemShut {NoStop}%
\bibitem [{\citenamefont {Coester}(1958)}]{Coester1958}%
  \BibitemOpen
  \bibfield  {author} {\bibinfo {author} {\bibfnamefont {F.}~\bibnamefont
  {Coester}},\ }\href {\doibase 10.1016/0029-5582(58)90280-3} {\bibfield
  {journal} {\bibinfo  {journal} {Nucl. Phys.}\ }\textbf {\bibinfo {volume}
  {7}},\ \bibinfo {pages} {421} (\bibinfo {year} {1958})}\BibitemShut {NoStop}%
\bibitem [{\citenamefont {{\v{C}}{\'\i}{\v{z}}ek}(1966)}]{Cizek1966}%
  \BibitemOpen
  \bibfield  {author} {\bibinfo {author} {\bibfnamefont {J.}~\bibnamefont
  {{\v{C}}{\'\i}{\v{z}}ek}},\ }\href {\doibase 10.1063/1.1727484} {\bibfield
  {journal} {\bibinfo  {journal} {J. Phys. Chem.}\ }\textbf {\bibinfo {volume}
  {45}},\ \bibinfo {pages} {4256} (\bibinfo {year} {1966})}\BibitemShut
  {NoStop}%
\bibitem [{\citenamefont {\v{C}\'{\i}\v{z}ek}(1969)}]{Cizek1969}%
  \BibitemOpen
  \bibfield  {author} {\bibinfo {author} {\bibfnamefont {J.}~\bibnamefont
  {\v{C}\'{\i}\v{z}ek}},\ }\href {\doibase 10.1002/9780470143599.ch2}
  {\bibfield  {journal} {\bibinfo  {journal} {Adv. Chem. Phys.}\ }\textbf
  {\bibinfo {volume} {14}},\ \bibinfo {pages} {35} (\bibinfo {year}
  {1969})}\BibitemShut {NoStop}%
\bibitem [{\citenamefont {{\v{C}}{\'\i}{\v{z}}ek}\ and\ \citenamefont
  {Paldus}(1971)}]{Cizek1971}%
  \BibitemOpen
  \bibfield  {author} {\bibinfo {author} {\bibfnamefont {J.}~\bibnamefont
  {{\v{C}}{\'\i}{\v{z}}ek}}\ and\ \bibinfo {author} {\bibfnamefont
  {J.}~\bibnamefont {Paldus}},\ }\href {\doibase 10.1002/qua.560050402}
  {\bibfield  {journal} {\bibinfo  {journal} {Int. J. Quantum Chem.}\ }\textbf
  {\bibinfo {volume} {5}},\ \bibinfo {pages} {359} (\bibinfo {year}
  {1971})}\BibitemShut {NoStop}%
\bibitem [{\citenamefont {Paldus}\ \emph {et~al.}(1972)\citenamefont {Paldus},
  \citenamefont {\ifmmode \check{C}\else
  \v{C}\fi{}\'{\i}\ifmmode~\check{z}\else \v{z}\fi{}ek},\ and\ \citenamefont
  {Shavitt}}]{PaldusPRA72}%
  \BibitemOpen
  \bibfield  {author} {\bibinfo {author} {\bibfnamefont {J.}~\bibnamefont
  {Paldus}}, \bibinfo {author} {\bibfnamefont {J.}~\bibnamefont {\ifmmode
  \check{C}\else \v{C}\fi{}\'{\i}\ifmmode~\check{z}\else \v{z}\fi{}ek}}, \ and\
  \bibinfo {author} {\bibfnamefont {I.}~\bibnamefont {Shavitt}},\ }\href
  {\doibase 10.1103/PhysRevA.5.50} {\bibfield  {journal} {\bibinfo  {journal}
  {Phys. Rev. A}\ }\textbf {\bibinfo {volume} {5}},\ \bibinfo {pages} {50}
  (\bibinfo {year} {1972})}\BibitemShut {NoStop}%
\bibitem [{\citenamefont {Bartlett}(1981)}]{Bartlett1981}%
  \BibitemOpen
  \bibfield  {author} {\bibinfo {author} {\bibfnamefont {R.~J.}\ \bibnamefont
  {Bartlett}},\ }\href {\doibase 10.1146/annurev.pc.32.100181.002043}
  {\bibfield  {journal} {\bibinfo  {journal} {Annu. Rev. Phys. Chem.}\ }\textbf
  {\bibinfo {volume} {32}},\ \bibinfo {pages} {359} (\bibinfo {year}
  {1981})}\BibitemShut {NoStop}%
\bibitem [{\citenamefont {Bartlett}(1989)}]{Bartlett1989}%
  \BibitemOpen
  \bibfield  {author} {\bibinfo {author} {\bibfnamefont {R.~J.}\ \bibnamefont
  {Bartlett}},\ }\href {\doibase 10.1021/j100342a008} {\bibfield  {journal}
  {\bibinfo  {journal} {J. Phys. Chem.}\ }\textbf {\bibinfo {volume} {93}},\
  \bibinfo {pages} {1697} (\bibinfo {year} {1989})}\BibitemShut {NoStop}%
\bibitem [{\citenamefont {Bishop}(1991)}]{Bishop1991}%
  \BibitemOpen
  \bibfield  {author} {\bibinfo {author} {\bibfnamefont {R.}~\bibnamefont
  {Bishop}},\ }\href {\doibase 10.1007/BF01119617} {\bibfield  {journal}
  {\bibinfo  {journal} {Theor. Chim. Acta}\ }\textbf {\bibinfo {volume} {80}},\
  \bibinfo {pages} {95} (\bibinfo {year} {1991})}\BibitemShut {NoStop}%
\bibitem [{\citenamefont {Paldus}\ and\ \citenamefont {Li}(1999)}]{Paldus1999}%
  \BibitemOpen
  \bibfield  {author} {\bibinfo {author} {\bibfnamefont {J.}~\bibnamefont
  {Paldus}}\ and\ \bibinfo {author} {\bibfnamefont {X.}~\bibnamefont {Li}},\
  }\href {\doibase 10.1002/9780470141694.ch1} {\bibfield  {journal} {\bibinfo
  {journal} {Adv. Chem. Phys.}\ }\textbf {\bibinfo {volume} {110}},\ \bibinfo
  {pages} {1} (\bibinfo {year} {1999})}\BibitemShut {NoStop}%
\bibitem [{\citenamefont {Bartlett}\ and\ \citenamefont
  {Musia{\l}}(2007)}]{Musial2007}%
  \BibitemOpen
  \bibfield  {author} {\bibinfo {author} {\bibfnamefont {R.~J.}\ \bibnamefont
  {Bartlett}}\ and\ \bibinfo {author} {\bibfnamefont {M.}~\bibnamefont
  {Musia{\l}}},\ }\href {\doibase 10.1103/RevModPhys.79.291} {\bibfield
  {journal} {\bibinfo  {journal} {Rev. Mod. Phys.}\ }\textbf {\bibinfo {volume}
  {79}},\ \bibinfo {pages} {291} (\bibinfo {year} {2007})}\BibitemShut
  {NoStop}%
\bibitem [{\citenamefont {Lyakh}\ \emph {et~al.}(2012)\citenamefont {Lyakh},
  \citenamefont {Musia{\l}}, \citenamefont {Lotrich},\ and\ \citenamefont
  {Bartlett}}]{Lyakh2012}%
  \BibitemOpen
  \bibfield  {author} {\bibinfo {author} {\bibfnamefont {D.~I.}\ \bibnamefont
  {Lyakh}}, \bibinfo {author} {\bibfnamefont {M.}~\bibnamefont {Musia{\l}}},
  \bibinfo {author} {\bibfnamefont {V.~F.}\ \bibnamefont {Lotrich}}, \ and\
  \bibinfo {author} {\bibfnamefont {R.~J.}\ \bibnamefont {Bartlett}},\ }\href
  {\doibase 10.1021/cr2001417} {\bibfield  {journal} {\bibinfo  {journal}
  {Chem. Rev.}\ }\textbf {\bibinfo {volume} {112}},\ \bibinfo {pages} {182}
  (\bibinfo {year} {2012})}\BibitemShut {NoStop}%
\bibitem [{\citenamefont {McGuyer}\ \emph {et~al.}(2013)\citenamefont
  {McGuyer}, \citenamefont {Osborn}, \citenamefont {McDonald}, \citenamefont
  {Reinaudi}, \citenamefont {Skomorowski}, \citenamefont {Moszynski},\ and\
  \citenamefont {Zelevinsky}}]{McGuyerPRL13}%
  \BibitemOpen
  \bibfield  {author} {\bibinfo {author} {\bibfnamefont {B.~H.}\ \bibnamefont
  {McGuyer}}, \bibinfo {author} {\bibfnamefont {C.~B.}\ \bibnamefont {Osborn}},
  \bibinfo {author} {\bibfnamefont {M.}~\bibnamefont {McDonald}}, \bibinfo
  {author} {\bibfnamefont {G.}~\bibnamefont {Reinaudi}}, \bibinfo {author}
  {\bibfnamefont {W.}~\bibnamefont {Skomorowski}}, \bibinfo {author}
  {\bibfnamefont {R.}~\bibnamefont {Moszynski}}, \ and\ \bibinfo {author}
  {\bibfnamefont {T.}~\bibnamefont {Zelevinsky}},\ }\href {\doibase
  10.1103/PhysRevLett.111.243003} {\bibfield  {journal} {\bibinfo  {journal}
  {Phys. Rev. Lett.}\ }\textbf {\bibinfo {volume} {111}},\ \bibinfo {pages}
  {243003} (\bibinfo {year} {2013})}\BibitemShut {NoStop}%
\bibitem [{\citenamefont {McGuyer}\ \emph {et~al.}(2014)\citenamefont
  {McGuyer}, \citenamefont {McDonald}, \citenamefont {Iwata}, \citenamefont
  {Tarallo}, \citenamefont {Skomorowski}, \citenamefont {Moszynski},\ and\
  \citenamefont {Zelevinsky}}]{McGuyerNaturePhys14}%
  \BibitemOpen
  \bibfield  {author} {\bibinfo {author} {\bibfnamefont {B.~H.}\ \bibnamefont
  {McGuyer}}, \bibinfo {author} {\bibfnamefont {M.}~\bibnamefont {McDonald}},
  \bibinfo {author} {\bibfnamefont {G.~Z.}\ \bibnamefont {Iwata}}, \bibinfo
  {author} {\bibfnamefont {M.~G.}\ \bibnamefont {Tarallo}}, \bibinfo {author}
  {\bibfnamefont {W.}~\bibnamefont {Skomorowski}}, \bibinfo {author}
  {\bibfnamefont {R.}~\bibnamefont {Moszynski}}, \ and\ \bibinfo {author}
  {\bibfnamefont {T.}~\bibnamefont {Zelevinsky}},\ }\href {\doibase
  10.1038/nphys3182} {\bibfield  {journal} {\bibinfo  {journal} {Nature Phys.}\
  }\textbf {\bibinfo {volume} {11}},\ \bibinfo {pages} {32} (\bibinfo {year}
  {2014})}\BibitemShut {NoStop}%
\bibitem [{\citenamefont {Bukowski}\ \emph {et~al.}(2007)\citenamefont
  {Bukowski}, \citenamefont {Szalewicz}, \citenamefont {Groenenboom},\ and\
  \citenamefont {Van Der~Avoird}}]{BukowskiScience07}%
  \BibitemOpen
  \bibfield  {author} {\bibinfo {author} {\bibfnamefont {R.}~\bibnamefont
  {Bukowski}}, \bibinfo {author} {\bibfnamefont {K.}~\bibnamefont {Szalewicz}},
  \bibinfo {author} {\bibfnamefont {G.~C.}\ \bibnamefont {Groenenboom}}, \ and\
  \bibinfo {author} {\bibfnamefont {A.}~\bibnamefont {Van Der~Avoird}},\ }\href
  {\doibase 10.1126/science.1136371} {\bibfield  {journal} {\bibinfo  {journal}
  {Science}\ }\textbf {\bibinfo {volume} {315}},\ \bibinfo {pages} {1249}
  (\bibinfo {year} {2007})}\BibitemShut {NoStop}%
\bibitem [{\citenamefont {Kennedy}\ \emph {et~al.}(2014)\citenamefont
  {Kennedy}, \citenamefont {McDonald}, \citenamefont {DePrince}, \citenamefont
  {Marshall}, \citenamefont {Podeszwa},\ and\ \citenamefont
  {Sherrill}}]{PodeszwaJCP14}%
  \BibitemOpen
  \bibfield  {author} {\bibinfo {author} {\bibfnamefont {M.~R.}\ \bibnamefont
  {Kennedy}}, \bibinfo {author} {\bibfnamefont {A.~R.}\ \bibnamefont
  {McDonald}}, \bibinfo {author} {\bibfnamefont {A.~E.}\ \bibnamefont
  {DePrince}}, \bibinfo {author} {\bibfnamefont {M.~S.}\ \bibnamefont
  {Marshall}}, \bibinfo {author} {\bibfnamefont {R.}~\bibnamefont {Podeszwa}},
  \ and\ \bibinfo {author} {\bibfnamefont {C.~D.}\ \bibnamefont {Sherrill}},\
  }\href {\doibase 10.1063/1.4869686} {\bibfield  {journal} {\bibinfo
  {journal} {J. Chem. Phys.}\ }\textbf {\bibinfo {volume} {140}},\ \bibinfo
  {eid} {121104} (\bibinfo {year} {2014})}\BibitemShut {NoStop}%
\bibitem [{\citenamefont {Cederbaum}\ \emph {et~al.}(2006)\citenamefont
  {Cederbaum}, \citenamefont {Alon},\ and\ \citenamefont
  {Streltsov}}]{Cederbaum2006}%
  \BibitemOpen
  \bibfield  {author} {\bibinfo {author} {\bibfnamefont {L.~S.}\ \bibnamefont
  {Cederbaum}}, \bibinfo {author} {\bibfnamefont {O.~E.}\ \bibnamefont {Alon}},
  \ and\ \bibinfo {author} {\bibfnamefont {A.~I.}\ \bibnamefont {Streltsov}},\
  }\href {\doibase 10.1103/PhysRevA.73.043609} {\bibfield  {journal} {\bibinfo
  {journal} {Phys. Rev. A}\ }\textbf {\bibinfo {volume} {73}},\ \bibinfo
  {pages} {043609} (\bibinfo {year} {2006})}\BibitemShut {NoStop}%
\bibitem [{\citenamefont {Alon}\ \emph {et~al.}(2006)\citenamefont {Alon},
  \citenamefont {Streltsov},\ and\ \citenamefont {Cederbaum}}]{Alon2006}%
  \BibitemOpen
  \bibfield  {author} {\bibinfo {author} {\bibfnamefont {O.~E.}\ \bibnamefont
  {Alon}}, \bibinfo {author} {\bibfnamefont {A.~I.}\ \bibnamefont {Streltsov}},
  \ and\ \bibinfo {author} {\bibfnamefont {L.~S.}\ \bibnamefont {Cederbaum}},\
  }\href {\doibase 10.1016/j.theochem.2006.05.026} {\bibfield  {journal}
  {\bibinfo  {journal} {J. Mol. Struct. THEOCHEM}\ }\textbf {\bibinfo {volume}
  {768}},\ \bibinfo {pages} {151} (\bibinfo {year} {2006})}\BibitemShut
  {NoStop}%
\bibitem [{\citenamefont {Bishop}\ \emph {et~al.}(1994)\citenamefont {Bishop},
  \citenamefont {Hale},\ and\ \citenamefont {Xian}}]{BishopPRL94}%
  \BibitemOpen
  \bibfield  {author} {\bibinfo {author} {\bibfnamefont {R.~F.}\ \bibnamefont
  {Bishop}}, \bibinfo {author} {\bibfnamefont {R.~G.}\ \bibnamefont {Hale}}, \
  and\ \bibinfo {author} {\bibfnamefont {Y.}~\bibnamefont {Xian}},\ }\href
  {\doibase 10.1103/PhysRevLett.73.3157} {\bibfield  {journal} {\bibinfo
  {journal} {Phys. Rev. Lett.}\ }\textbf {\bibinfo {volume} {73}},\ \bibinfo
  {pages} {3157} (\bibinfo {year} {1994})}\BibitemShut {NoStop}%
\bibitem [{\citenamefont {Bishop}\ and\ \citenamefont
  {Li}(2011)}]{BishopPRA11}%
  \BibitemOpen
  \bibfield  {author} {\bibinfo {author} {\bibfnamefont {R.~F.}\ \bibnamefont
  {Bishop}}\ and\ \bibinfo {author} {\bibfnamefont {P.~H.~Y.}\ \bibnamefont
  {Li}},\ }\href {\doibase 10.1103/PhysRevA.83.042111} {\bibfield  {journal}
  {\bibinfo  {journal} {Phys. Rev. A}\ }\textbf {\bibinfo {volume} {83}},\
  \bibinfo {pages} {042111} (\bibinfo {year} {2011})}\BibitemShut {NoStop}%
\bibitem [{\citenamefont {Wilson}\ and\ \citenamefont
  {Silver}(1976)}]{Wilson1976}%
  \BibitemOpen
  \bibfield  {author} {\bibinfo {author} {\bibfnamefont {S.}~\bibnamefont
  {Wilson}}\ and\ \bibinfo {author} {\bibfnamefont {D.~M.}\ \bibnamefont
  {Silver}},\ }\href {\doibase 10.1103/PhysRevA.14.1949} {\bibfield  {journal}
  {\bibinfo  {journal} {Phys. Rev. A}\ }\textbf {\bibinfo {volume} {14}},\
  \bibinfo {pages} {1949} (\bibinfo {year} {1976})}\BibitemShut {NoStop}%
\bibitem [{\citenamefont {Busch}\ \emph {et~al.}(1998)\citenamefont {Busch},
  \citenamefont {Englert}, \citenamefont {Rza{\.z}ewski},\ and\ \citenamefont
  {Wilkens}}]{Busch1998}%
  \BibitemOpen
  \bibfield  {author} {\bibinfo {author} {\bibfnamefont {T.}~\bibnamefont
  {Busch}}, \bibinfo {author} {\bibfnamefont {B.-G.}\ \bibnamefont {Englert}},
  \bibinfo {author} {\bibfnamefont {K.}~\bibnamefont {Rza{\.z}ewski}}, \ and\
  \bibinfo {author} {\bibfnamefont {M.}~\bibnamefont {Wilkens}},\ }\href
  {\doibase 10.1023/A:1018705520999} {\bibfield  {journal} {\bibinfo  {journal}
  {Found. Phys.}\ }\textbf {\bibinfo {volume} {28}},\ \bibinfo {pages} {549}
  (\bibinfo {year} {1998})}\BibitemShut {NoStop}%
\bibitem [{\citenamefont {{Franke-Arnold, S.}}\ \emph
  {et~al.}(2003)\citenamefont {{Franke-Arnold, S.}}, \citenamefont {{Barnett,
  S. M.}}, \citenamefont {{Huyet, G.}},\ and\ \citenamefont {{Sailliot,
  C.}}}]{FrankeArnold2003}%
  \BibitemOpen
  \bibfield  {author} {\bibinfo {author} {\bibnamefont {{Franke-Arnold, S.}}},
  \bibinfo {author} {\bibnamefont {{Barnett, S. M.}}}, \bibinfo {author}
  {\bibnamefont {{Huyet, G.}}}, \ and\ \bibinfo {author} {\bibnamefont
  {{Sailliot, C.}}},\ }\href {\doibase 10.1140/epjd/e2003-00036-6} {\bibfield
  {journal} {\bibinfo  {journal} {Eur. Phys. J. D}\ }\textbf {\bibinfo {volume}
  {22}},\ \bibinfo {pages} {373} (\bibinfo {year} {2003})}\BibitemShut
  {NoStop}%
\bibitem [{\citenamefont {Abramowitz}\ and\ \citenamefont
  {Stegun}()}]{AbramowitzStegun}%
  \BibitemOpen
  \bibfield  {author} {\bibinfo {author} {\bibfnamefont {M.}~\bibnamefont
  {Abramowitz}}\ and\ \bibinfo {author} {\bibfnamefont {I.}~\bibnamefont
  {Stegun}},\ }\href@noop {} {\emph {\bibinfo {title} {Handbook of Mathematical
  Functions with Formulas, Graphs, and Mathematical Tables}}},\ \bibinfo
  {edition} {9th}\ ed.\ (\bibinfo  {publisher} {Dover Publications})\
  Chap.~\bibinfo {chapter} {13}\BibitemShut {NoStop}%
\bibitem [{\citenamefont {Hill}(1985)}]{Hill1985}%
  \BibitemOpen
  \bibfield  {author} {\bibinfo {author} {\bibfnamefont {R.~N.}\ \bibnamefont
  {Hill}},\ }\href {\doibase 10.1063/1.449481} {\bibfield  {journal} {\bibinfo
  {journal} {J. Chem. Phys.}\ }\textbf {\bibinfo {volume} {83}},\ \bibinfo
  {pages} {1173} (\bibinfo {year} {1985})}\BibitemShut {NoStop}%
\bibitem [{\citenamefont {Helgaker}\ \emph {et~al.}(1997)\citenamefont
  {Helgaker}, \citenamefont {Klopper}, \citenamefont {Koch},\ and\
  \citenamefont {Noga}}]{HelgakerJCP97}%
  \BibitemOpen
  \bibfield  {author} {\bibinfo {author} {\bibfnamefont {T.}~\bibnamefont
  {Helgaker}}, \bibinfo {author} {\bibfnamefont {W.}~\bibnamefont {Klopper}},
  \bibinfo {author} {\bibfnamefont {H.}~\bibnamefont {Koch}}, \ and\ \bibinfo
  {author} {\bibfnamefont {J.}~\bibnamefont {Noga}},\ }\href {\doibase
  10.1063/1.473863} {\bibfield  {journal} {\bibinfo  {journal} {J. Chem.
  Phys.}\ }\textbf {\bibinfo {volume} {106}},\ \bibinfo {pages} {9639}
  (\bibinfo {year} {1997})}\BibitemShut {NoStop}%
\bibitem [{\citenamefont {Halkier}\ \emph {et~al.}(1998)\citenamefont
  {Halkier}, \citenamefont {Helgaker}, \citenamefont {Jørgensen},
  \citenamefont {Klopper}, \citenamefont {Koch}, \citenamefont {Olsen},\ and\
  \citenamefont {Wilson}}]{HalkierCPL98}%
  \BibitemOpen
  \bibfield  {author} {\bibinfo {author} {\bibfnamefont {A.}~\bibnamefont
  {Halkier}}, \bibinfo {author} {\bibfnamefont {T.}~\bibnamefont {Helgaker}},
  \bibinfo {author} {\bibfnamefont {P.}~\bibnamefont {Jørgensen}}, \bibinfo
  {author} {\bibfnamefont {W.}~\bibnamefont {Klopper}}, \bibinfo {author}
  {\bibfnamefont {H.}~\bibnamefont {Koch}}, \bibinfo {author} {\bibfnamefont
  {J.}~\bibnamefont {Olsen}}, \ and\ \bibinfo {author} {\bibfnamefont {A.~K.}\
  \bibnamefont {Wilson}},\ }\href {\doibase 10.1016/S0009-2614(98)00111-0}
  {\bibfield  {journal} {\bibinfo  {journal} {Chem. Phys. Lett.}\ }\textbf
  {\bibinfo {volume} {286}},\ \bibinfo {pages} {243} (\bibinfo {year}
  {1998})}\BibitemShut {NoStop}%
\bibitem [{\citenamefont {Szabo}\ and\ \citenamefont
  {Ostlund}(1996)}]{Szabo1996}%
  \BibitemOpen
  \bibinfo {editor} {\bibfnamefont {A.}~\bibnamefont {Szabo}}\ and\ \bibinfo
  {editor} {\bibfnamefont {N.}~\bibnamefont {Ostlund}},\ eds.,\ \href@noop {}
  {\emph {\bibinfo {title} {Modern Quantum Chemistry}}}\ (\bibinfo  {publisher}
  {Dover Publishing, Mineola, New York},\ \bibinfo {year} {1996})\BibitemShut
  {NoStop}%
\bibitem [{\citenamefont {Reed}\ and\ \citenamefont {Simon}(1980)}]{Reed}%
  \BibitemOpen
  \bibfield  {author} {\bibinfo {author} {\bibfnamefont {M.}~\bibnamefont
  {Reed}}\ and\ \bibinfo {author} {\bibfnamefont {B.}~\bibnamefont {Simon}},\
  }\href@noop {} {\emph {\bibinfo {title} {Methods of modern mathematical
  physics: Functional analysis}}},\ Vol.~\bibinfo {volume} {1}\ (\bibinfo
  {publisher} {Academic Press, Inc.},\ \bibinfo {year} {1980})\ Chap.~\bibinfo
  {chapter} {2}\BibitemShut {NoStop}%
\bibitem [{\citenamefont {Paldus}\ and\ \citenamefont
  {{\v{C}}{\'\i}{\v{z}}ek}(1975)}]{PaldusAQC75}%
  \BibitemOpen
  \bibfield  {author} {\bibinfo {author} {\bibfnamefont {J.}~\bibnamefont
  {Paldus}}\ and\ \bibinfo {author} {\bibfnamefont {J.}~\bibnamefont
  {{\v{C}}{\'\i}{\v{z}}ek}},\ }\href@noop {} {\bibfield  {journal} {\bibinfo
  {journal} {Adv. Quantum. Chem.}\ }\textbf {\bibinfo {volume} {9}},\ \bibinfo
  {pages} {105} (\bibinfo {year} {1975})}\BibitemShut {NoStop}%
\bibitem [{\citenamefont {Olsen}\ \emph {et~al.}(1988)\citenamefont {Olsen},
  \citenamefont {Roos}, \citenamefont {Jo̸rgensen},\ and\ \citenamefont
  {Jensen}}]{Olsen1988}%
  \BibitemOpen
  \bibfield  {author} {\bibinfo {author} {\bibfnamefont {J.}~\bibnamefont
  {Olsen}}, \bibinfo {author} {\bibfnamefont {B.~O.}\ \bibnamefont {Roos}},
  \bibinfo {author} {\bibfnamefont {P.}~\bibnamefont {Jo̸rgensen}}, \ and\
  \bibinfo {author} {\bibfnamefont {H.~J.~A.}\ \bibnamefont {Jensen}},\ }\href
  {\doibase 10.1063/1.455063} {\bibfield  {journal} {\bibinfo  {journal} {J.
  Chem. Phys.}\ }\textbf {\bibinfo {volume} {89}},\ \bibinfo {pages} {2185}
  (\bibinfo {year} {1988})}\BibitemShut {NoStop}%
\bibitem [{\citenamefont {Freeman}(1978)}]{Freeman1978}%
  \BibitemOpen
  \bibfield  {author} {\bibinfo {author} {\bibfnamefont {D.}~\bibnamefont
  {Freeman}},\ }\href {\doibase 10.1016/0038-1098(78)91095-5} {\bibfield
  {journal} {\bibinfo  {journal} {Solid State Commun.}\ }\textbf {\bibinfo
  {volume} {26}},\ \bibinfo {pages} {289} (\bibinfo {year} {1978})}\BibitemShut
  {NoStop}%
\bibitem [{\citenamefont {Purvis}\ and\ \citenamefont
  {Bartlett}(1982)}]{Purvis1982}%
  \BibitemOpen
  \bibfield  {author} {\bibinfo {author} {\bibfnamefont {G.~D.}\ \bibnamefont
  {Purvis}}\ and\ \bibinfo {author} {\bibfnamefont {R.~J.}\ \bibnamefont
  {Bartlett}},\ }\href {\doibase 10.1063/1.443164} {\bibfield  {journal}
  {\bibinfo  {journal} {J. Chem. Phys.}\ }\textbf {\bibinfo {volume} {76}},\
  \bibinfo {pages} {1910} (\bibinfo {year} {1982})}\BibitemShut {NoStop}%
\bibitem [{\citenamefont {Noga}\ and\ \citenamefont
  {Bartlett}(1987)}]{Noga1987}%
  \BibitemOpen
  \bibfield  {author} {\bibinfo {author} {\bibfnamefont {J.}~\bibnamefont
  {Noga}}\ and\ \bibinfo {author} {\bibfnamefont {R.~J.}\ \bibnamefont
  {Bartlett}},\ }\href {\doibase 10.1063/1.455742} {\bibfield  {journal}
  {\bibinfo  {journal} {J. Chem. Phys.}\ }\textbf {\bibinfo {volume} {86}},\
  \bibinfo {pages} {7041} (\bibinfo {year} {1987})}\BibitemShut {NoStop}%
\bibitem [{\citenamefont {Kucharski}\ and\ \citenamefont
  {Bartlett}(1991)}]{Kucharski1991}%
  \BibitemOpen
  \bibfield  {author} {\bibinfo {author} {\bibfnamefont {S.~A.}\ \bibnamefont
  {Kucharski}}\ and\ \bibinfo {author} {\bibfnamefont {R.~J.}\ \bibnamefont
  {Bartlett}},\ }\href {\doibase 10.1007/BF01117419} {\bibfield  {journal}
  {\bibinfo  {journal} {Theor. Chim. Acta}\ }\textbf {\bibinfo {volume} {80}},\
  \bibinfo {pages} {387} (\bibinfo {year} {1991})}\BibitemShut {NoStop}%
\bibitem [{\citenamefont {Oliphant}\ and\ \citenamefont
  {Adamowicz}(1991)}]{Oliphant1991}%
  \BibitemOpen
  \bibfield  {author} {\bibinfo {author} {\bibfnamefont {N.}~\bibnamefont
  {Oliphant}}\ and\ \bibinfo {author} {\bibfnamefont {L.}~\bibnamefont
  {Adamowicz}},\ }\href {\doibase 10.1063/1.461534} {\bibfield  {journal}
  {\bibinfo  {journal} {J. Chem. Phys.}\ }\textbf {\bibinfo {volume} {95}},\
  \bibinfo {pages} {6645} (\bibinfo {year} {1991})}\BibitemShut {NoStop}%
\bibitem [{\citenamefont {Kucharski}\ and\ \citenamefont
  {Bartlett}(1992)}]{Kucharski1992}%
  \BibitemOpen
  \bibfield  {author} {\bibinfo {author} {\bibfnamefont {S.~A.}\ \bibnamefont
  {Kucharski}}\ and\ \bibinfo {author} {\bibfnamefont {R.~J.}\ \bibnamefont
  {Bartlett}},\ }\href {\doibase 10.1063/1.463930} {\bibfield  {journal}
  {\bibinfo  {journal} {J. Chem. Phys.}\ }\textbf {\bibinfo {volume} {97}},\
  \bibinfo {pages} {4282} (\bibinfo {year} {1992})}\BibitemShut {NoStop}%
\bibitem [{\citenamefont {Raghavachari}\ \emph {et~al.}(1989)\citenamefont
  {Raghavachari}, \citenamefont {Trucks}, \citenamefont {Pople},\ and\
  \citenamefont {Head-Gordon}}]{Raghavachari1989}%
  \BibitemOpen
  \bibfield  {author} {\bibinfo {author} {\bibfnamefont {K.}~\bibnamefont
  {Raghavachari}}, \bibinfo {author} {\bibfnamefont {G.~W.}\ \bibnamefont
  {Trucks}}, \bibinfo {author} {\bibfnamefont {J.~A.}\ \bibnamefont {Pople}}, \
  and\ \bibinfo {author} {\bibfnamefont {M.}~\bibnamefont {Head-Gordon}},\
  }\href {\doibase 10.1016/j.cplett.2013.08.064} {\bibfield  {journal}
  {\bibinfo  {journal} {Chem. Phys. Lett.}\ }\textbf {\bibinfo {volume}
  {589}},\ \bibinfo {pages} {37} (\bibinfo {year} {1989})}\BibitemShut
  {NoStop}%
\bibitem [{\citenamefont {M{\o}ller}\ and\ \citenamefont
  {Plesset}(1934)}]{Møller1934}%
  \BibitemOpen
  \bibfield  {author} {\bibinfo {author} {\bibfnamefont {C.}~\bibnamefont
  {M{\o}ller}}\ and\ \bibinfo {author} {\bibfnamefont {M.~S.}\ \bibnamefont
  {Plesset}},\ }\href {\doibase 10.1103/PhysRev.46.618} {\bibfield  {journal}
  {\bibinfo  {journal} {Phys.}\ }\textbf {\bibinfo {volume} {46}},\ \bibinfo
  {pages} {618} (\bibinfo {year} {1934})}\BibitemShut {NoStop}%
\bibitem [{\citenamefont {Krishnan}\ \emph {et~al.}(1980)\citenamefont
  {Krishnan}, \citenamefont {Frisch},\ and\ \citenamefont
  {Pople}}]{Krishnan1980}%
  \BibitemOpen
  \bibfield  {author} {\bibinfo {author} {\bibfnamefont {R.}~\bibnamefont
  {Krishnan}}, \bibinfo {author} {\bibfnamefont {M.~J.}\ \bibnamefont
  {Frisch}}, \ and\ \bibinfo {author} {\bibfnamefont {J.~A.}\ \bibnamefont
  {Pople}},\ }\href {\doibase 10.1063/1.439657} {\bibfield  {journal} {\bibinfo
   {journal} {J. Chem. Phys.}\ }\textbf {\bibinfo {volume} {72}},\ \bibinfo
  {pages} {4244} (\bibinfo {year} {1980})}\BibitemShut {NoStop}%
\bibitem [{\citenamefont {Kucharski}\ \emph {et~al.}(1989)\citenamefont
  {Kucharski}, \citenamefont {Noga},\ and\ \citenamefont
  {Bartlett}}]{Kucharski1989}%
  \BibitemOpen
  \bibfield  {author} {\bibinfo {author} {\bibfnamefont {S.~A.}\ \bibnamefont
  {Kucharski}}, \bibinfo {author} {\bibfnamefont {J.}~\bibnamefont {Noga}}, \
  and\ \bibinfo {author} {\bibfnamefont {R.~J.}\ \bibnamefont {Bartlett}},\
  }\href {\doibase 10.1063/1.456206} {\bibfield  {journal} {\bibinfo  {journal}
  {J. Chem. Phys.}\ }\textbf {\bibinfo {volume} {90}},\ \bibinfo {pages} {7282}
  (\bibinfo {year} {1989})}\BibitemShut {NoStop}%
\bibitem [{\citenamefont {Jeziorski}\ and\ \citenamefont
  {Monkhorst}(1981)}]{Jeziorski1981}%
  \BibitemOpen
  \bibfield  {author} {\bibinfo {author} {\bibfnamefont {B.}~\bibnamefont
  {Jeziorski}}\ and\ \bibinfo {author} {\bibfnamefont {H.~J.}\ \bibnamefont
  {Monkhorst}},\ }\href {\doibase 10.1103/PhysRevA.24.1668} {\bibfield
  {journal} {\bibinfo  {journal} {Phys. Rev. A}\ }\textbf {\bibinfo {volume}
  {24}},\ \bibinfo {pages} {1668} (\bibinfo {year} {1981})}\BibitemShut
  {NoStop}%
\bibitem [{\citenamefont {Jeziorski}\ and\ \citenamefont
  {Moszynski}(1993)}]{Jeziorski1993}%
  \BibitemOpen
  \bibfield  {author} {\bibinfo {author} {\bibfnamefont {B.}~\bibnamefont
  {Jeziorski}}\ and\ \bibinfo {author} {\bibfnamefont {R.}~\bibnamefont
  {Moszynski}},\ }\href {\doibase 10.1002/qua.560480303} {\bibfield  {journal}
  {\bibinfo  {journal} {Int. J. Quantum Chem.}\ }\textbf {\bibinfo {volume}
  {48}},\ \bibinfo {pages} {161} (\bibinfo {year} {1993})}\BibitemShut
  {NoStop}%
\bibitem [{\citenamefont {Przybytek}()}]{HECTOR}%
  \BibitemOpen
  \bibfield  {author} {\bibinfo {author} {\bibfnamefont {M.}~\bibnamefont
  {Przybytek}},\ }\href@noop {} {\enquote {\bibinfo {title} {{FCI program
  HECTOR; University of Warsaw, (unpublished)}},}\ }\BibitemShut {NoStop}%
\bibitem [{\citenamefont {Stanton}\ \emph {et~al.}()\citenamefont {Stanton},
  \citenamefont {Gauss}, \citenamefont {Watts}, \citenamefont {Lauderdale},\
  and\ \citenamefont {Bartlett}}]{ACESII}%
  \BibitemOpen
  \bibfield  {author} {\bibinfo {author} {\bibfnamefont {J.~F.}\ \bibnamefont
  {Stanton}}, \bibinfo {author} {\bibfnamefont {J.}~\bibnamefont {Gauss}},
  \bibinfo {author} {\bibfnamefont {J.~D.}\ \bibnamefont {Watts}}, \bibinfo
  {author} {\bibfnamefont {W.~J.}\ \bibnamefont {Lauderdale}}, \ and\ \bibinfo
  {author} {\bibfnamefont {R.~J.}\ \bibnamefont {Bartlett}},\ }\href@noop {}
  {\enquote {\bibinfo {title} {{ACES II Program System Release 2.0 QTP;
  University of Florida: Gainesville, FL, 1994.}}}\ }\BibitemShut {NoStop}%
\bibitem [{\citenamefont {Schwartz}(1962)}]{Schwartz1962}%
  \BibitemOpen
  \bibfield  {author} {\bibinfo {author} {\bibfnamefont {C.}~\bibnamefont
  {Schwartz}},\ }\href {\doibase 10.1103/PhysRev.126.1015} {\bibfield
  {journal} {\bibinfo  {journal} {Phys. Rev.}\ }\textbf {\bibinfo {volume}
  {126}},\ \bibinfo {pages} {1015} (\bibinfo {year} {1962})}\BibitemShut
  {NoStop}%
\bibitem [{\citenamefont {Cencek}\ \emph {et~al.}(2012)\citenamefont {Cencek},
  \citenamefont {Przybytek}, \citenamefont {Komasa}, \citenamefont {Mehl},
  \citenamefont {Jeziorski},\ and\ \citenamefont {Szalewicz}}]{Cencek2012}%
  \BibitemOpen
  \bibfield  {author} {\bibinfo {author} {\bibfnamefont {W.}~\bibnamefont
  {Cencek}}, \bibinfo {author} {\bibfnamefont {M.}~\bibnamefont {Przybytek}},
  \bibinfo {author} {\bibfnamefont {J.}~\bibnamefont {Komasa}}, \bibinfo
  {author} {\bibfnamefont {J.~B.}\ \bibnamefont {Mehl}}, \bibinfo {author}
  {\bibfnamefont {B.}~\bibnamefont {Jeziorski}}, \ and\ \bibinfo {author}
  {\bibfnamefont {K.}~\bibnamefont {Szalewicz}},\ }\href {\doibase
  10.1063/1.4712218} {\bibfield  {journal} {\bibinfo  {journal} {J. Chem.
  Phys.}\ }\textbf {\bibinfo {volume} {136}},\ \bibinfo {pages} {224303}
  (\bibinfo {year} {2012})}\BibitemShut {NoStop}%
\bibitem [{\citenamefont {Kutzelnigg}\ and\ \citenamefont {{Morgan
  III}}(1992)}]{Kutzelnigg1992}%
  \BibitemOpen
  \bibfield  {author} {\bibinfo {author} {\bibfnamefont {W.}~\bibnamefont
  {Kutzelnigg}}\ and\ \bibinfo {author} {\bibfnamefont {J.~D.}\ \bibnamefont
  {{Morgan III}}},\ }\href {\doibase 10.1063/1.463358} {\bibfield  {journal}
  {\bibinfo  {journal} {J. Chem. Phys.}\ }\textbf {\bibinfo {volume} {96}},\
  \bibinfo {pages} {4484} (\bibinfo {year} {1992})}\BibitemShut {NoStop}%
\bibitem [{\citenamefont {Schmidt}\ and\ \citenamefont
  {Hirschhausen}(1983)}]{Schmidt1983}%
  \BibitemOpen
  \bibfield  {author} {\bibinfo {author} {\bibfnamefont {H.~M.}\ \bibnamefont
  {Schmidt}}\ and\ \bibinfo {author} {\bibfnamefont {H.~v.}\ \bibnamefont
  {Hirschhausen}},\ }\href {\doibase 10.1103/PhysRevA.28.3179} {\bibfield
  {journal} {\bibinfo  {journal} {Phys. Rev. A}\ }\textbf {\bibinfo {volume}
  {28}},\ \bibinfo {pages} {3179} (\bibinfo {year} {1983})}\BibitemShut
  {NoStop}%
\bibitem [{\citenamefont {King}(1996)}]{King1996}%
  \BibitemOpen
  \bibfield  {author} {\bibinfo {author} {\bibfnamefont {H.}~\bibnamefont
  {King}},\ }\href {\doibase 10.1007/BF00186448} {\bibfield  {journal}
  {\bibinfo  {journal} {Theor. Chim. Acta}\ }\textbf {\bibinfo {volume} {94}},\
  \bibinfo {pages} {345} (\bibinfo {year} {1996})}\BibitemShut {NoStop}%
\bibitem [{\citenamefont {Fischer}\ and\ \citenamefont
  {Parish}(2013)}]{Fischer2014}%
  \BibitemOpen
  \bibfield  {author} {\bibinfo {author} {\bibfnamefont {A.~M.}\ \bibnamefont
  {Fischer}}\ and\ \bibinfo {author} {\bibfnamefont {M.~M.}\ \bibnamefont
  {Parish}},\ }\href {\doibase 10.1103/PhysRevA.88.023612} {\bibfield
  {journal} {\bibinfo  {journal} {Phys. Rev. A}\ }\textbf {\bibinfo {volume}
  {88}},\ \bibinfo {pages} {023612} (\bibinfo {year} {2013})}\BibitemShut
  {NoStop}%
\bibitem [{\citenamefont {Rotureau}(2013)}]{Rotureau2013}%
  \BibitemOpen
  \bibfield  {author} {\bibinfo {author} {\bibfnamefont {J.}~\bibnamefont
  {Rotureau}},\ }\href {\doibase 10.1140/epjd/e2003-00036-6} {\bibfield
  {journal} {\bibinfo  {journal} {Eur. Phys. J. D}\ }\textbf {\bibinfo {volume}
  {67}},\ \bibinfo {pages} {1} (\bibinfo {year} {2013})}\BibitemShut {NoStop}%
\bibitem [{\citenamefont {Christiansen}\ \emph {et~al.}(1996)\citenamefont
  {Christiansen}, \citenamefont {Olsen}, \citenamefont {J{\o}rgensen},
  \citenamefont {Koch},\ and\ \citenamefont {Malmqvist}}]{Koch1996}%
  \BibitemOpen
  \bibfield  {author} {\bibinfo {author} {\bibfnamefont {O.}~\bibnamefont
  {Christiansen}}, \bibinfo {author} {\bibfnamefont {J.}~\bibnamefont {Olsen}},
  \bibinfo {author} {\bibfnamefont {P.}~\bibnamefont {J{\o}rgensen}}, \bibinfo
  {author} {\bibfnamefont {H.}~\bibnamefont {Koch}}, \ and\ \bibinfo {author}
  {\bibfnamefont {P.-{\AA}.}\ \bibnamefont {Malmqvist}},\ }\href {\doibase
  http://dx.doi.org/10.1016/0009-2614(96)00974-8} {\bibfield  {journal}
  {\bibinfo  {journal} {Chem. Phys. Lett.}\ }\textbf {\bibinfo {volume}
  {261}},\ \bibinfo {pages} {369 } (\bibinfo {year} {1996})}\BibitemShut
  {NoStop}%
\bibitem [{\citenamefont {Olsen}\ \emph {et~al.}(1996)\citenamefont {Olsen},
  \citenamefont {Christiansen}, \citenamefont {Koch},\ and\ \citenamefont
  {J{\o}rgensen}}]{Olsen1996}%
  \BibitemOpen
  \bibfield  {author} {\bibinfo {author} {\bibfnamefont {J.}~\bibnamefont
  {Olsen}}, \bibinfo {author} {\bibfnamefont {O.}~\bibnamefont {Christiansen}},
  \bibinfo {author} {\bibfnamefont {H.}~\bibnamefont {Koch}}, \ and\ \bibinfo
  {author} {\bibfnamefont {P.}~\bibnamefont {J{\o}rgensen}},\ }\href {\doibase
  10.1063/1.472352} {\bibfield  {journal} {\bibinfo  {journal} {J. Chem.
  Phys.}\ }\textbf {\bibinfo {volume} {105}},\ \bibinfo {pages} {5082}
  (\bibinfo {year} {1996})}\BibitemShut {NoStop}%
\bibitem [{\citenamefont {Leininger}\ \emph {et~al.}(2000)\citenamefont
  {Leininger}, \citenamefont {Allen}, \citenamefont {Schaefer},\ and\
  \citenamefont {Sherrill}}]{Leininger2000}%
  \BibitemOpen
  \bibfield  {author} {\bibinfo {author} {\bibfnamefont {M.~L.}\ \bibnamefont
  {Leininger}}, \bibinfo {author} {\bibfnamefont {W.~D.}\ \bibnamefont
  {Allen}}, \bibinfo {author} {\bibfnamefont {H.~F.}\ \bibnamefont {Schaefer}},
  \ and\ \bibinfo {author} {\bibfnamefont {C.~D.}\ \bibnamefont {Sherrill}},\
  }\href {\doibase 10.1063/1.481764} {\bibfield  {journal} {\bibinfo  {journal}
  {J. Chem. Phys.}\ }\textbf {\bibinfo {volume} {112}},\ \bibinfo {pages}
  {9213} (\bibinfo {year} {2000})}\BibitemShut {NoStop}%
\bibitem [{\citenamefont {\'{C}wiok}\ \emph
  {et~al.}(1992{\natexlab{a}})\citenamefont {\'{C}wiok}, \citenamefont
  {Jeziorski}, \citenamefont {Ko{\l}os}, \citenamefont {Moszynski},
  \citenamefont {Rychlewski},\ and\ \citenamefont {Szalewicz}}]{Cwiok1992}%
  \BibitemOpen
  \bibfield  {author} {\bibinfo {author} {\bibfnamefont {T.}~\bibnamefont
  {\'{C}wiok}}, \bibinfo {author} {\bibfnamefont {B.}~\bibnamefont
  {Jeziorski}}, \bibinfo {author} {\bibfnamefont {W.}~\bibnamefont {Ko{\l}os}},
  \bibinfo {author} {\bibfnamefont {R.}~\bibnamefont {Moszynski}}, \bibinfo
  {author} {\bibfnamefont {J.}~\bibnamefont {Rychlewski}}, \ and\ \bibinfo
  {author} {\bibfnamefont {K.}~\bibnamefont {Szalewicz}},\ }\href {\doibase
  10.1016/0009-2614(92)85912-T} {\bibfield  {journal} {\bibinfo  {journal}
  {Chem. Phys. Lett.}\ }\textbf {\bibinfo {volume} {195}},\ \bibinfo {pages}
  {67} (\bibinfo {year} {1992}{\natexlab{a}})}\BibitemShut {NoStop}%
\bibitem [{\citenamefont {\'{C}wiok}\ \emph
  {et~al.}(1992{\natexlab{b}})\citenamefont {\'{C}wiok}, \citenamefont
  {Jeziorski}, \citenamefont {Ko{\l}os}, \citenamefont {Moszynski},\ and\
  \citenamefont {Szalewicz}}]{Cwiok1992a}%
  \BibitemOpen
  \bibfield  {author} {\bibinfo {author} {\bibfnamefont {T.}~\bibnamefont
  {\'{C}wiok}}, \bibinfo {author} {\bibfnamefont {B.}~\bibnamefont
  {Jeziorski}}, \bibinfo {author} {\bibfnamefont {W.}~\bibnamefont {Ko{\l}os}},
  \bibinfo {author} {\bibfnamefont {R.}~\bibnamefont {Moszynski}}, \ and\
  \bibinfo {author} {\bibfnamefont {K.}~\bibnamefont {Szalewicz}},\ }\href
  {\doibase 10.1063/1.463475} {\bibfield  {journal} {\bibinfo  {journal} {J.
  Chem. Phys.}\ }\textbf {\bibinfo {volume} {97}},\ \bibinfo {pages} {7555}
  (\bibinfo {year} {1992}{\natexlab{b}})}\BibitemShut {NoStop}%
\bibitem [{\citenamefont {Lesiuk}\ \emph {et~al.}(2015)\citenamefont {Lesiuk},
  \citenamefont {Przybytek}, \citenamefont {Musia{\l}}, \citenamefont
  {Jeziorski},\ and\ \citenamefont {Moszynski}}]{Lesiuk2015}%
  \BibitemOpen
  \bibfield  {author} {\bibinfo {author} {\bibfnamefont {M.}~\bibnamefont
  {Lesiuk}}, \bibinfo {author} {\bibfnamefont {M.}~\bibnamefont {Przybytek}},
  \bibinfo {author} {\bibfnamefont {M.}~\bibnamefont {Musia{\l}}}, \bibinfo
  {author} {\bibfnamefont {B.}~\bibnamefont {Jeziorski}}, \ and\ \bibinfo
  {author} {\bibfnamefont {R.}~\bibnamefont {Moszynski}},\ }\href {\doibase
  10.1103/PhysRevA.91.012510} {\bibfield  {journal} {\bibinfo  {journal} {Phys.
  Rev. A}\ }\textbf {\bibinfo {volume} {91}},\ \bibinfo {pages} {012510}
  (\bibinfo {year} {2015})}\BibitemShut {NoStop}%
\bibitem [{\citenamefont {Grining}\ \emph {et~al.}(2015)\citenamefont
  {Grining}, \citenamefont {Tomza}, \citenamefont {Lesiuk}, \citenamefont
  {Przybytek}, \citenamefont {Musia\l{}}, \citenamefont {Moszynski},
  \citenamefont {Lewenstein},\ and\ \citenamefont {Massignan}}]{Grining2015a}%
  \BibitemOpen
  \bibfield  {author} {\bibinfo {author} {\bibfnamefont {T.}~\bibnamefont
  {Grining}}, \bibinfo {author} {\bibfnamefont {M.}~\bibnamefont {Tomza}},
  \bibinfo {author} {\bibfnamefont {M.}~\bibnamefont {Lesiuk}}, \bibinfo
  {author} {\bibfnamefont {M.}~\bibnamefont {Przybytek}}, \bibinfo {author}
  {\bibfnamefont {M.}~\bibnamefont {Musia\l{}}}, \bibinfo {author}
  {\bibfnamefont {R.}~\bibnamefont {Moszynski}}, \bibinfo {author}
  {\bibfnamefont {M.}~\bibnamefont {Lewenstein}}, \ and\ \bibinfo {author}
  {\bibfnamefont {P.}~\bibnamefont {Massignan}},\ }\href {\doibase
  10.1103/PhysRevA.92.061601} {\bibfield  {journal} {\bibinfo  {journal} {Phys.
  Rev. A}\ }\textbf {\bibinfo {volume} {92}},\ \bibinfo {pages} {061601}
  (\bibinfo {year} {2015})}\BibitemShut {NoStop}%
\bibitem [{\citenamefont {Mitroy}\ \emph {et~al.}(2013)\citenamefont {Mitroy},
  \citenamefont {Bubin}, \citenamefont {Horiuchi}, \citenamefont {Suzuki},
  \citenamefont {Adamowicz}, \citenamefont {Cencek}, \citenamefont {Szalewicz},
  \citenamefont {Komasa}, \citenamefont {Blume},\ and\ \citenamefont
  {Varga}}]{Mitroy2013}%
  \BibitemOpen
  \bibfield  {author} {\bibinfo {author} {\bibfnamefont {J.}~\bibnamefont
  {Mitroy}}, \bibinfo {author} {\bibfnamefont {S.}~\bibnamefont {Bubin}},
  \bibinfo {author} {\bibfnamefont {W.}~\bibnamefont {Horiuchi}}, \bibinfo
  {author} {\bibfnamefont {Y.}~\bibnamefont {Suzuki}}, \bibinfo {author}
  {\bibfnamefont {L.}~\bibnamefont {Adamowicz}}, \bibinfo {author}
  {\bibfnamefont {W.}~\bibnamefont {Cencek}}, \bibinfo {author} {\bibfnamefont
  {K.}~\bibnamefont {Szalewicz}}, \bibinfo {author} {\bibfnamefont
  {J.}~\bibnamefont {Komasa}}, \bibinfo {author} {\bibfnamefont
  {D.}~\bibnamefont {Blume}}, \ and\ \bibinfo {author} {\bibfnamefont
  {K.}~\bibnamefont {Varga}},\ }\href {\doibase 10.1103/RevModPhys.85.693}
  {\bibfield  {journal} {\bibinfo  {journal} {Rev. Mod. Phys.}\ }\textbf
  {\bibinfo {volume} {85}},\ \bibinfo {pages} {693} (\bibinfo {year}
  {2013})}\BibitemShut {NoStop}%
\bibitem [{\citenamefont {Carroll}\ \emph {et~al.}(1979)\citenamefont
  {Carroll}, \citenamefont {Silverstone},\ and\ \citenamefont
  {Metzger}}]{Carroll1979}%
  \BibitemOpen
  \bibfield  {author} {\bibinfo {author} {\bibfnamefont {D.~P.}\ \bibnamefont
  {Carroll}}, \bibinfo {author} {\bibfnamefont {H.~J.}\ \bibnamefont
  {Silverstone}}, \ and\ \bibinfo {author} {\bibfnamefont {R.~M.}\ \bibnamefont
  {Metzger}},\ }\href {\doibase 10.1063/1.438187} {\bibfield  {journal}
  {\bibinfo  {journal} {J. Chem. Phys.}\ }\textbf {\bibinfo {volume} {71}},\
  \bibinfo {pages} {4142} (\bibinfo {year} {1979})}\BibitemShut {NoStop}%
\end{thebibliography}%

\end{document}